\documentclass[fleqn,usenatbib]{mnras}

\usepackage{newtxtext,newtxmath}

\usepackage[T1]{fontenc}
\usepackage{ae,aecompl}

\usepackage{graphicx}	
\usepackage{amsmath}	
\usepackage{amssymb}	
\usepackage{hyperref}

\newcommand{\brooks}{g14 and Marvel/Justice League }
\newcommand{\updated}[1]{#1}

\title[SFH variability in different models]{The Diversity and Variability of Star Formation Histories in Models of Galaxy Evolution}

\author[K. G. Iyer et al.]{Kartheik G. Iyer,$^{1,2}$\thanks{E-mail: kartheik.iyer@dunlap.utoronto.ca}
Sandro Tacchella,$^{3}$\thanks{E-mail: sandro.tacchella@cfa.harvard.edu}
Shy Genel,$^{4,5}$\newauthor
Christopher C. Hayward,$^{4}$
Lars Hernquist,$^{3}$
Alyson M. Brooks,$^{1}$
Neven Caplar,$^{6}$\newauthor
Romeel Dav\'e,$^{7,8,9}$
Benedikt Diemer,$^{3,10}$
John C. Forbes,$^{4}$
Eric Gawiser,$^{1}$\newauthor
Rachel S. Somerville,$^{1,4}$
Tjitske K. Starkenburg,$^{4,11}$ \vspace{0.2cm}
\\
$^{1}$Dept. of Physics and Astronomy, Rutgers, The State University of New Jersey, 136 Frelinghuysen Road, Piscataway, NJ 08854, USA\\
$^{2}$Dunlap Institute for Astronomy and Astrophysics, University of Toronto, 50 St George St, Toronto, ON M5S 3H4, Canada\\
$^{3}$Center for Astrophysics | Harvard \& Smithsonian, 60 Garden St, Cambridge, MA 02138, USA\\
$^{4}$Center for Computational Astrophysics, Flatiron Institute, 162 5th Ave, New York, NY 10010, USA\\
$^{5}$Columbia Astrophysics Laboratory, Columbia University, 550 West 120th Street, New York, NY 10027, USA\\
$^{6}$Department of Astrophysical Sciences, Princeton University, 4 Ivy Ln., Princeton, NJ 08544, USA\\
$^{7}$South African Astronomical Observatories, Observatory, Cape Town 7925, South Africa\\
$^{8}$University of the Western Cape, Bellville, Cape Town 7535, South Africa\\
$^{9}$Institute for Astronomy, Royal Observatory, University of Edinburgh, Edinburgh EH9 3HJ, UK\\
$^{10}$NHFP Einstein Fellow, Department of Astronomy, University of Maryland, College Park, MD 20742, USA\\
$^{11}$Department of Physics \& Astronomy and CIERA, Northwestern University, 2145 Sheridan Rd., Evanston, IL 60208, USA\\
}

\pubyear{2020}

\begin{document}
\label{firstpage}
\pagerange{\pageref{firstpage}--\pageref{lastpage}}
\maketitle

\begin{abstract}
Understanding the variability of galaxy star formation histories (SFHs) across a range of timescales provides insight into the underlying physical processes that regulate star formation within galaxies.
We compile the SFHs of galaxies at $z=0$ from an extensive set of models, ranging from cosmological hydrodynamical simulations (Illustris, IllustrisTNG, Mufasa, Simba, EAGLE), zoom simulations (FIRE-2, g14, and Marvel/Justice League), semi-analytic models (Santa Cruz SAM) and empirical models (UniverseMachine), and quantify the variability of these SFHs on different timescales using the power spectral density (PSD) formalism.
We find that the PSDs are well described by broken power-laws, and variability on long timescales ($\gtrsim1$ Gyr) accounts for most of the power in galaxy SFHs.
Most hydrodynamical models show increased variability on shorter timescales ($\lesssim300$ Myr) with decreasing stellar mass.
Quenching can induce $\sim0.4-1$ dex of additional power on timescales $>1$ Gyr.
The dark matter accretion histories of galaxies
have remarkably self-similar PSDs and are coherent with the in-situ star formation on timescales $>3$ Gyr.
There is considerable diversity among the different models in their (i) power due to SFR variability at a given timescale, (ii) amount of correlation with adjacent timescales (PSD slope), (iii) evolution of median PSDs with stellar mass, and (iv) presence and locations of breaks in the PSDs.
The PSD framework is a useful space to study the SFHs of galaxies since model predictions vary widely. Observational constraints in this space will help constrain the relative strengths of the physical processes responsible for this variability.

\end{abstract}




\section{Introduction}

Galaxies in the observable universe show a remarkable diversity in their structure and properties. This diversity can be understood in the context of the many different pathways that exist for galaxies to form stars, grow and eventually cease their star formation (`quench').

The broad features of galaxy assembly have been found to correlate with the assembly of their dark matter haloes \citep[see reviews by][]{2018ARA&A..56..435W}.
Galaxy growth can happen through the smooth accretion of gas, through gas-rich and gas-poor mergers, and can be prolonged by inefficient star formation due to turbulence and feedback in the interstellar (ISM) and circum-galactic media (CGM) \citep{white1978core, somerville2008semi, tacchella2018redshift, behroozi2018universemachine}.
Galaxy quenching, on the other hand, involves mechanisms that either heat the gas in galaxies or remove it entirely, so that it can no longer form stars. The processes involved in this are thought to be a combination of `halo quenching', arising from halo gas being shock heated over time, stellar feedback,
winds from exploding supernovae,
thermal and kinetic feedback from Active Galactic Nuclei (AGN), as well as external factors such as mergers and interactions \citep{scannapieco2005agn, dekel2006galaxy, kaviraj2007uv, bell2008galaxy, kimm2009correlation, kerevs2009galaxies, woo2012dependence, bundy2008aegis,  weinberger2016simulating}. In between these states, galaxies are affected by the interplay of these different processes and are also found to rejuvenate after periods of relative quiescence \citep{fang2012slow, pandya2017massivetransition}.

\begin{figure*}
    \centering
    \includegraphics[width=\textwidth, trim={6cm 6cm 4cm 0cm},clip]{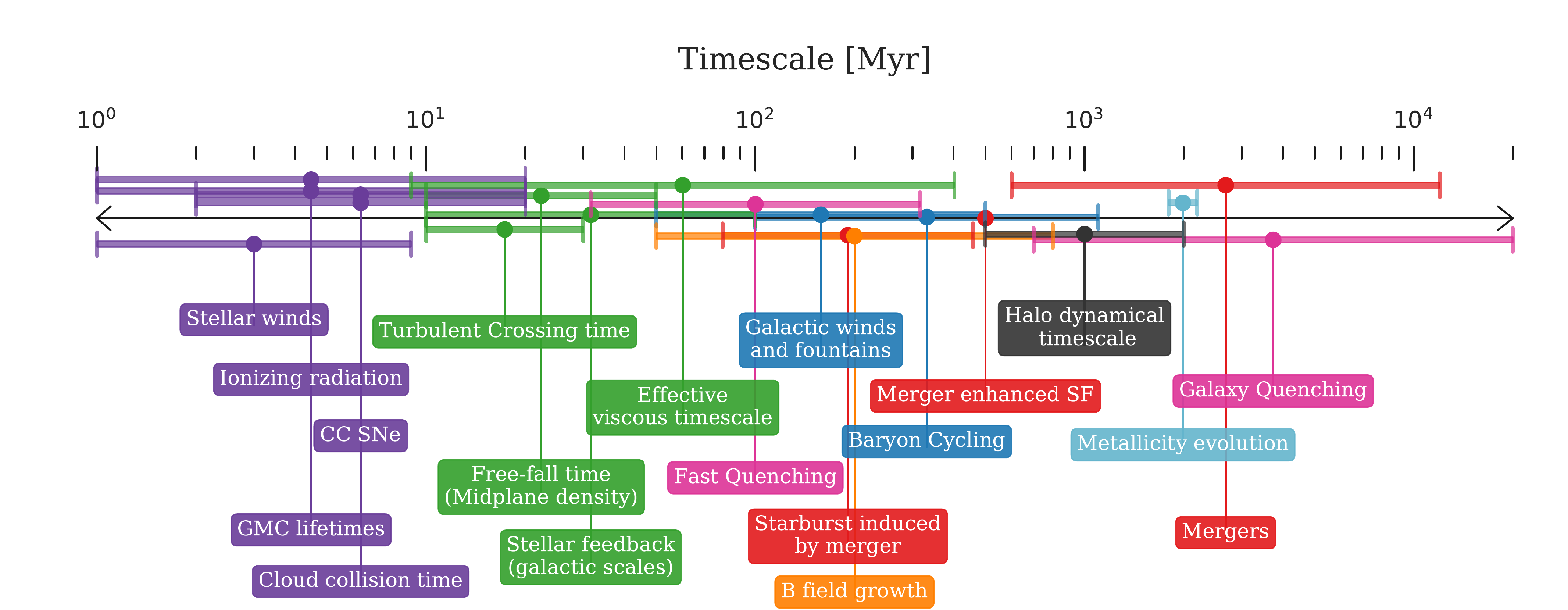}
    \caption{A summary of current estimates in the literature for the timescales on which different physical processes regulate the growth of galaxies.
    These timescales are estimated from theoretical models and simulations of galaxy evolution, with the corresponding references listed in Appendix \ref{sec:lit_timescales_table}.
    The different colours highlight the different scales and types of physical processes, ranging from processes regulating the creation and destruction of GMCs (purple; \citealt{starburst99, tan2000star, tasker2011star, faucher2017model, benincasa2019live}), dynamical processes within galaxies (green; \citealt{krumholz2010dynamics, hopkins2014galaxies, forbes2014origin, semenov2017physical}, the cycling of baryons in the ISM and CGM (blue; \citealt{marcolini2004three, angles2017cosmic}), the growth of magnetic fields (yellow; \citealt{hanasz2004amplification, pakmor2017magnetic}), metallicityi evolution (cyan; \citealt{torrey2018similar}), mergers and merger induced star formation (red; \citealt{robertson2006merger, jiang2008fitting, boylan2008dynamical, hani2020interacting}), environmental factors (grey; \citealt{mo2010galaxy, lilly2013gas}) and galaxy quenching (pink; \citealt{sales2015colours, nelson2018abundance, wright2019quenching, rodriguez2019mergers}).
    While the figure shows the large range of estimated timescales for different processes, it also encodes the diversity in the estimated timescales of individual processes (e.g., quenching timescales) across different models in the literature.
    While this is not an exhaustive list of timescales, it is intended to be a fairly representative subset of the range and diversity in current estimates.}
    \label{fig:literature_timescales}
\end{figure*}

The spatial and temporal scales involved in these processes differ by several orders of magnitude, and yet all we see in galaxy surveys are their cumulative effects on the entire observable population of galaxies at any given epoch. These processes act over timescales ranging from $< 1$ Myr to over a Hubble time, and can regulate star formation either locally within a giant molecular cloud or across the entire galaxy. Figure \ref{fig:literature_timescales} shows a summary of various physical processes and estimates of the timescales they are estimated to act upon in contemporary literature at $z\sim 0$. As the figure shows, while different physical processes are estimated to act over characteristic timescales, they can extend over multiple orders of magnitude and overlap with other processes. This enormously complicates the process of understanding the effect of any individual process on galaxy evolution. Even within a model where it is possible to turn a certain process off or modulate its strengths, the corresponding effects are difficult to generalize, and might change in response to other variables.

Explaining the observed diversity of galaxies today is thus one of the key challenges facing theories of how galaxies form and evolve. Since physical processes regulate star formation over characteristic timescales, it should be possible to study their effects on galaxy evolution using the imprints they leave on the star formation histories (SFHs) of galaxies. Specifically, studying the variability of galaxy SFHs over different timescales provides a useful space to quantify and understand the cumulative effects of different processes driving or suppressing star formation on that timescale.
The key open questions can therefore be phrased in terms of SFH variability on different timescales, as,
\begin{itemize}
    \item What drives the variability of galaxy SFHs on different timescales? Do different models of galaxy evolution predict different amounts of variability at a given timescale?
    \item Is there a relation between the variability on different timescales? How does this change as a function of galaxy properties?
\end{itemize}

This approach towards understanding galaxy evolution through timescales is particularly informative since the SFHs of galaxies contain a wealth of observationally accessible information about the timescales of mergers, of bursts of star formation and quenching, baryon cycling and short timescale burstiness\footnote{The term `burstiness' is loosely used to denote variability in SFR across a range of timescales in the literature \citep{weisz2011modeling, guo2016bursty, matthee2018origin, wang2020var}. With this in mind, we preface the term with an appropriate timescale range whenever used.} - relating them to the strengths of AGN and stellar feedback as well as the dark matter accretion histories of their parent galaxies. This information is encoded in the form of the overall SFH shape, as well as fluctuations on different timescales. The strength of SFH fluctuations on short timescales is tied to the formation and destruction of giant molecular clouds (GMCs) due to supernovae explosions, cosmic rays and photoionization feedback \citep{gnedin2008escape, parrish2009anisotropic, hopkins2014galaxies, faucher2017model, tacchella2020stochastic}. On intermediate timescales it is thought to arise from a variety of sources, like mergers, stellar winds, and AGN feedback \citep{mihos1994triggering, thomas1999probing, di2005energy, robertson2006fundamental, mcquinn2010nature, robaina2010merger, tacchella2016confinement}.
On the largest timescales it is dictated by the behaviour of their parent haloes, and by processes like AGN feedback that drive quenching \citep{scannapieco2005agn, kaviraj2007uv, bell2008galaxy, kimm2009correlation, woo2012dependence, bundy2008aegis,  weinberger2016simulating, angles2017black}.
While the longest and shortest timescales have been extensively studied in theory and have observational constraints, the strength of fluctuations on intermediate timescales remains of prime interest since they are difficult to constrain observationally and experience contributions from a variety of different processes with overlapping timescales.

As observations continue to grow in quality, techniques that reconstruct the SFHs from observations are able to extract more robust constraints on the SFHs of individual galaxies and ensembles (\citealt{pacifici}; \citealt{smith2015deriving}; \citealt{leja2017deriving}; \citealt{carnall2018inferring};  \citealt{iyer2019gpsfh}). We now approach the point where we can compare observational distributions of galaxy SFHs to those from simulations to obtain constraints on intermediate-to-long timescales. Performing this analysis for mass-complete samples across a range of redshifts will allow us to understand and constrain the strengths of the various feedback processes that regulate star formation within and across galaxies.

On shorter timescales, a multitude of papers study the `burstiness' of star formation \citep{weisz2011modeling, sparre2015star, dominguez2015consequences, guo2016bursty, sparre2017starbursts, emami2018closer, broussard2019stochastic, caplar2019sfhpsd, hahn2019constraining}.
There exist many definitions for burstiness in the literature, with most using some ratio of H$\alpha$ or UV-based SFR measurements, which are averaged over timescales of $\sim 4-10$ Myr and $\sim 20-100$ Myr respectively. Comparing distributions of SFRs measured using these two indicators affords a probe of the increase or decrease in the SFR over the recent past, with a distribution therefore affording a statistical view of how the galaxy population is behaving. However, such analysis is extremely difficult due to inherent uncertainties in SFR measurements, assumptions about the monotonicity of SFRs over different timescales, and degeneracies with a stochastic IMF and dust properties \citep{johnson2013measuring, shivaei2018mosdef}. \citet{caplar2019sfhpsd} undertook an effort to quantify the variability of the SFH on short, intermediate and long timescales by constraining the Power Spectral Density (PSD) from the scatter of the star-forming sequence (SFS). \citet{wang2020var, wang2020vartwo} complement this by using the PSD formalism to obtain constraints on the ratio of the burstiness of SFRs on 10 Myr to 1 Gyr timescales using resolved SDSS-IV MaNGA observations.

In this paper, we build on this to establish a framework for understanding the fluctuations in galaxy SFHs using the PSD formalism \citep{caplar2019sfhpsd}. The PSD at any timescale is a measure of the amount of power contained in SFR fluctuations on that timescale, and therefore encodes the variability or `burstiness' on that timescale. This provides us with a view of the relative power across different frequencies (and therefore across different timescales) in a galaxy's SFH, and therefore a first step towards tying the signatures in SFHs to the underlying physical implementations of feedback in the different models. Using this formalism, we compare the star formation histories of galaxies across different models, ranging from empirical models to full numerical magnetohydrodynamical (MHD) simulations. This is important towards understanding how the SFHs of galaxies may be affected by the input numerical methods, sub-grid prescriptions, and resolution effects, and can be seen in comparisons between different models that are calibrated to reproduce the same observations. In the current work we consider five cosmological hydrodynamical simulations (Illustris, \citealt{vogelsberger2014properties, vogelsberger2014introducing, genel2014introducing, nelson2015illustris}; IllustrisTNG, \citealt{pillepich2017simulating, weinberger2018supermassive, springel2018first, naiman2018first, marinacci2018first, nelson2019illustristng}; Mufasa, \citealt{dave2016Mufasa}; Simba, \citealt{dave2019Simba}; EAGLE,  \citealt{schaye2014Eagle, crain2015eagle, mcalpine2016eagle}), three suites of zoom simulations (FIRE-2, \citealt{hopkins2014galaxies, hopkins2018fire}; g14 \citet{governato2012cuspy, munshi2013reproducing, brooks2014baryons} and Marvel/Justice League \citealt[Munshi et al. \textit{in prep}.]{bellovary2019multimessenger}), a semi-analytic model (Santa Cruz SAM, \citealt{somerville2008semi, somerville2015star, yung2018semi, brennan2016relationship}) and an empirical model (UniverseMachine, \citealt{behroozi2013average, behroozi2018universemachine}).

While this paper introduces and applies the PSD formalism to galaxy SFHs from simulations, it is outside the scope of the current work to conclusively correlate PSD features with their underlying physical mechanisms. The main focus of this work lies in comparing PSDs of different models. Future work will examine individual models in more detail and introduce observational constraints in PSD space, using the full extent of observationally recoverable temporal information to validate and constrain theories of galaxy evolution.

Section \ref{sec:data_full} briefly describes the various models we consider in the current analysis, how we extract SFH information from these models and compute their PSDs.
Section \ref{sec:results} presents the the SFHs and corresponding PSDs of galaxies from the various models as a function of stellar mass. It also considers the effects of galaxy quenching on PSDs, and the ties between SFHs and the dark matter accretion histories of their parent haloes.
Section \ref{sec:discussion} ties the results from this paper to estimates from the current literature of the timescales on which physical processes affect galaxy growth, and sources of observational constraints in PSD space.
We summarize and conclude in Section \ref{sec:conclusions}. The appendices provide additional tailored validation tests for the shortest timescale that can be probed by the PSD of a simulation with a given resolution (Appendix \ref{sec:timescale_tests}), plot SFH parameters and covariances at $z\sim 0$ for the various models (Appendix \ref{sec:sfh_prior_dists}), and collect references for various timescales estimated in the literature (Appendix \ref{sec:lit_timescales_table}).

\section{Dataset and Methodology}
\label{sec:data_full}

In this section, we set ourselves up to compute the PSDs of galaxy SFHs from different models and provide context for interpreting them.
Section \ref{sec:data} describes the various models for galaxy evolution we consider in the current analysis.
Section \ref{sec:get_sfhs_from_models} describes how we extract SFHs from these models, and
Section \ref{sec:psd_description} describes the PSD and how we compute it.
Sections \ref{sec:timestep} and \ref{sec:validation_simple} address the problems due to quenching and discrete star particles in computing the PSDs for SFHs from hydrodynamical simulations, with a more detailed forward-modeling approach given in Appendix \ref{sec:timescale_tests}.

\subsection{Models simulating galaxy evolution}
\label{sec:data}

\begin{figure}
\begin{center}
\includegraphics[width=240px]{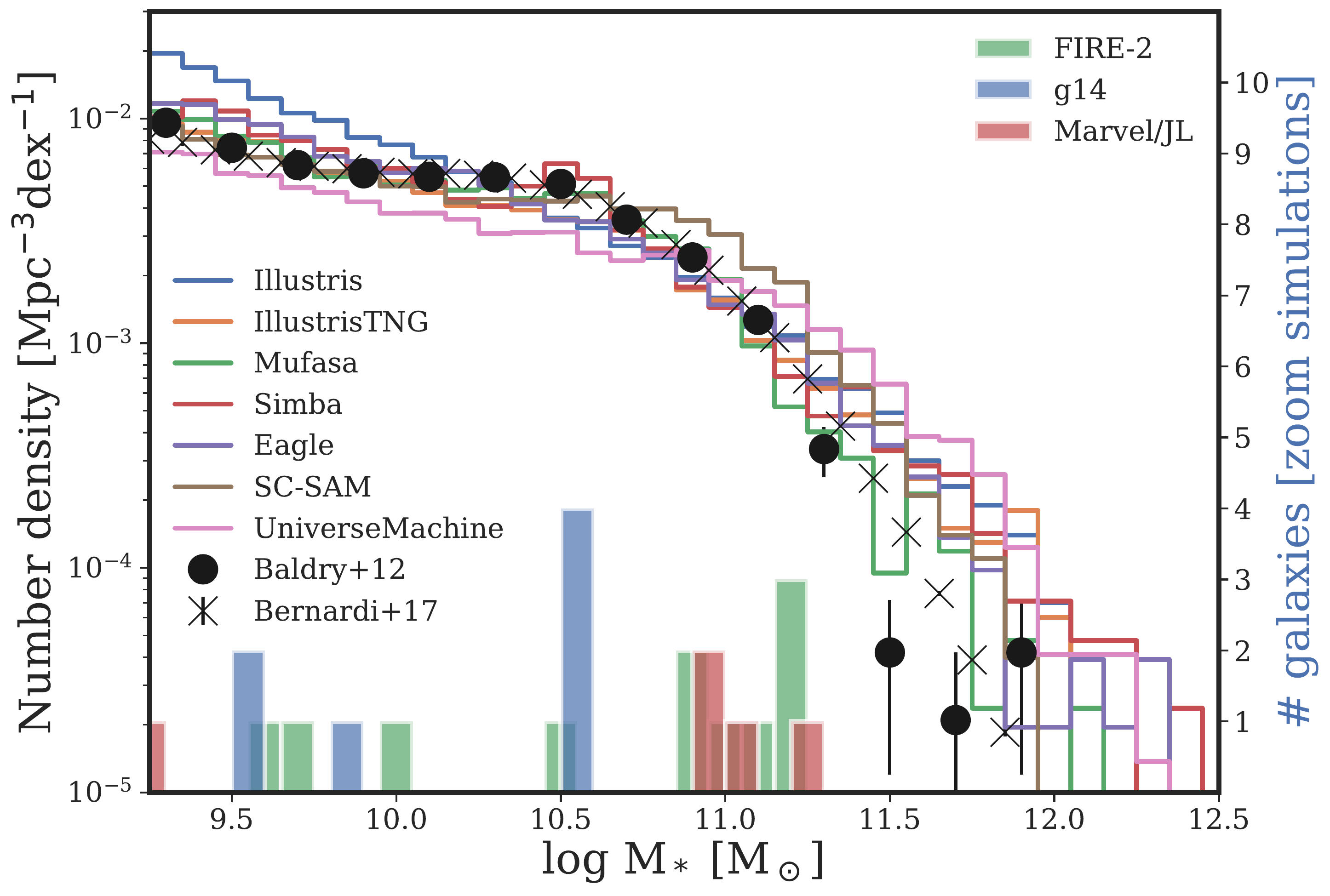}
\caption[[$z\sim 0$ Stellar Mass distributions from different cosmological models]{The stellar mass function of $z\sim 0$ galaxies from the large-volume models we consider: Illustris, IllustrisTNG, Mufasa, Simba, EAGLE, the Santa Cruz semi-analytic model, and UniverseMachine.
The black points with error bars provide a comparison to observations. The solid histograms in the bottom and the corresponding y-axis on the right show the distribution of stellar masses for the 14 galaxies from FIRE-2 (green), 8 galaxies from g14 (blue), and 5 galaxies from Marvel/Justice League (red) that we consider.}
\label{fig:mstar_allsims_z0}
\end{center}
\end{figure}

We consider the star formation histories from a wide range of galaxy evolution models, ranging from hydrodynamical simulations (Illustris, IllustrisTNG, Mufasa, Simba, EAGLE), a semi-analytic model (Santa-Cruz SAM), an empirical model tuned to match observations across a range of observations (UniverseMachine), and three suites of zoom simulations (FIRE-2, \brooks) with a higher resolution and more explicit prescriptions for the interstellar medium (ISM) and stellar feedback \citep[see reviews by][for a summary of the individual components of these various models]{somerville2015davereview, vogelsberger2020cosmological}.

For simplicity, in the current analysis we limit ourselves to (i) considering only a fiducial run from each model, since some models have multiple runs varying the parameters of various sub-grid recipes, (ii) considering the SFHs of only central galaxies above a stellar mass threshold of $10^9 $M$_\odot$ to partially mitigate resolution effects, and (iii) studying galaxies at $z\sim 0$, with model variants and redshift evolution to be considered in further work.
Figure \ref{fig:mstar_allsims_z0} shows the normalised distributions of stellar mass for galaxies from each model at $z\sim 0$, used in the current analysis.
The FIRE-2, g14 and Marvel/JL zoom simulations have a much smaller sample of 14, 8 and 5 galaxies, respectively, spanning a range of stellar masses from $\sim 10^9 $M$_\odot$ to $\sim 10^{11.5} $M$_\odot$. While these zoom simulations allow us to probe SFH fluctuations to shorter timescales compared to the large-volume models \updated{due to their much finer spatiotemporal resolution, which allows them to resolve GMC-scale structures and treat feedback more explicitly}, these galaxies are not representative of a cosmological sample.
Caution should therefore be employed in generalizing trends in their variability.

Each model of galaxy evolution is described briefly below, with references to relevant papers containing more detailed descriptions. Since the current analysis deals with galaxy SFHs, the descriptions focus on how each simulation implements star formation and feedback, and Table \ref{tab:sim_details} contains a summary of the resolution, box size and number of galaxies from each simulation.

\begin{itemize}
    \item \textbf{Illustris} \citep{vogelsberger2014properties, genel2014introducing}

    The Illustris project\footnote{\url{https://www.illustris-project.org/}} is a large-scale hydrodynamical simulation of galaxy formation using the moving mesh code AREPO \citep{springel2010moving}. The model includes recipes for primordial and metal-line cooling, stellar evolution and feedback, gas recycling, chemical enrichment, supermassive black hole (BH) growth and AGN feedback \citep{springel2003cosmological, vogelsberger2013model}.
    Given the spatial resolution of $\simeq 1$ kpc, giant molecular clouds are not resolved. A sub-resolution model for an effective equation-of-state
    \citep{springel2003cosmological}
    is implemented where a star particle is stochastically produced
    above the critical hydrogen number density of
    $n_{\mathrm{SF}} = 0.13 \mathrm{cm}^{-3}$
    on a density-dependent timescale that reproduces the observed Kennicutt-Schmidt relation \citep{schmidt1959rate, kennicutt1989star}.
    Star formation results in supernovae, which release kinetic winds that expel gas from their surroundings and chemically enrich the ISM. \updated{These winds are implemented by launching hydrodynamically decoupled `wind particles' that recouple to the gas when they leave the dense local ISM and reach a cell with a density $< 0.05 n_{\mathrm{SF}}$ \citep{springel2003cosmological, pillepich2017simulating}.
    This results in a non-local coupling of the stellar wind feedback to the gas, in contrast to the local feedback from AGN.}
    Feedback from AGN can be either thermal or kinetic, following the model of \citet{springel2005modelling, sijacki2007unified}.
    Galaxies in the simulation are quenched primarily due to radio mode feedback from AGN, with an expanding jet induced bubble transferring energy from the BH to the halo and heating the gas.
    Parameters of the Illustris model have been chosen to roughly reproduce the cosmic star formation rate density (SFRD), and the galaxy stellar mass function (SMF), the stellar mass-halo mass relation (SMHM), and the stellar mass-black hole mass relation (SMBH) at $z=0$.

    \item \textbf{IllustrisTNG} \citep{pillepich2017simulating, weinberger2016simulating}

    A significantly updated version of the original Illustris project, IllustrisTNG\footnote{\url{https://www.tng-project.org/}} carries over recipes for star formation and evolution, chemical enrichment, cooling, feedback with outflows, growth and multi-mode feedback from Illustris
    with substantial updates
    \citep{pillepich2017simulating, weinberger2016simulating, nelson2018first}. In addition to this, it incorporates new black hole driven kinetic feedback at low accretion rates, magnetohydrodynamics and improvements to the numerical scheme.
    Unlike Illustris, TNG injects winds isotropically with a modified wind speed that depends on the local 1D dark matter (DM) velocity dispersion, with a redshift dependence that matches the growth of the virial halo mass.
    AGN feedback is modeled using two modes: a pure thermal mode at high accretion rate (quasar mode) and a pure kinetic mode at low accretion rate (radio mode), with a kinetic wind feedback model  \citep{weinberger2016simulating} responsible for quenching galaxies \citep{weinberger2018supermassive}.
    In addition to the observations used with Illustris, the TNG simulation parameters are also chosen to reproduce galaxy sizes and halo gas fractions at $z=0$.

    \item \textbf{Mufasa} \citep{dave2016Mufasa}

    The Mufasa meshless hydrodynamic simulations
    use the GIZMO code \citep{hopkins2015new}, prescriptions for cooling and heating with Grackle \citep{smith2017grackle}, and star formation and feedback from massive stars using scalings from FIRE \citep{hopkins2014galaxies, muratov2015gusty}.
    Star formation is implemented stochastically from gas particles using the \citet{krumholz2009star} formalism to estimate the H$_2$ formation at coarse resolution accounting for sub-grid clumping. Then, for densities $\geq 0.13 \mathrm{cm}^{-3}$, stars are formed stochastically over local dynamical timescales ($t_{\mathrm{dyn}} = 1/\sqrt{G\rho}$) with $\sim 2\%$ efficiency, following \citet{kennicutt1989star}.
    Sub-grid recipes for feedback from massive stars launch two-phase winds that drive material out of galaxies through a combination of type-II supernovae, radiation pressure and stellar winds. These winds are parametrized using a mass loading factor and wind speed, and scaling relations for these parameters based on galaxy properties are adopted from the FIRE simulations \citep{muratov2015gusty} instead of being tuned to reproduce observations.
    Since Mufasa does not explicitly model AGN, quenching is accomplished
    by keeping all the gas in massive haloes heated (except gas that is self-shielded) to reproduce the effects of `maintenance mode' feedback from long lived and AGB stars \citep{gabor2015hot}.
    Parameters in Mufasa have been chosen to reproduce the galaxy SMF at $z=0$.

    \item \textbf{Simba} \citep{dave2019Simba}

    The Simba cosmological galaxy formation simulations are built on the Mufasa simulations including black hole growth and feedback, using the GIZMO cosmological gravity+hydrodynamics code with its Meshless Finite Mass (MFM) solver \citep{hopkins2014gizmo,hopkins2017new}.
    Similar to Mufasa, Simba uses a stochastic H$_2$ based star formation model, with the SFR given by the H$_2$ density divided by the local dynamical timescale.
    Simba also uses two-phase winds with updated mass loading factor scalings from FIRE \citep{angles2017cosmic}, which is similar to those adopted by IllustrisTNG but with slightly lower wind velocities.
    Simba implements a torque limited BH accretion model along with a kinetic subgrid model for BH feedback similar to \citet{angles2017black}, but with a variable outflow velocity. Wind particles are decoupled for a short amount of time ($10^{-4} \tau_\mathrm{H}$, where $\tau_{\mathrm{H}}$ is the Hubble time) from hydrodynamics and radiative cooling. The BH feedback is overall similar to the two-mode model in IllustrisTNG, with some differences detailed in \citet{dave2019Simba}. The majority of galaxy quenching occurs due to the AGN jet mode feedback, with a bimodal distribution of quenching timescales found in \citet{rodriguez2019mergers}.
    Parameters in the Simba model have been chosen to reproduce the $M_{\mathrm{BH}}-\sigma$ relation and the galaxy SMF at $z=0$.

    \item \textbf{EAGLE} \citep{schaye2014Eagle, crain2015eagle, schaller2015eagle, mcalpine2016eagle}

    The Evolution and Assembly of GaLaxies and their Environments (EAGLE)\footnote{\url{http://icc.dur.ac.uk/Eagle/}} is a set of cosmological hydrodynamic simulations of galaxy formation using a modified version of the Tree-PM smoothed particle hydrodynamics (SPH) code GADGET-3 \citep{springel2005cosmological}. EAGLE does not resolve molecular clouds for accurate modeling of the warm gas within galaxies, and implements sub-grid recipes for stellar evolution, cooling and heating of gas due to stars and other emission, metal enrichment of ISM gas and energy injection due supernovae, and the formation, accretion and feedback of AGN.
    Star formation occurs via gas particles that are stochastically converted into star particles at a pressure-dependent rate that reproduces the observed Kennicutt-Schmidt law \citep{schaye2008relation}. A metallicity-dependent density threshold \citep{crain2015eagle} is adopted to ensure that star formation happens in cold, dense gas.
    The local ISM is heated stochastically due to feedback from massive stars and supernovae with a fixed temperature increment \citep{dalla2012simulating}. At high SFR, this feedback can lead to large-scale galactic outflows \citep{crain2015eagle}.
    Similar to feedback from star formation, AGN feedback is implemented using a single-mode thermal feedback model. The fraction of radiated energy that couples to the ISM is calibrated to reproduce the stellar mass-black hole mass relation at $z=0$, and mimics the `radio'- and `quasar'-like modes depending on the BH accretion rate \citep{crain2015eagle}. Quenching is thought to happen on long timescales ($\sim 3-4$ Gyr) for low mass central galaxies due to stellar feedback, and high-mass centrals on shorter timescales due to AGN feedback and environmental quenching \citep{trayford2016green, wright2019quenching}.
    Parameters in the EAGLE suite are chosen to reproduce the galaxy SMF at $z=0.1$ and the disc galaxy size-mass relation.

    \item \textbf{Santa-Cruz SAM} \citep{somerville2008semi, somerville2015star, porter2014understanding, brennan2016relationship}

    The Santa-Cruz Semi-Analytic Model contains a number of well motivated semi-analytic prescriptions for the hierarchical growth of structure, gas heating and cooling, star formation and stellar evolution, supernova feedback and its effect on the ISM and ICM, AGN feedback, starbursts and morphological transformations due to mergers and disc instabilities that are used in conjunction with the Bolshoi-Planck \citep{klypin2011dark, rodriguez2016main, klypin2016multidark} dark matter simulation merger trees to construct populations of galaxies that are tuned to match observations at $z=0$.
    The model implements two modes of star formation: a `normal' disc mode following the Schmidt-Kennicutt relation, along with exploding supernovae which drive outflows with recycling that occurs in isolated discs, and a `starburst' mode that occurs as a result of a merger or internal disc instability.
    The SAM implements a multi-phase gas model for the ISM. Cold gas can be ejected from galaxies by winds driven by SN feedback. Heated gas is either trapped within the DM halo potential well, or ejected from the halo into the diffuse IGM.
    \citet{brennan2016relationship} and \citet{somerville2015davereview} find that virial shock heating due to massive haloes alone is not enough to quench massive galaxies, with a significant role played by feedback from AGN activity, driven by galaxy mergers or in-situ processes like disc instabilities.
    Model parameters such as the strengths of stellar and AGN feedback are calibrated using the observed SMF at $z=0$, with further details in \citet{porter2014understanding}.

    \item \textbf{UniverseMachine} \citep{behroozi2013average, behroozi2018universemachine}

    The UniverseMachine is an empirical model that determines the SFRs of galaxies as a function of their host haloes' potential well depths, assembly histories, and redshifts.
    The model uses halo properties and assembly histories from the Bolshoi-Planck dark matter simulation \citep{klypin2011dark, rodriguez2016halo, klypin2016multidark} in conjunction with a variety of observational constraints including the cosmic SFRD, observed SMFs, specific SFR functions, quenched fractions, UV luminosity functions, UV-stellar mass relations, IRX-UV relations, autocorrelation and cross-correlation functions, and the dependence of quenching on environment across $0<z<10$ to constrain its free parameters (see Table 1 in \citealt{behroozi2018universemachine}).
    Star formation rates are parametrized in terms of redshift and halo properties, with the list of parameters in Table 2 of \citet{behroozi2018universemachine}, which include the scatter in the SFRs of star forming galaxies, a model for the SFR-$v_{M,\mathrm{peak}}$ relation\footnote{Where  $v_{M,\mathrm{peak}}$ is the maximum circular velocity of the halo at peak halo mass.}, quenched fraction properties and random errors in measuring stellar masses and star formation rates. The parameters are tuned using Markov Chain Monte Carlo optimization to match the observational constraints. In the current analysis we use SFHs from the public Data Release 1 of UniverseMachine.

    \item \textbf{FIRE-2} \citep{hopkins2018fire}

    The Feedback In Realistic Environments (FIRE)\footnote{\url{http://fire.northwestern.edu}} simulations considers a fully explicit treatment of the multi-phase ISM, and stellar feedback.
    The simulations in this work are specificially part of the ``FIRE-2'' version of the code; all details of the methods are described in \citet{hopkins2018fire}, Section~2. The simulations use the code GIZMO \citep{hopkins2015new},\footnote{\url{http://www.tapir.caltech.edu/~phopkins/Site/GIZMO.html}}, with hydrodynamics solved using the mesh-free Lagrangian Godunov ``MFM'' method.
    Gas dynamics and radiative cooling
    from a meta-galactic background and local sources
    are incorporated using tabulated cooling rates from CLOUDY \citep{ferland2017cloudy}. Stars form by stochastically turning gas particles into stellar particles in dense, self-shielding molecular, self-gravitating regions above a density threshold.
    The stellar feedback prescription includes radiation pressure from massive stars, local photoionization and multi-wavelength photoelectric heating, core-collapse and type Ia supernovae with appropriate momentum and thermal energy injection, and stellar winds. The FIRE physics, source code, and all numerical parameters are identical to those in \citet{hopkins2018fire}.
    The higher resolution of the FIRE simulations resolves the ISM to a larger extent than the large-volume simulations.  \citealt{hopkins2014galaxies} find that supernova feedback alone is not enough, radiative feedback (photo-heating and radiation pressure) is needed to destroy GMCs and enable efficient coupling of later supernovae to gas. Multiple feedback mechanisms are also responsible for regulating the ISM: supernovae regulate stellar masses/winds; stellar mass-loss fuels late star formation; radiative feedback suppresses accretion on to dwarfs and instantaneous star formation in discs.
    Feedback from supermassive black holes is not included in the simulations \citep{hopkins2018fire}.
    While there are approximations for the momentum and energy deposition from SNe when the cooling radius is not resolved, the simulations are not explicitly tuned.

    \item \textbf{g14} \citep{governato2012cuspy, munshi2013reproducing, brooks2014baryons, christensen2016n, brooks2016bulge, brooks2017reconcile, christensen2018tracing}

    The g14 suite of cosmological zoom simulations are run using the N-body+SPH code Gasoline \citep{wadsley2004gasoline} within a WMAP3 cosmology.
    The galaxies are chosen to have a range of merger histories and spin values. The g14 simulations follow the non-equilibrium formation and destruction of molecular hydrogen, and allow stars to form in the presence of H$_2$, with resolution high enough to resolve the disks of galaxies and the GMCs in which stars form \citep{christensen2012implementing}. Stars are born with a \citet{kroupa1993distribution} IMF, mass and metals are returned in stellar winds as star particles evolve and SN Ia and II return thermal energy to the \updated{surrounding gas} (see \citealt{stinson2006star} for details). For SN II, 10$^{51}$ erg of energy are injected per SN. Metal diffusion occurs in the ISM \citep{shen2010enrichment}, and a cosmic UV background is included following \citet{haardt1996radiative}.  The g14 suite was calibrated to match the SMHM relation of \citet{moster2013galactic}.

    \item \textbf{Marvel/Justice League} \citep[Munshi et al., in prep.]{bellovary2019multimessenger}

    The Marvel/Justice League simulations  are run using ChaNGa \citep{menon2015adaptive}, the successor to Gasoline. The Marvel-ous dwarfs (henceforth Marvel) are a sample of field dwarfs (4-11 Mpc from a Milky Way-mass galaxy) at 65pc force resolution, while the DC Justice League (henceforth JL) are zooms of MW-mass disk galaxies and their surrounding environments at 170pc resolution.
    Many of the physics modules in ChaNGa remain the same as in Gasoline, such as the star formation and stellar feedback schemes, with the exception that 1.5$\times$10$^{51}$ erg of thermal energy is injected per SN II. This increase is motivated by the fact that ChaNGa contains an improved implementation of Kelvin-Helmholtz instabilities compared to Gasoline \citep{wadsley2017gasoline2}, which leads to more efficient accretion onto the disk.  An updated UV background is adopted, based on \citet{haardt2012radiative}. In addition, supermassive black hole growth and feedback is implemented using the models described in \citet{tremmel2017romulus}. Parameters in the simulations were calibrated to reproduce the SMHM, SMBH, and SFRs of galaxies at $z = 0$.

\end{itemize}

\begin{table*}
    \centering
    \begin{tabular}{c|c|c|c|c|c|c}
    \hline
         Simulation Name & Type & Box Length & m$_{\mathrm{DM}}$  & m$_{\mathrm{sp}}$  & n$_{\mathrm{galaxies}}$ & $f_{\mathrm{SFR} \leq 10^{-3} \mathrm{M}_\odot/\mathrm{yr}}$ \\
         &  & [Mpc] & [$10^6 $M$_\odot$] & [$10^6 $M$_\odot$] & [M$_*>10^9$M$_\odot$] & [$\Delta t = 100 \mathrm{Myr}$]   \\
         \hline \hline
        Illustris & Hydro & 106.5 & 6.26 & 1.26 & 19354 & 0.02 \\
        IllustrisTNG & Hydro & 110.7 & 7.5  & 1.4  & 12220 & 0.03 \\
        Mufasa & Hydro & 50 & 96  & 48 & 3042 & 0.18 \\
        Simba & Hydro & 100 & 96 & 18 & 11300 & 0.13 \\
        EAGLE & Hydro & 100 & 9.7  & 1.81 & 7482 & 0.04\\
        \hline
        Santa-Cruz SAM & SAM & 100 & 203.7 & N/A & 12821 & 0.04 \\
        \hline
        UniverseMachine & Empirical & \updated{70.3} & 203.7 & N/A & 7361 & 0.05 \\
        \hline
        FIRE-2 & Zoom & N/A & 1.3($10^{-3}$)-0.28 & 2.5($10^{-4}$)-5.6($10^{-2}$) & 14 & 0.0 \\
        g14 & Zoom & N/A & 0.126 & 8.0($10^{-3}$) & 8 & 0.0  \\
        Marvel-ous dwarfs & Zoom & N/A & 0.0067 &  4.23($10^{-4}$) & 1 & 0.0\\
        DC Justice League & Zoom & N/A & 0.042 & 8.0($10^{-3}$) & 4 & 0.0
    \end{tabular}
    \caption{Details of the various models compared in this paper. The box length for UniverseMachine denotes the subset of the full 250$/h$ Mpc box used in the current analysis. The number of galaxies reported is the subset of central galaxies with stellar masses $> 10^9 $M$_\odot$. References for each simulation from which these parameters are taken can be found in Section \ref{sec:data}. m$_{\mathrm{DM}}$ and m$_{\mathrm{sp}}$ denote the masses of DM and stellar particles, respectively, at the time of formation. n$_{\mathrm{galaxies}}$ is the number of galaxies in our $z=0$ sample above M$_*\sim 10^{9}$M$_\odot$ used in the current analysis, and $f_{\mathrm{SFR} \leq 10^{-3} \mathrm{M}_\odot/\mathrm{yr}}$  is the fraction of the total sample for which SFR$=0$ due to discrete star particles in the hydrodynamical simulations and is set to $10^{-3}$M$_\odot$ to compute PSDs in log SFR space, and the fraction of time when SFR$< 10^{-3}$M$_\odot$ for the SAM and empirical model, with SFHs binned in $100$ Myr intervals.}
    \label{tab:sim_details}
\end{table*}

\subsection{Extracting star formation histories}
\label{sec:get_sfhs_from_models}

We compute SFHs for each galaxy in the hydrodynamical simulations under consideration (Illustris, IllustrisTNG, Mufasa, Simba, EAGLE, FIRE-2, \brooks) by performing a mass-weighted binning of the star particles with $\Delta t = 100$ Myr. The choice of time bin is further explored in Section \ref{sec:timestep}. For models where we only have access to the masses of the star particles at the time of observation, we account for mass-loss using the FSPS \citep{conroy2009propagation, conroy2010propagation} stellar population synthesis code, adopting the initial mass function (IMF) of the stellar particles in the simulation. In this procedure, we consider all stellar particles belonging to a galaxy at $z\sim 0$ instead of tracing the gas-phase SFR as a function of time. The reason for this is twofold: (i) since the hydrodynamical simulations trace the times when star particles were formed, this gives us finer time-resolution than the snapshots saved for the different simulations, (ii)  since the SFHs we observationally reconstruct are the sum over all the progenitors, this archaeological approach therefore allows us to compare directly with observations. Both UniverseMachine and the Santa Cruz SAM track the SFR for each galaxy, so we simply interpolate these to match the same time grid with $100$ Myr steps as the hydrodynamical simulations. In both cases, the resolution is fine enough that the interpolation does not need to up-sample the SFR. Since the UniverseMachine SFHs are stored in terms of scale factor instead of absolute time, an additional periodogram is computed using the uneven spacing to check that the PSDs are not significantly affected by the interpolation. Additional fine-resolution SFHs are also computed for the galaxies from the zoom simulations, with a timestep $\Delta t = 1$ Myr.

Since the SFHs span a large dynamic range, we work in log SFR space in order to be able to better quantify the relative strengths of SFR fluctuations. Analyzing the SFHs in linear SFR space effectively amounts to a different weighting scheme. This choice is motivated by physical considerations, since the SFRs of star forming galaxies are often found to be distributed  normally in log SFR space, with a tail towards low SFRs from passive galaxies that do not have ongoing star formation \citep{feldmann2017star, hahn2018iq, caplar2019sfhpsd}.

\subsection{The Power Spectral Density (PSD)}
\label{sec:psd_description}

The variability or `burstiness' of galaxy SFRs is a topic of much interest, and has been studied in a variety of ways - using burstiness indicators based on the timescales of different SFR tracers \citep{guo2016bursty, emami2018closer, broussard2019stochastic}, fitting an exponential to the Pearson correlation coefficient of SFRs as a function of time-separation to quantify an `SFR evolution timescale' \citep{torrey2018similar}, quantifying the scatter in SFRs smoothing on different timescales \citep{hopkins2014galaxies, matthee2018origin}, using power spectral densities (PSDs) to quantify the variability in Fourier space \citep{caplar2019sfhpsd, wang2020var, wang2020vartwo}, or performing a PCA decomposition of SFHs to get estimate the fraction of variance accounted for by different timescales \citep{matthee2018origin}. In other studies involving timeseries data, the structure function \citep{hughes1992university, macleod2010modeling, kozlowski2016revisiting, caplar2017optical} has also been used as a metric to quantify variability on different timescales in quasar and AGN studies.

In the current analysis, we choose to quantify the variability of galaxy SFHs using the PSD formalism, since
\begin{itemize}
    \item the PSD formalism is well studied and easily interpretable, and Fourier space provides an excellent domain to quantify and compare the variability of SFHs across different timescales;
    \item the decomposition of variability into different frequencies, and therefore different timescales, allows us to understand the relative contribution to the overall burstiness from each timescale. This takes us one step closer toward relating this variability to the underlying physical processes responsible; and
    \item evolving analysis techniques coupled with upcoming observational datasets will make it possible to obtain observational constraints in PSD space.
\end{itemize}

\begin{figure*}
    \centering
    \includegraphics[width=0.9\textwidth]{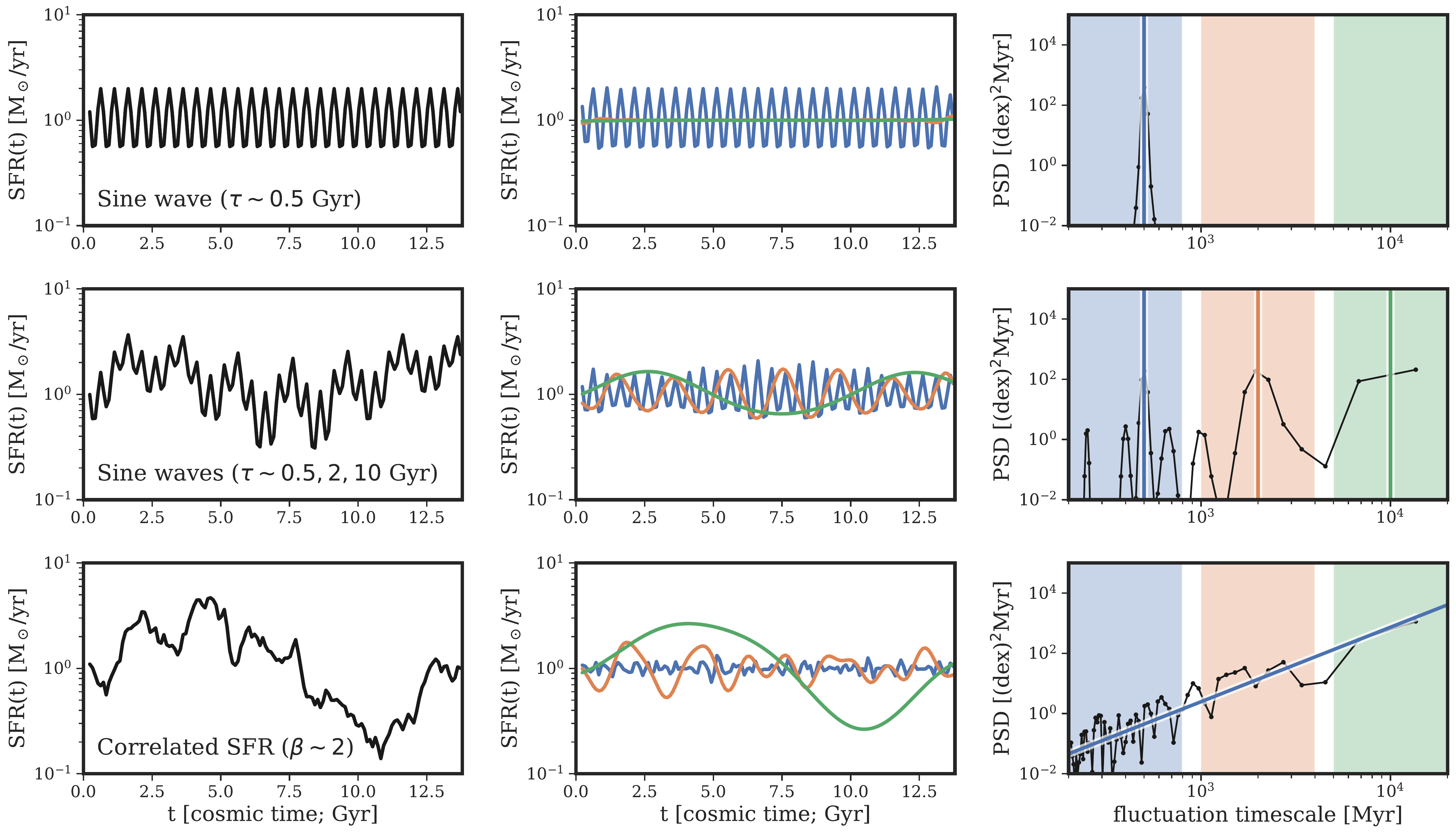}
    \caption[Interpreting galaxy SFH power spectral densities]{Illustrating the power spectral density (PSD) computation using three example SFHs: (\textbf{top:}) A simple sine wave with a timescale of $500$ Myr, (\textbf{middle row:}) a combination of three sine waves, with timescales: $500$ Myr, $2$ Gyr and $10$ Gyr, and (\textbf{bottom:}) a stochastic SFH with a spectral slope of $\beta=2$. (\textbf{Left:}) The individual galaxy SFHs, in log SFR space. (\textbf{Middle column:}) SFH fluctuations on short, intermediate and long timescales isolated using a band-pass filter in Fourier space - the green curves show the power arising due to the long timescales ($> 4$ Gyr), orange curves show the power contribution from intermediate timescales ($1-3$ Gyr) and blue from relatively shorter timescales ($< 0.9$ Gyr). (\textbf{Right:}) The PSD (black lines) corresponding to each SFH from the left panels, while the three coloured ranges correspond to the band-passes used to isolate the Fourier modes in the middle panel. The PSD in each band pass is proportional to the net strength of the fluctuations contained in the coloured curves from the middle column averaged over all phases. For the sine wave, the PSD is well localized at a single frequency. With multiple sine waves, it is harder to separate the contributions from individual components. For a stochastic process with spectral slope $\beta\sim 2$, the power is distributed accross a range of timescales.}
    \label{fig:psd_explanation_simple}
\end{figure*}

\begin{figure*}
    \centering
    \includegraphics[width=0.9\textwidth]{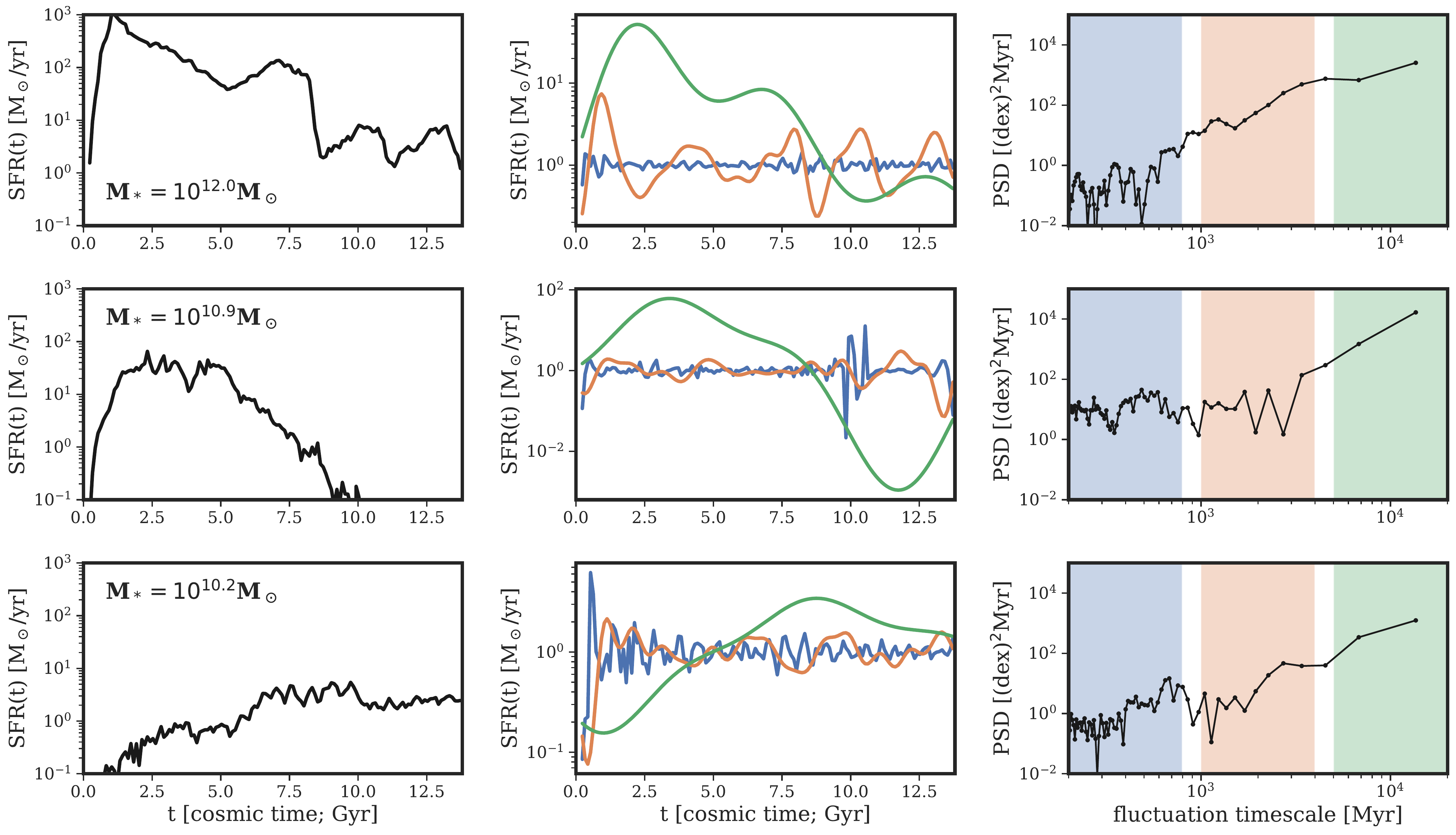}
    \caption[Interpreting galaxy SFH power spectral densities]{Similar to Figure \ref{fig:psd_explanation_simple}, showing the (PSD) computation using three galaxies from the IllustrisTNG simulation. (\textbf{Top row:}) A green-valley galaxy, (\textbf{middle row:}) a quiescent galaxy with no star formation in the last $\sim 3$ Gyr, and (\textbf{bottom row:}) an actively star forming galaxy building up its stellar mass. (\textbf{Left column:}) The individual galaxy SFHs, obtained by binning mass-weighted star particles in 100 Myr bins. (\textbf{Middle column:})  Log SFR fluctuations on short, intermediate and long timescales isolated using a band-pass filter in Fourier space - the long timescales contain the most power and capture the overall shape of the SFH, while the shorter timescales capture fluctuations around it. (\textbf{Right column:}) The black line shows the full PSD, and the the integral of the coloured curves in the middle columns sets the strength of the PSD in the corresponding coloured timescale ranges. As seen in the middle panel,overall trends in the SFH can be described by the contribution from the longest timescales, similar to the stochastic process in Figure \ref{fig:psd_explanation_simple}. However, depending on the shape of the SFH, the distribution of power on shorter timescales can change significantly.}
    \label{fig:psd_explanation}
\end{figure*}

For a continuous time series $\psi(t)$, the PSD is defined in terms of the Fourier transform $f(k) = \int dt ~e^{-ikt} \psi(t)$ as $\mathrm{PSD}(k) = |f(k)|^2$. In practice, we compute the PSD for each SFH using Welch's method \citep{welch1967use}, implemented in the \verb|scipy.signal.welch| module.

The PSD corresponding to the SFH for an individual galaxy reports a phase-averaged estimate of the strength of SFR fluctuations at a given frequency\footnote{Or, inverting it, at a given timescale.}.

For a sinusoidal signal with a frequency $\nu$, the corresponding PSD is given by a delta function at the frequency $\nu$, shown in the top panel of Figure \ref{fig:psd_explanation_simple}.
Generalized to more complicated timeseries, the PSD therefore provides a way to disentangle and interpret the strength of the fluctuations on different timescales, as previously done in studies of AGN variability and theoretically with SFHs \citep{macleod2010modeling, macleod2012description, caplar2017optical,sartori2018model, caplar2019sfhpsd}. A sharp peak in the PSD would indicate strong SFR fluctuations at a given timescale, possibly driven by a physical process. However, physical processes acting over a range of timescales spread out the peaks and make it more difficult to isolate the effects of individual processes. An example of this is shown in the middle column of Figure \ref{fig:psd_explanation_simple}, where the sum of three sinusoidal curves produces three peaks in the PSD, along with additional artifacts due to the finite length of the time series. Processes like hierarchical structure formation and correlated stochastic star formation additionally link short timescales to longer ones, creating an overall spectral slope to the PSD, shown in the bottom panel of Figure \ref{fig:psd_explanation_simple}. Physical processes can additionally drive features at certain characteristic timescales, for e.g., the regulator model (\citealt{lilly2013gas}, see also \citealt{bouche2010impact, dave2012analytic, forbes2014origin}) predicts SFHs correlated below an equilibrium timescale of a galaxy's gas reservoir, with the slope at timescales below the break steeper by 2 than the slope above it (\citealt{wang2020vartwo, tacchella2020stochastic}).
Such features can be seen as breaks in the PSD \citep{caplar2019sfhpsd}. The PSDs of galaxy SFHs therefore contain a wealth of information about the different physical processes responsible for its shape.

Examples of this procedure are shown in Figure \ref{fig:psd_explanation}, which shows SFHs for galaxies from the IllustrisTNG simulation (left column) as well as their corresponding PSDs (right column). The contribution to the PSDs at three different timescales due to the strength of SFH fluctuations are highlighted in different colours in the middle panels.
Unlike the case for the sine wave, the power in these PSDs is spread over a large dynamic range, indicative of the stochastic nature of star formation and the wide range of timescales over which physical processes in galaxies induce variability in the star formation rates.
With a thorough understanding of a galaxy's evolution and merger history, it might be possible to interpret its individual PSD. However, in the current work we focus on studying the broader trends in a sample of galaxy SFHs and their corresponding PSDs as a way to compare different models of galaxy evolution on the same footing. In doing so, we examine the variability of galaxy SFHs on intermediate ($\sim 200$ Myr) to long ($\sim 10$ Gyr) timescales, and
study the evolution in the PSDs as a function of stellar mass and star forming state (star forming vs quiescent). We choose stellar mass since it is a good tracer for the overall state of a galaxy, correlating well with a wide range of other physical properties including halo mass, SFR, metallicity and BH mass, and can be calculated self-consistently for all the models directly from the SFHs after accounting for mass-loss.

\subsubsection{Choosing a minimum time interval and SFR=0}
\label{sec:timestep}

In choosing the $\Delta t$ for our time bins, we need to consider the effects of the discreteness of individual star particles, since the SFR will be zero in bins that do not contain any star particles. This effect is particularly important for low-mass galaxies, where the number of star particles is $\mathcal{O}(10^2-10^3)$ depending on the model resolution. If not accounted for, these bins lead to shot noise in log SFR space, biasing the computed PSDs. We avoid this by increasing the size of the time bins until the fraction of our data with $\mathrm{SFR} = 0$ is significantly reduced. We also verified that the PSDs at timescales longer than our adopted bin size $\Delta t$ are insensitive to the choice of binning. In practice, we find that with time bins of $100$ Myr, the percentage of bins where $\mathrm{SFR} = 0$ is $\sim 3 - 6\%$ across the various models. The only notable exceptions are Mufasa and Simba, which have poorer resolution. The fraction of the total SFRs that are $\leq 10^{-3}$M$_\odot /yr$ for each model are given in Table \ref{tab:sim_details}. Finally, we set values of $\mathrm{SFR} = 0$ to $\mathrm{SFR} = \mathrm{SFR}_{\mathrm{min}} = 10^{-3} $M$_\odot yr^{-1}$ for a given model to avoid values of $-\infty$ in the PSD computation. We tested the procedure to ensure that this does not significantly affect the PSDs of quiescent galaxies by broadening the time-bins (increasing $\Delta t$) to reduce the number of bins with $\mathrm{SFR} = 0$ and comparing the PSDs for longer timescales.
An example of the PSD for a fully quenched galaxy can be seen in the middle row of Figure \ref{fig:psd_explanation}, which shows the SFH for a single quenched galaxy from IllustrisTNG.

\subsubsection{Shot noise due to discrete star particles}
\label{sec:validation_simple}

In the hydrodynamical simulations we consider, gas is turned into a star particle probabilistically, depending on whether certain temperature and/or density conditions are met. This introduces a $\mathcal{O}(1)$ fluctuation in a given time bin (width $\Delta t$) based on whether the N+1$^{th}$ star particle is created. In log SFR space, the sudden conversion of a gas particle into a star particle creates large fluctuations when the SFR is low, i.e., there are only a few star particles in a given time bin. To avoid this, we only consider the portion of the PSD on timescales ($\Delta t > \Delta t_{min}$) that are large enough that there are enough star particles in a bin to minimize the effects of discrete star particles.

Since we are working with log SFR, the biggest fluctuations due to discrete star particles will be in bins that contain $\mathcal{O}(1)$ star particle. Given a galaxy with mass M$_*$ and resolution such that a star particle is of mass $m_{\mathrm{sp}}$, this effect becomes more likely when the number of time bins ($\tau_{\mathrm{H}} / \Delta t$) is comparable to the number of star particles. Therefore, we would like to avoid the limit M$_{*}/m_{\mathrm{sp}} \lesssim \tau_{\mathrm{H}} / \Delta t$. For a simulation with resolution $m_{\mathrm{sp}}$, we therefore require: $\Delta t \gg \tau_{\mathrm{H}} m_{\mathrm{sp}} / $M$_* $. For a galaxy with M$_* \sim 10^{10} $M$_\odot$, with resolution $m_{\mathrm{sp}} \sim 10^6 $M$_\odot$, this means that the time-bin width at $z\sim 0$ has to be $\gg 1.3$ Myr.

However, this is significantly complicated by the fact that the SFHs of galaxies tend to rise and fall, which means that star particles are not uniformly distributed across time. Moreover, an $\mathcal{O}(1)$ fluctuation causes different contributions depending on what the SFR is in a given bin. To account for all of these effects, we forward-model the contribution of discrete star particles in Appendix \ref{sec:timescale_tests}, by creating realistic SFHs corresponding to various stellar masses and then discretizing them to match the resolution of the models we consider. We then compute the power spectra of the true and discretized SFHs, to determine the lowest timescales to which we can accurately probe the PSD at a given resolution and stellar mass. The PSDs below these thresholds have been shown as dashed black lines in Section \ref{sec:psd_sim_results}.
In practice, this means that to probe fluctuations on timescales below 1 Gyr, we need galaxies that have a stellar mass of at least $10^{8.5}$, $10^{8.6}$, $10^{9.9}$, $10^{9.5}$ and $10^{8.7} $M$_\odot$ for Illustris, IllustrisTNG, Mufasa, Simba and Eagle respectively. As we go to shorter timescales the threshold goes up, e.g., to probe fluctuations below 300 Myr, the minimum stellar mass of galaxies needed is $10^{10.1}$, $10^{10.1}$, $10^{11.0}$, $10^{10.7}$ and $10^{10.2} $M$_\odot$ respectively.

Having established the procedures for extracting SFHs from the various models and studying them in PSD space, we now look at the PSDs of galaxies across the different models.

\section{Star-formation diversity and variability in different models}
\label{sec:results}

The variability of galaxy SFHs on different timescales are linked to the underlying processes that regulate star formation across galaxies. The strength of this variability, i.e., the amount of power in the PSD at a given timescale is therefore a useful constraint regarding the cumulative effect of all the processes that contribute to the PSD at that timescale. Since the shapes of the SFHs are intimately linked by scaling relations to the other physical properties of galaxies like stellar mass, environment and morphology \citep{kauffmann2003stellar, whitaker2014constraining, iyer2019gpsfh, tacchella2019morphology}, understanding the link between SFHs and the power on different timescales acts as a step towards linking these properties to the underlying physical processes responsible.

For all the models, Section \ref{sec:psd_sim_results} reports the median SFHs and PSDs in bins of stellar mass, and examines their characteristics.
Sections \ref{sec:sfh_diversity_shapes}
compares the diversity of SFHs predicted by the different models we consider, while
Section \ref{sec:sfh_diversity_psd_slopes} examines the diversity in the PSDs on particular timescales of interest. Section \ref{sec:psd_quenching} looks at the difference in the PSDs based on whether galaxies are actively star-forming or quiescent. Finally, the relation between galaxy SFHs and the dark matter accretion histories (DMAHs) of their parent halos is studied in Section \ref{sec:dmfh_psds}.

\subsection{Variability in the different models at z=0}
\label{sec:psd_sim_results}

\subsubsection{Large-volume simulations}

\begin{figure*}
    \centering
    \includegraphics[width=0.9\textwidth]{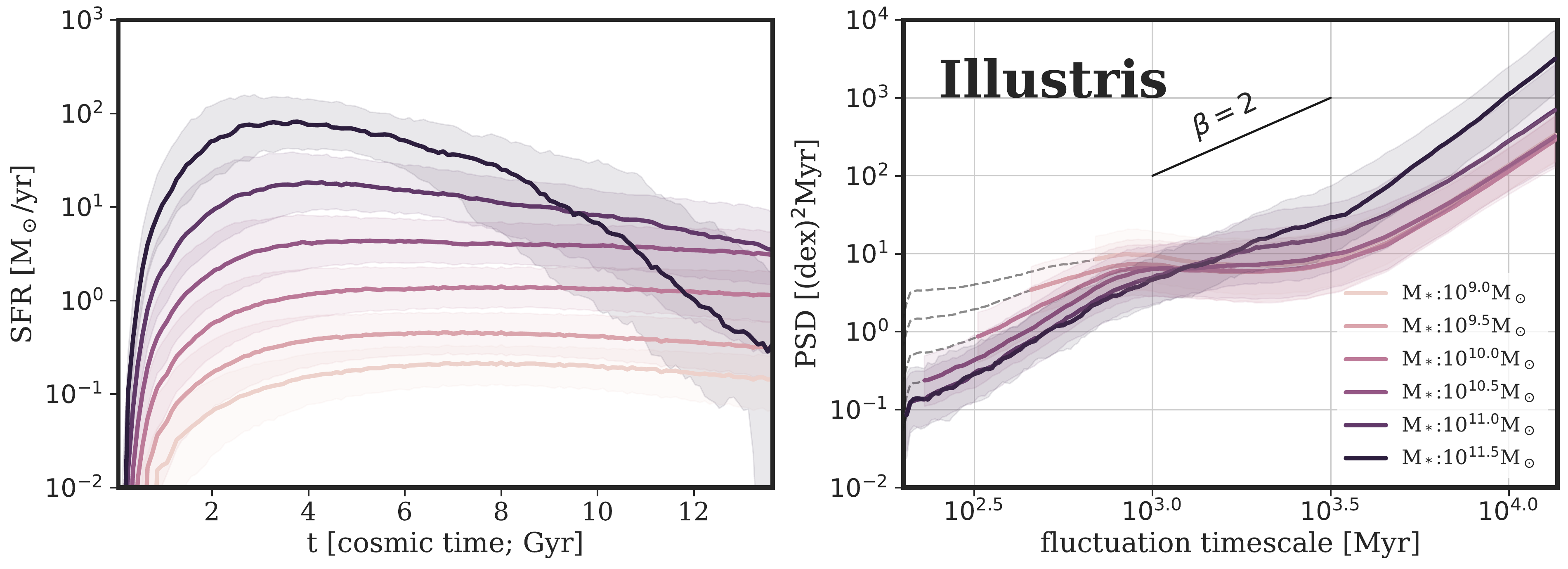}
    \includegraphics[width=0.9\textwidth]{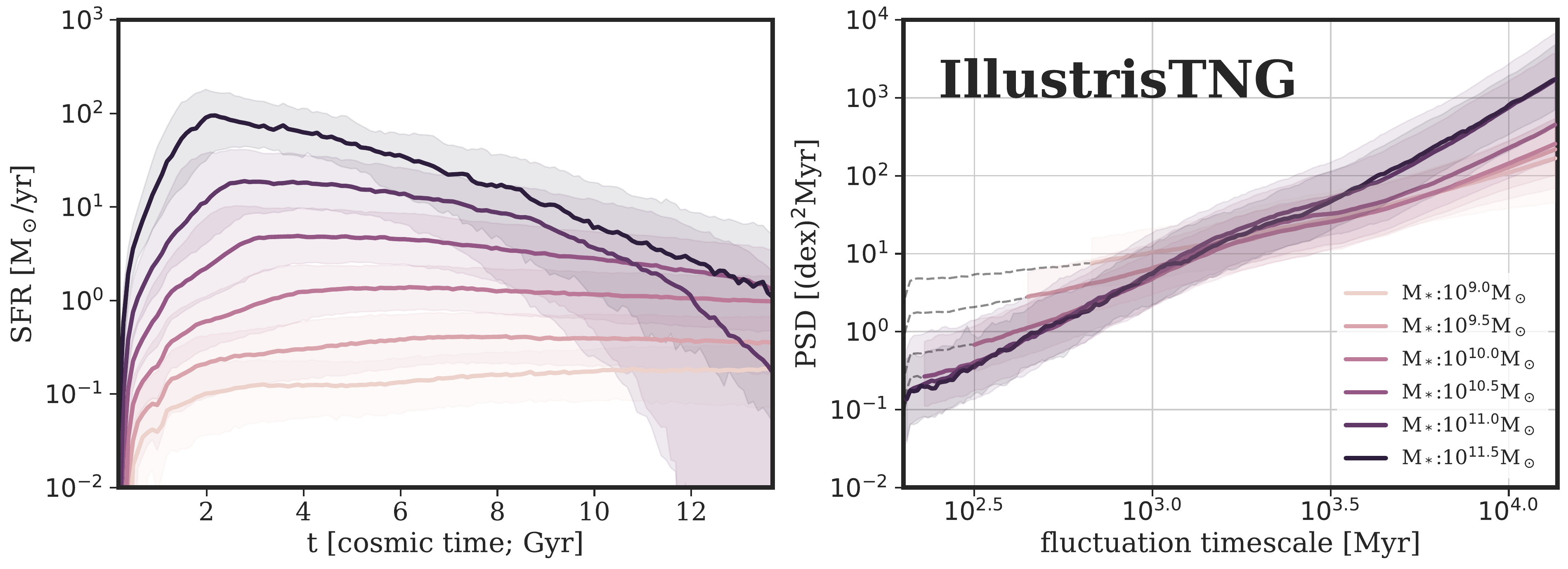}
    \caption[Median SFHs and PSDs in bins of stellar mass for Illustris and IllustrisTNG]{The median star formation histories (\textbf{SFHs; left}) and corresponding power spectral densities (\textbf{PSDs; right}) of galaxies from the Illustris and IllustrisTNG cosmological hydrodynamical simulations, shown here in 0.5 dex bins of stellar mass, centered on the values given in the legend. PSDs are computed from individual SFHs prior to taking the median. The shaded regions show the 16$^{th}$-84$^{th}$ percentile of the distribution in a given mass bin at each point in time (left) and fluctuation timescale (right). Dashed lines indicate regions where shot noise due to discrete star particles may contaminate the PSDs according to our validation tests (see Appendix \ref{sec:timescale_tests}). The PSDs of low- and intermediate-mass galaxies in Illustris and IllustrisTNG show a break at $1-2$ Gyr (more prominent in Illustris than IllustrisTNG), which disappears in higher-mass galaxies, i.e. the PSD of the most massive galaxies is nearly scale-free.
    }
    \label{fig:psd_hydrosims}
\end{figure*}

\begin{figure*}
    \centering
    \includegraphics[width=0.9\textwidth]{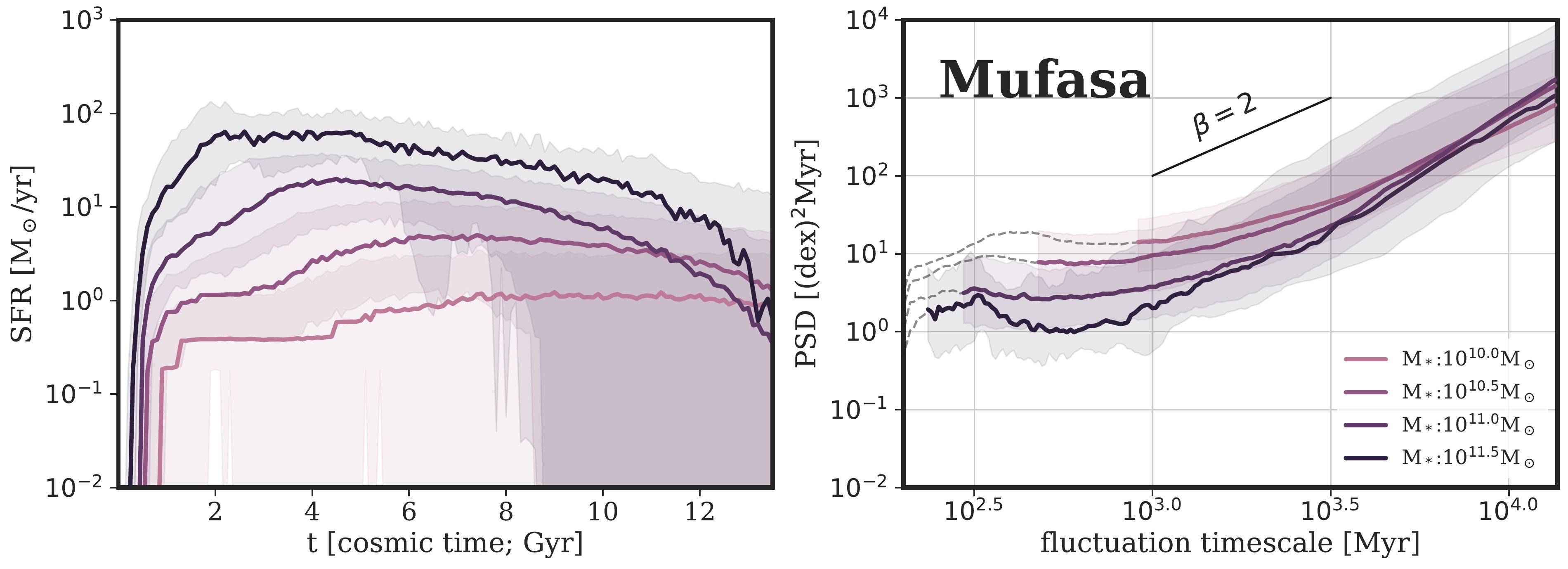}
    \includegraphics[width=0.9\textwidth]{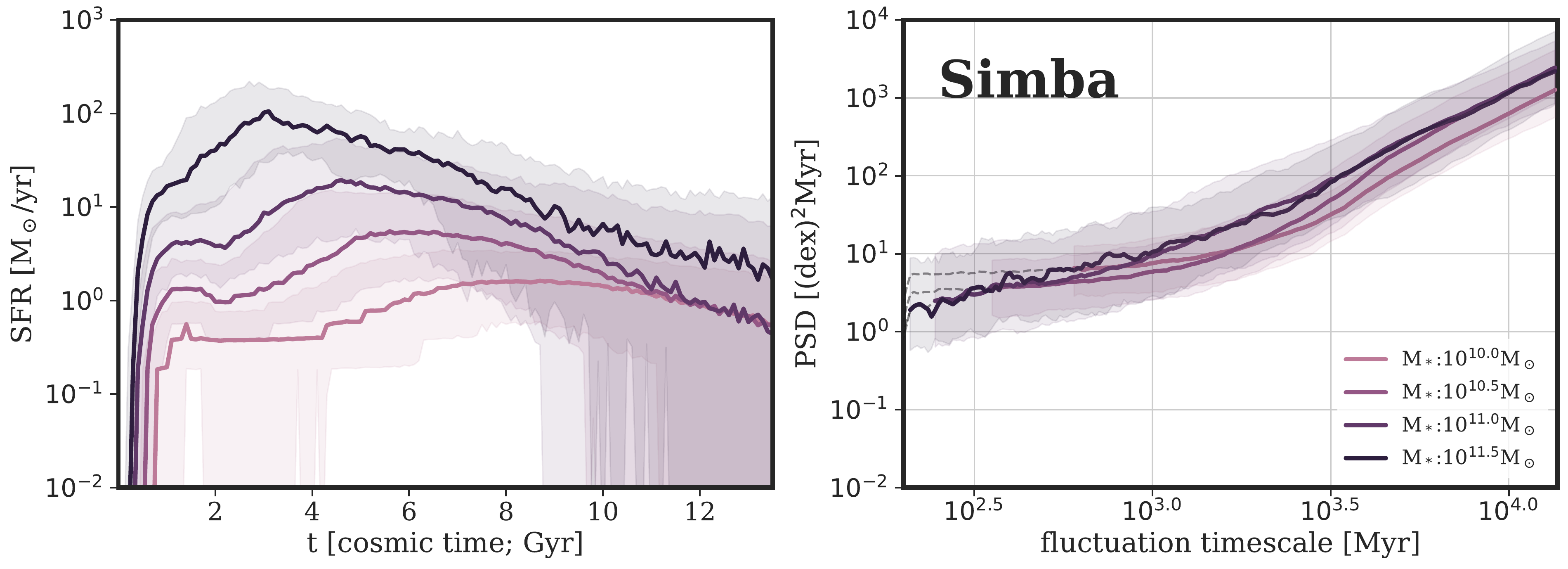}
    \includegraphics[width=0.9\textwidth]{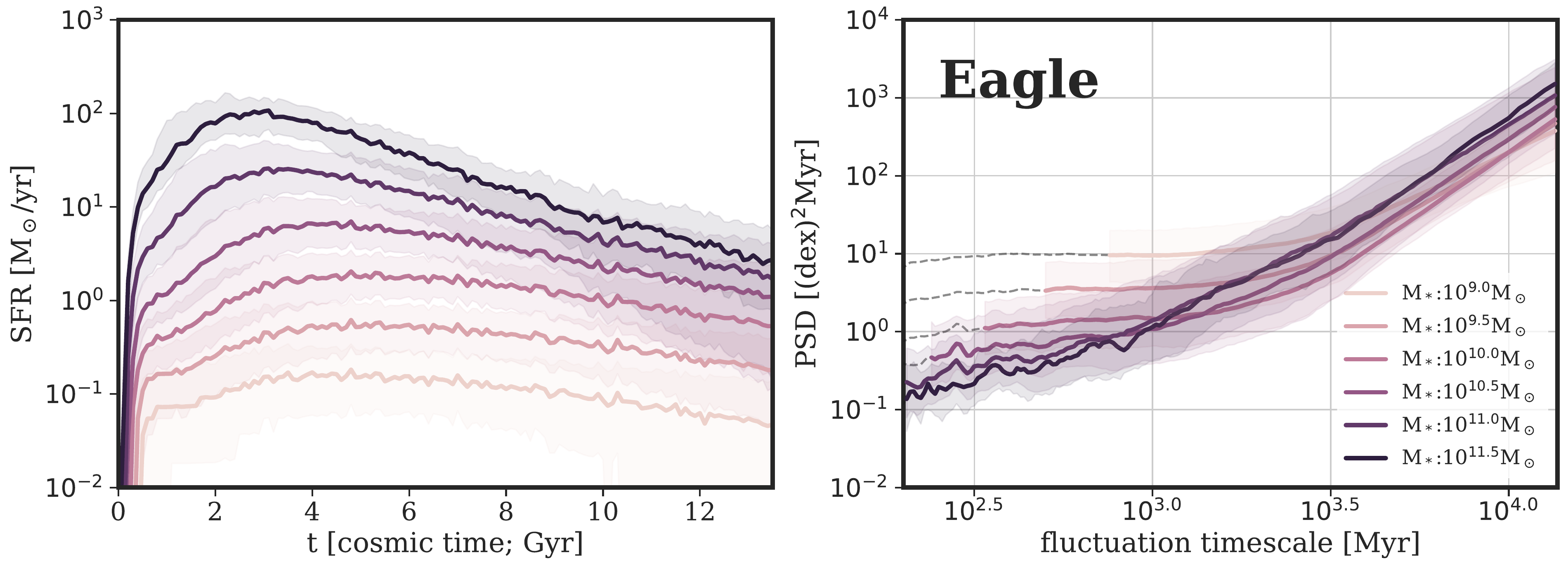}
    \caption[Median SFHs and PSDs in bins of stellar mass for Mufasa, Simba and EAGLE]{Same as Figure \ref{fig:psd_hydrosims}, but for the Mufasa, Simba and EAGLE cosmological hydrodynamical simulations. Due to the lower resolution of the Mufasa and Simba simulations, we only show galaxies with ${\rm M}_*> 10^{10}{\rm M}_\odot$. The PSDs of these simulations show a smoothly increasing PSD slope toward longer timescales, i.e. they show less significant breaks than PSDs in Illustris and IllustrisTNG.}
    \label{fig:psd_hydrosims2}
\end{figure*}

\begin{figure*}
    \centering
    \includegraphics[width=0.9\textwidth]{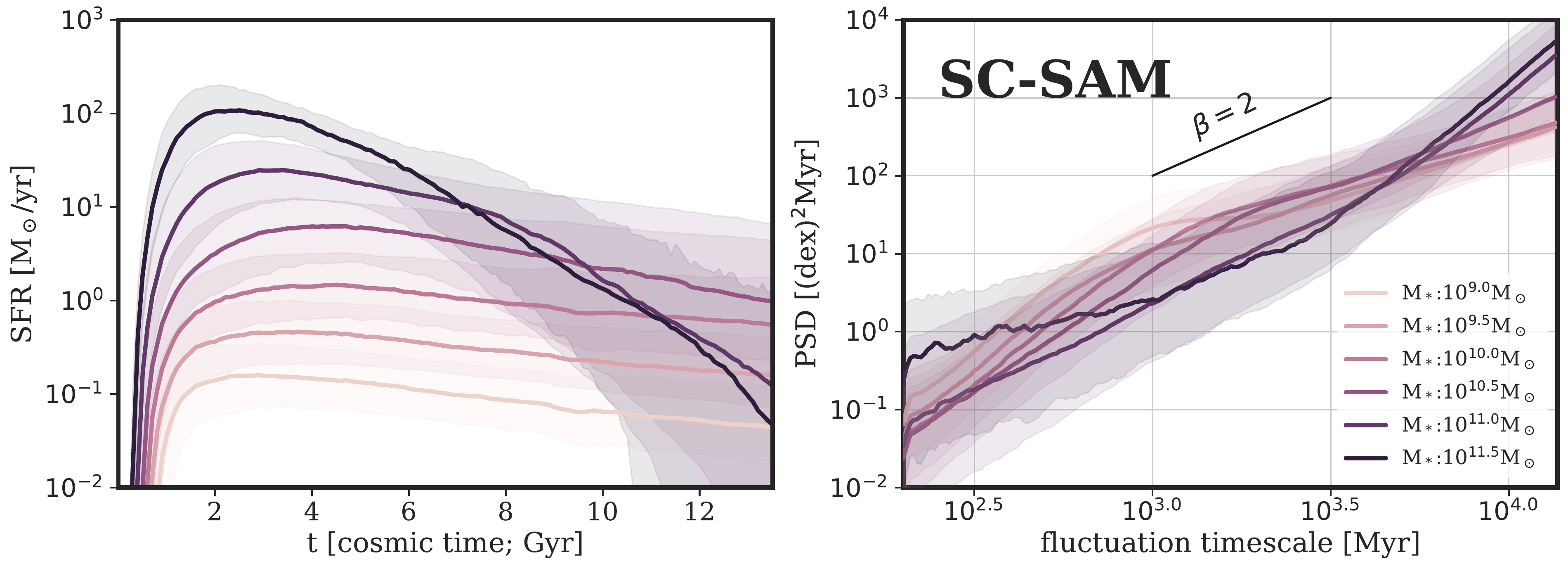}
    \includegraphics[width=0.9\textwidth]{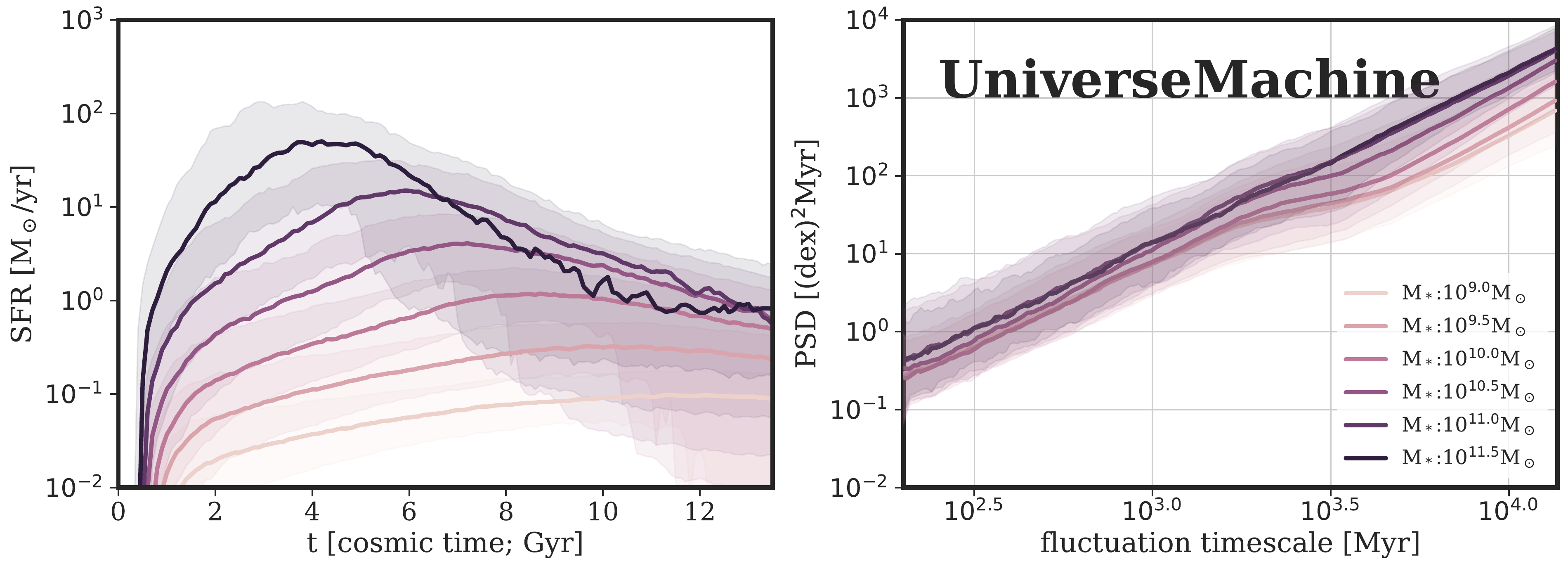}
    \caption[Median SFHs and PSDs in bins of stellar mass for SC-SAM and UniverseMachine]{Same as Figure \ref{fig:psd_hydrosims}, but for the Santa Cruz semi-analytic model and the UniverseMachine empirical model. Tying galaxy SFRs to the dark matter accretion histories of their parent halos without explicit prescriptions for dynamical processes in UniverseMachine manifests as a lack of features in the PSDs that is similar to IllustrisTNG at long timescales and high stellar masses.}
    \label{fig:psd_sam_um}
\end{figure*}

In Figures \ref{fig:psd_hydrosims}, \ref{fig:psd_hydrosims2} and \ref{fig:psd_sam_um} we show the SFHs and corresponding PSDs for galaxies binned in intervals of stellar mass for the Illustris, IllustrisTNG, Mufasa, Simba, and EAGLE hydrodynamical simulations, the Santa-Cruz semi-analytic model, and the UniverseMachine empirical model. Binning in stellar mass allows us to study the coherent features in the PSDs of similar demographics of galaxies across the various models. In a given mass bin, we plot the median SFH and median PSD; the median PSD is obtained from the PSDs of individual SFHs (i.e. not from the median SFH). We see that there is a large amount of diversity in both the star formation histories and the PSDs of galaxies from the various models, although some broad trends can be observed.
Overall, the SFHs of galaxies tend to rise and fall \citep{pacifici, pacifici2016timing}, with this behaviour accentuated as we go to higher stellar masses where the fraction of quenched galaxies is higher \citep{peng2010mass, whitaker2014constraining, schreiber2015herschel}. The times at which the median SFHs in a given mass bin peak and the rate at which they fall differ widely across the different models.
For example, the median SFHs for MW-like galaxies (M$_* \sim 10^{10.5}$M$_\odot$) peak at epochs ranging from $z\sim 1.75$ ($t=3.8$ Gyr) for IllustrisTNG to $z\sim 0.75$ ($t=7.1$ Gyr) for UniverseMachine, with the other models falling somewhere in between.
$10^{9}$M$_\odot$ galaxies in IllustrisTNG and UniverseMachine do not appear to fall on average, contrasted with the decline observed for the EAGLE and SC-SAM models. Due to the coarser resolution of Mufasa and Simba, we are unable to probe this mass range.

For all the models, the PSDs generally rise towards longer timescales, i.e., the dominant contribution to the overall shape of the SFH comes from fluctuations on the longest timescales.
More massive galaxies show a slight increase in the overall normalisation.
This increase in power on the longest timescales traces the increasing contribution on long timescales from quenched galaxies at higher stellar masses, and is discussed further in Section \ref{sec:psd_quenching}.

The PSDs can locally be described using a power-law, with the slopes varying across the models and also within models as a function of stellar mass and timescale. For the median PSDs, the spectral slopes range between $\beta \sim 0$ to $\beta \sim 4$, where the former implies that the \updated{strength} of fluctuations on adjacent timescales are uncorrelated, while the latter implies that the strength of fluctuations on adjacent timescales are highly correlated. Similar to the PSD power, the slope generally rises towards higher masses and longer timescales.
In conjunction with the SFHs, we see that this is tied to the quenching of galaxies, which selectively adds power on longer timescales, leading to an increase in the long-timescale slope. This can also be seen comparing the bottom to the middle panel of Figure \ref{fig:psd_explanation}. We discuss this in more detail in Section \ref{sec:psd_quenching}.

Apart from these overall similarities, the PSDs and corresponding SFHs display a lot of variety across the various models, with Mufasa and Simba showing greater variability on short timescales compared to Illustris, IllustrisTNG, \updated{EAGLE,} Santa Cruz SAM and UniverseMachine. For the most massive galaxies, this corresponds to a nearly 1 dex increase in the power on $\sim 200$ Myr timescales.

A possible concern is that this effect is in part due to resolution effects, since Mufasa and Simba star particles are $\sim 10\times$ those of Illustris, IllustrisTNG and EAGLE. While our forward-modeling of shot-noise accounts for this, we also consider the PSDs of three different IllustrisTNG runs with varying resolution in Section \ref{sec:psd_resolution} which shows that while there is a slight increase in the power on short timescales due to poorer resolution, this is an actual phenomenon due to the galaxies evolving differently and quenching faster, as evidenced by the difference in their median SFHs. In addition, the increase is not enough to completely account for the higher power found in Mufasa and Simba ($\sim 0.5$ dex due to resolution vs the $\sim 1$ dex difference between the shortest timescales for the most massive Mufasa/Simba and IllustrisTNG galaxies).

There are several notable breaks in the PSDs for particular models. In general, we see that the breaks generally decrease in strength toward higher stellar masses, tending to resemble an overall scale-free PSD with slope $\beta \sim 2$.
The timescales and number of breaks can vary significantly across the different models, and are briefly summarised below.

\begin{itemize}
    \item Illustris has two breaks - an intermediate-timescale break around $\sim 0.6-1$ Gyr and a longer-timescale $\sim 2.6-4.2$ Gyr timescales.
    These breaks are prominent at low and intermediate stellar masses. For the most massive galaxies, the breaks nearly disappear and the PSD is close to scale free.
    \item The breaks in IllustrisTNG are similar to the breaks in Illustris, but overall less pronounced.
    Furthermore, the break at $\sim 0.6-1$ Gyr in Illustris moves to longer timescales ($\sim 1.1-2.6$ Gyr) in IllustrisTNG. Again, the PSD becomes nearly scale-free at M$_*>10^{11}$M$_\odot$.
    \item Both Mufasa and Simba have no clear breaks, and instead show a gradual increase in PSD slope from $\beta \sim 0$ to $\beta \sim 2$ toward longer timescales. The highest mass bin in Mufasa shows a slight peak at $\sim 300$ Myr timescales. Above $\sim 3$ Gyr, the slopes in Mufasa stabilise at a constant value, and slopes in Simba approach $\beta\sim 2$.
    \item The PSDs in EAGLE show a smooth increase in slope similar to Mufasa. This increase in slope continues till $\sim 3$ Gyr timescales, beyond which the PSD slopes stay constant.
    \item The Santa-Cruz SAM has a clear break at low and intermediate masses: the break timescale increases from  $\sim 400-600$ Myr to $\sim 1-1.6$ Gyr from M$_*\sim 10^9$M$_\odot$ to M$_*\sim 10^{10.5}$M$_\odot$. At M$_*\sim 10^{11}$M$_\odot$, the PSD is nearly scale free. For the most massive galaxies, the PSDs resemble those of Simba and EAGLE, showing a smooth increase in slope toward long timescales.
    \item UniverseMachine shows the least variation in slope compared to the other models, with $\beta \in (1,2.5)$. It also contains a break at $\sim 1.5-3$ Gyr where the slope decreases with timescale, followed by a shallower break over $\sim 3-10$ Gyr where it rises again toward longer timescales (similar to IllustrisTNG). The break decreases in strength slightly with increasing stellar mass, and is probably tied to the inferred quenching behaviour learned from tying halo accretion to observed galaxy properties.
\end{itemize}

\subsubsection{Zoom simulations}
\label{sec:sfh_psds_diffsims}

\begin{figure*}
    \centering
    \includegraphics[width=0.9\textwidth]{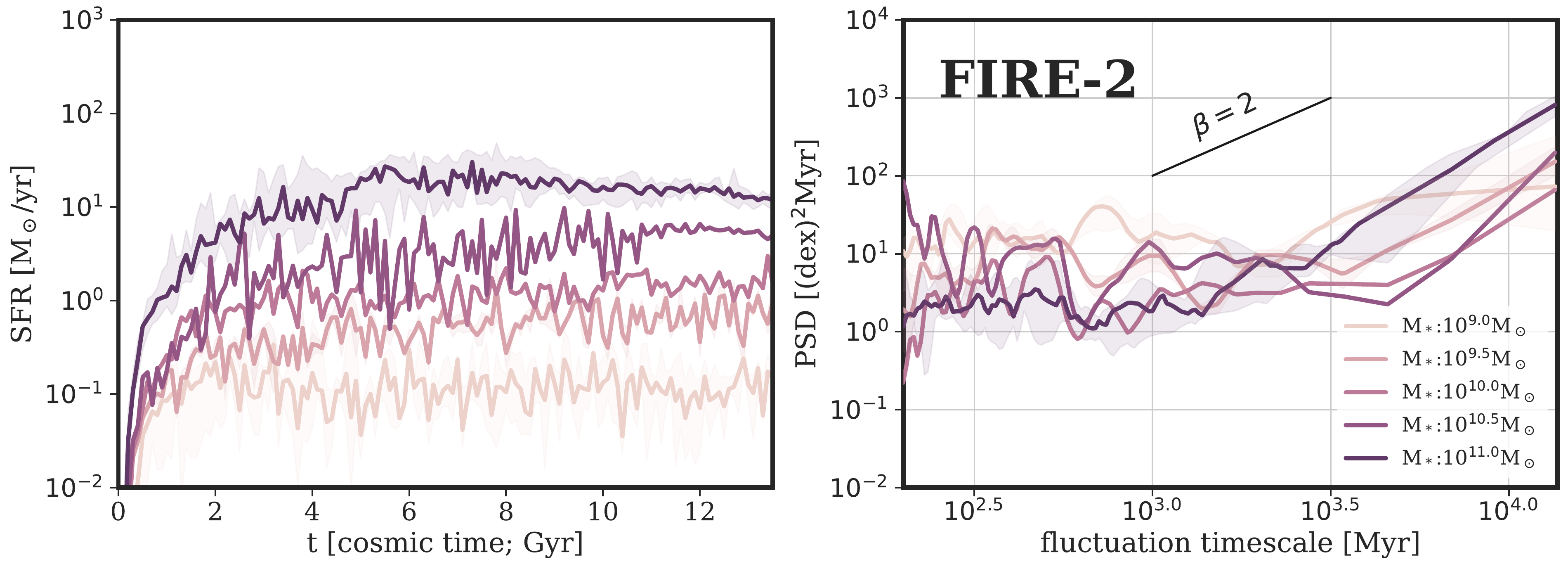}
    \includegraphics[width=0.9\textwidth]{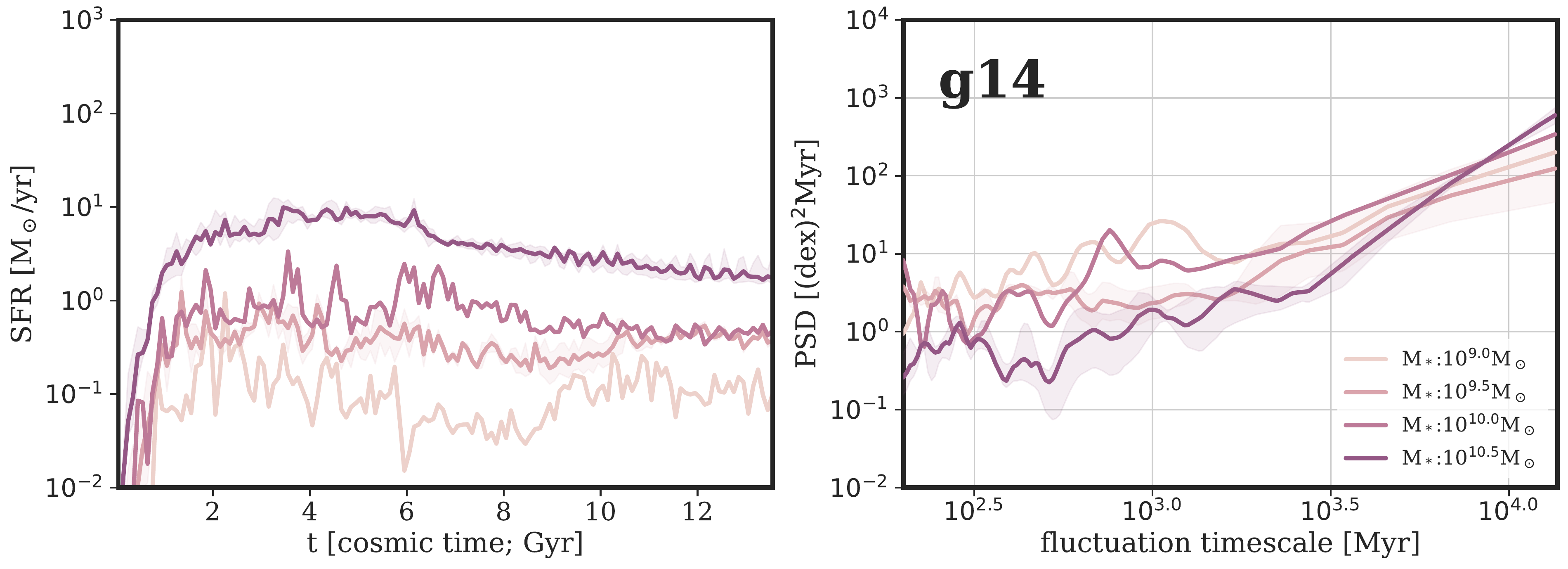}
    \includegraphics[width=0.9\textwidth]{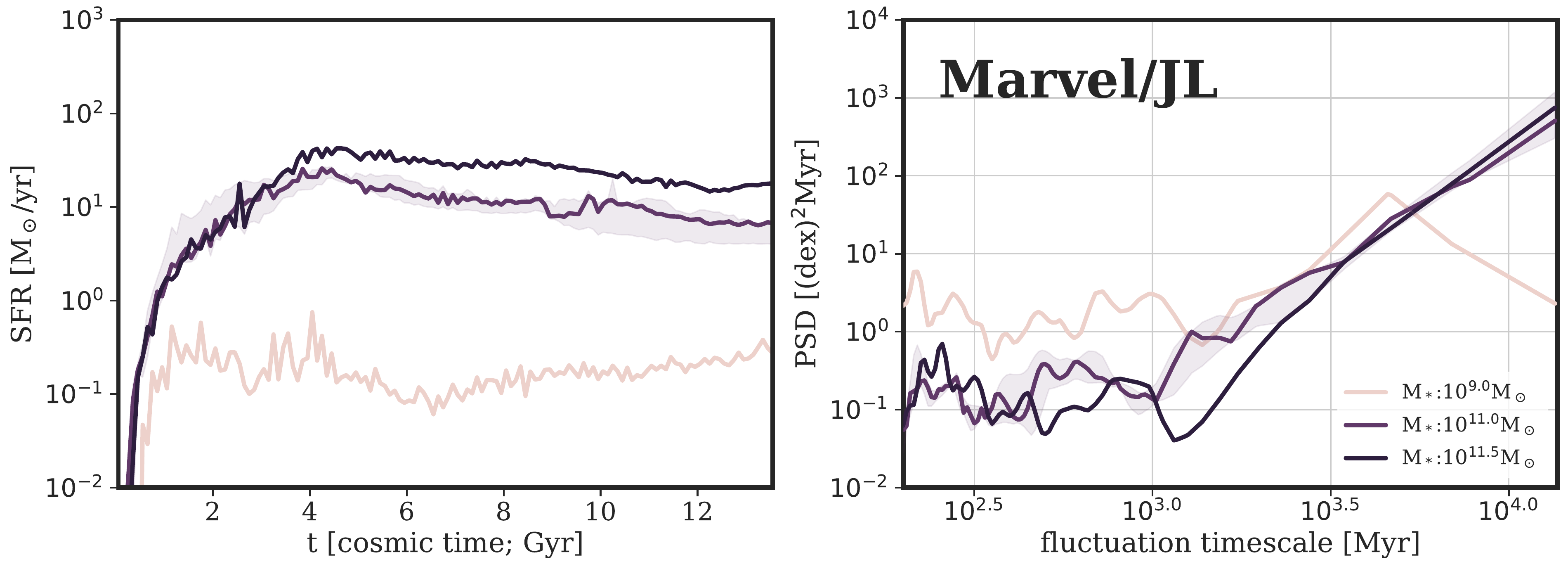}
    \caption[Median SFHs and PSDs for FIRE-2 galaxies]{Same as Figure \ref{fig:psd_hydrosims}, but for smaller samples of galaxies from the FIRE-2 and \brooks zoom hydrodynamical simulations.
    The FIRE-2 galaxies exhibit higher values of PSD at short timescales compared to the other models.
    The \brooks simulation shows lesser power on shorter timescales than FIRE-2 at a given mass. In addition, there is a stronger trend of increasing variability on shorter timescales as we go to lower masses.}
    \label{fig:psd_fire2}
\end{figure*}

\begin{figure*}
    \centering
    \includegraphics[width=0.9\textwidth]{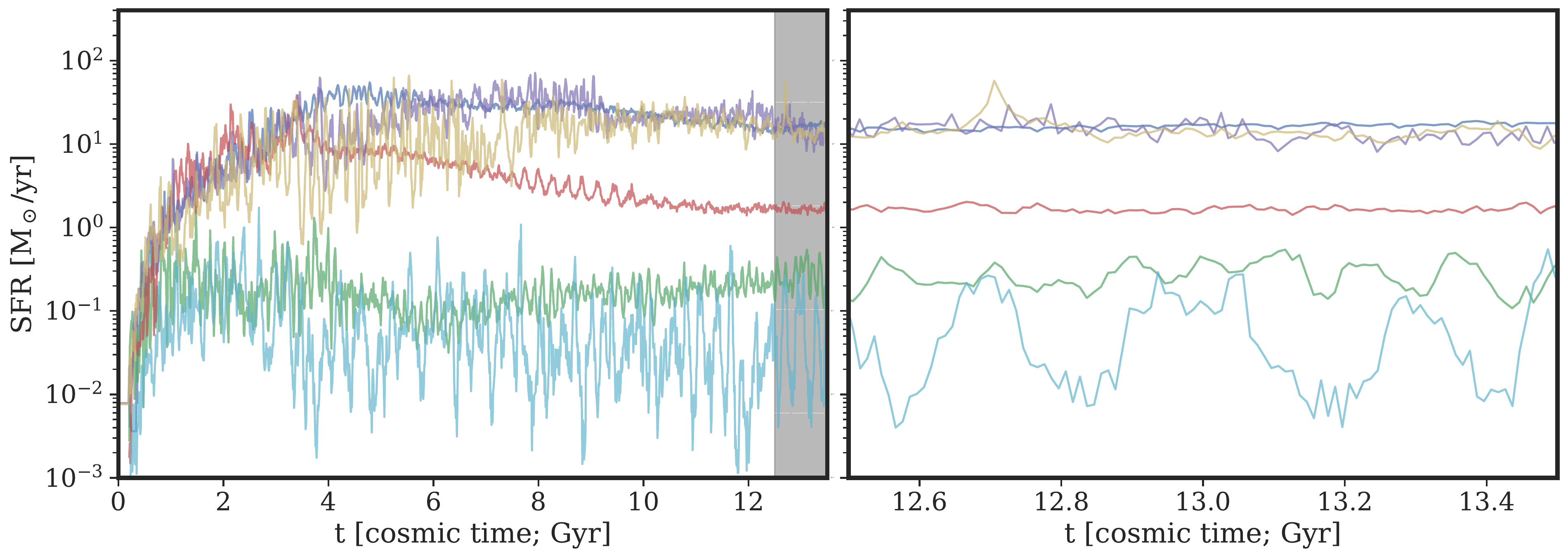}
    \includegraphics[width=0.92\textwidth]{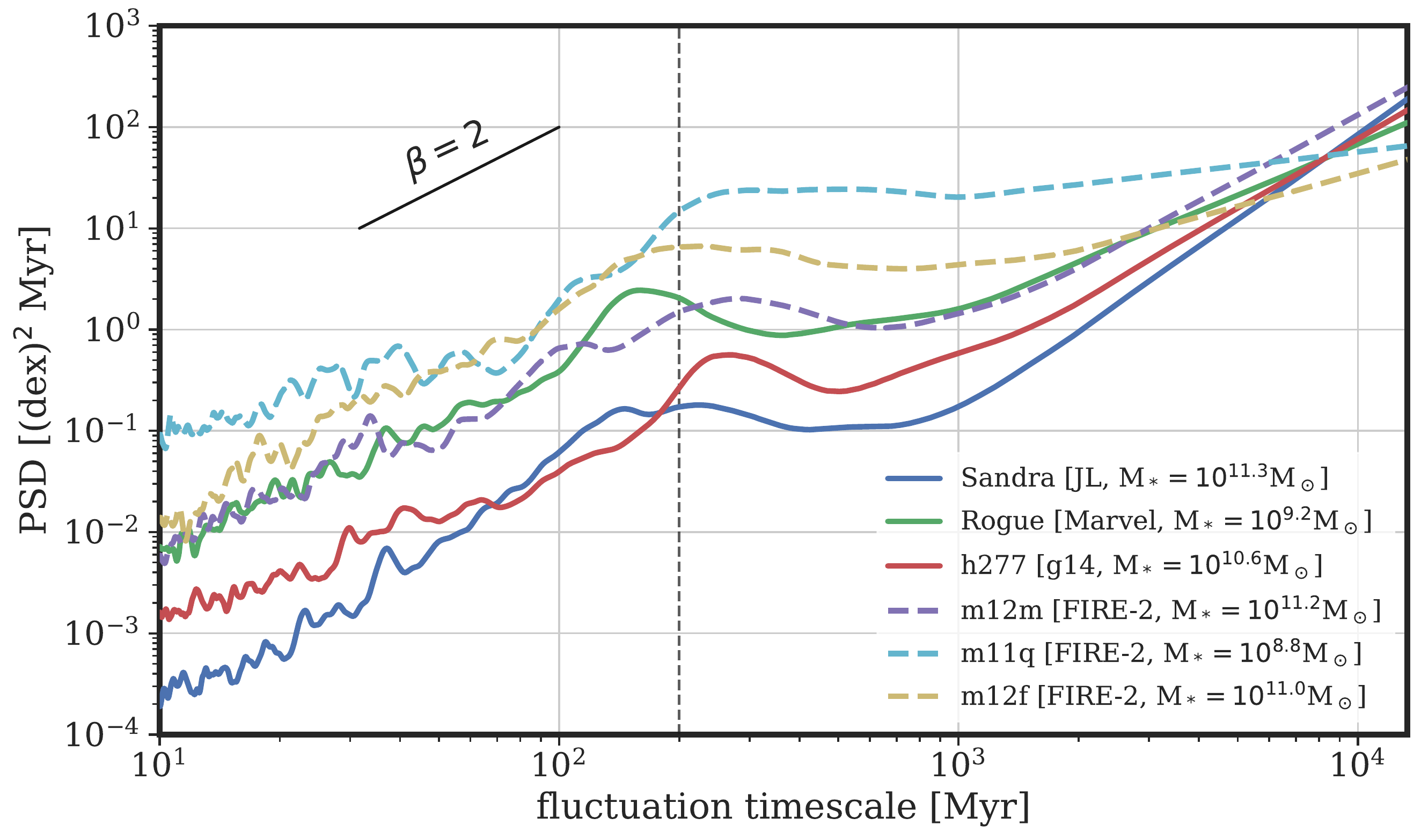}
    \caption[Probing short timescales with zoom sims]{The much finer resolution of  galaxies from the zoom simulations allows us to probe the PSDs of individual galaxy SFHs to much finer timescales than the large-volume models. We compute the PSDs for individual galaxies from the g14 (h277), Marvel (Rogue), JL (Sandra) and FIRE-2 (m12m - closest in stellar mass to Sandra, m11q -  an SMC-mass dwarf, m12f - a MW-like halo) suites of zoom simulations. The top panels show the full galaxy SFHs (left) and the SFHs over a period of $1$ Gyr (right), corresponding to the shaded region in the left panel. The bottom panel shows the corresponding PSDs. The vertical dashed line in the PSD plot shows the shortest timescales we probe with the large-volume simulations in Figures \ref{fig:psd_hydrosims}, \ref{fig:psd_hydrosims2} and \ref{fig:psd_sam_um}, an order of magnitude above what is possible with the zoom simulations. The overall slope of the PSDs continues down to shorter timescales, with the FIRE-2 galaxies showing more power on short timescales compared to the g14 and Marvel/JL galaxies.
    The PSDs of Rogue and h277 show a notable excess in the PSD at $\sim 100-300$ Myr timescales, while Sandra, m12m and m12f appear to show broader, less-prominent peaks spread over a longer range of timescales ($\sim 60-200$ Myr). m12f and m11q display a break in the PSD at $\sim 100$ Myr timescales, with a flattening of the PSD beyond that.
    Several galaxies also show distinct temporal dependence on variability, with m12f showing increased burstiness at earlier epochs, and Rogue showing oscillatory features at $t=7-10$ Gyr.}
    \label{fig:psd_fire2_single_galaxy}
\end{figure*}

In addition to the large-volume models, the FIRE-2, \brooks suites of zoom simulations, with star particles of $10^2 - 10^4 $M$_\odot$, allow us to (i) test the effect of much finer spatiotemporal resolution
that enables the simulations to resolve GMC-scale structures and
treat feedback more explicitly compared to the large-volume simulations
and (ii) probe specific parts of the PSD parameter space  (for e.g., shorter timescales) that are not accessible at present with large-volume cosmological models.

In Figure \ref{fig:psd_fire2}, we show the PSDs of 14 galaxies in FIRE-2, 8 galaxies in g14, and 5 galaxies in Marvel/JL that have $M_*>10^9$M$_\odot$. All the zoom simulations agree qualitatively with each other: the PSD is roughly constant between a timescale of $\sim 300$ Myr to $2-3$ Gyr and then increases toward longer timescales. Furthermore, the power around 1 Gyr increases toward lower masses in all three simulations, consistent with the idea that lower mass galaxies have burstier star formation than higher mass galaxies. Quantitatively, the zoom simulations show a few differences: galaxies in \brooks show less power on shorter timescales compared to FIRE-2 at a given stellar mass, indicating that they are less bursty in general. However, they show a stronger trend of increasing burstiness (i.e., power on shorter timescales) with decreasing stellar mass.

This behaviour of increasing power on short timescales toward lower mass galaxies can also been seen in the large-volume models like Illustris, IllustrisTNG, EAGLE and Mufasa. However, the presence of shot-noise at short timescales ($\lesssim 300$ Myr) portions of the PSDs makes this conclusion more difficult to draw. FIRE-2 shows a higher contribution to the power from shorter timescales compared to most large-volume simulations, with uniformly high power at all timescales $\lesssim 3$ Gyr that is comparable to Mufasa and Simba.

In Figure \ref{fig:psd_fire2_single_galaxy}, we show the PSDs of six individual galaxies from the three zoom suites --- h277 from g14 \citep{zolotov2012baryons, loebman2014milky, o2017effects}, Sandra from Justice League, Rogue from Marvel \citep[Munshi et al., in prep.]{bellovary2019multimessenger}, and m11q, m12f, and m12m from FIRE-2 \citep{hopkins2018fire}. m12m is a an early-forming halo hosting a MW-mass galaxy, and is closest in stellar mass to Sandra, and has a similar overall shape for the SFH. m12f is a MW-like galaxy. h277 is a MW analogue with no major mergers since $z=3$. Rogue and m11q are both SMC-mass dwarfs. More information about these galaxies and their physical properties can be found in the cited papers.

The increased resolution of the zoom simulations allow us to probe the PSDs down to much shorter timescales ($\sim 10$ Myr) than currently possible with the large-volume models.
The bottom panel of Figure \ref{fig:psd_fire2_single_galaxy} provides our first view of the PSD of simulated galaxies down to these timescales. We find that:
\begin{itemize}
    \item The broken power-law behaviour found in the PSDs on longer timescales continues down to the timescales of $\sim 10-30$ Myr.
    \item On short timescales, the PSDs show a slope of $\beta \sim 1-2$, with FIRE-2 tending towards a shallower slope with more overall power, consistent with increased burstiness.
    \item On timescales $\sim 100-300$ Myr, some PSDs show distinct peaks (h277 and Rogue). The absence of major mergers could play a part in setting the strength of this peak for h277 since h258, a similar g14 galaxy with a more active merger history does not display such a prominent peak and instead shows an elevated PSD overall.
    \item On timescales of $\sim 0.2-1$ Gyr, the PSDs flatten out ($\beta \sim 0$), before converging to a power-law with slope $\beta \sim 2-3$ on long timescales.
    \item The slopes of the FIRE-2 galaxies are generally shallower and have less power compared to galaxies in \brooks.
    \item Overall, lower mass galaxies like m11q and Rogue can sometimes display considerably higher power than their higher mass counterparts on timescales $\lesssim 6$ Gyr, in keeping with the trend of increasing burstiness with decreasing stellar mass.
\end{itemize}

The rich PSDs of these zoom simulations provide an excellent dataset to test and validate theories that connect physical processes to features in the PSDs. Specifically, these short timescales ($<100$ Myr) probe the gas cycle within galaxies, including the formation and disruption of star-forming clouds \citep{faucher2017model, jeffreson2018general, kruijssen2019fast}. Therefore, we might be able to use the PSD on these timescales to constrain the lifecycle of star-forming clouds \citep{tacchella2020stochastic}. Furthermore, the PSD is accessible from observations, since star formation rates estimated from $\mathrm{H\alpha}$ and the UV can allow us to constrain the slope of the PSD in this regime \citep{caplar2019sfhpsd}.

\subsection{The diversity in SFH shapes}
\label{sec:sfh_diversity_shapes}

\begin{figure*}
    \centering
    \includegraphics[width=\textwidth]{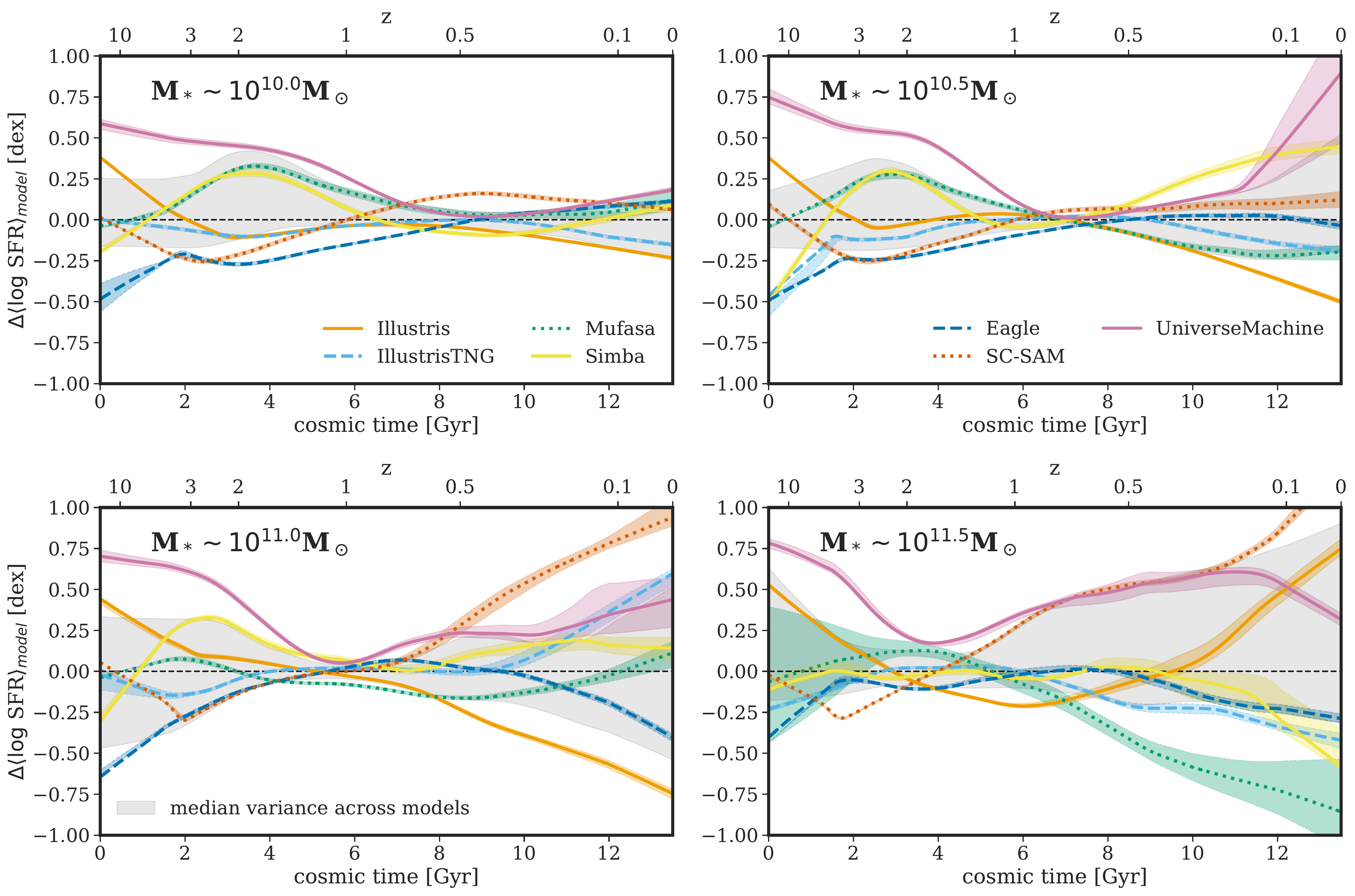}
    \caption{The diversity in the median SFHs for the different models. The dashed black line at $0$ dex corresponds to the median SFH of all galaxies in that mass bin across all the models, accounting for the differing number of galaxies from each model. Coloured solid lines show difference in log SFR space between this and the the median SFH for all galaxies from individual models. The shaded region shows the median variance ($84^\mathrm{th} - 16^\mathrm{th}$ percentile)/2 in the SFHs across all the models. At high redshifts, modeling differences give rise to high amounts of variability in galaxy SFHs. The differences between the different models is small for $z\lesssim 1$ in the $10^{10}$M$\odot$ bin, but rises in the higher mass bins as galaxies begin to quench and mass growth through merging gets more important.}
    \label{fig:sfh_diversity_allmodels}
\end{figure*}

In this section, we study how the different models deviate from the overall sample behaviour (and from each other) by quantifying the overall SFH diversity as a function of time and stellar mass. We compute the median SFH in a given mass bin for individual models
and compare it to the median SFH in a given mass bin across all the models.
To account for the differing number of galaxies in a given mass bin across the different models, we randomly sample 1000 SFHs with replacement from the available SFHs at each step of the calculation. We repeat this sampling and calculation 100 times to adequately capture small ($\sim 0.02-0.1$ dex) fluctuations due to random seeds.

The result is shown in Figure \ref{fig:sfh_diversity_allmodels}, which shows the difference between the median SFHs in different bins of stellar mass.
It should be noted that the median SFH across all models is not the `correct' SFH, but merely a guide to the eye. Therefore, instead of comparing the deviation from the median for any given model, it is more instructive to (i) look at the differences between the models themselves, and (ii) use the median to get an idea of the overall variance among models at a given mass and epoch (shown as shaded grey regions).
Although there is a considerable diversity across the different models, the largest differences occur when the SFR is low -- at early epochs when galaxies are beginning to assemble their mass and when they are quenching. A locus of agreement across the various models exists in each mass bin, moving to higher redshifts with increasing mass. This is correlated with the epoch when the median SFHs peak in their SFR, as seen in Section \ref{sec:sfh_psds_diffsims}. This means that despite these differences, the overall picture of galaxy mass assembly described by the models is similar.

At late times (low redshifts), there is an increase in the overall variance between the different models with increasing mass, ranging from $\sim 0.3$ dex at M$_*\sim 10^{10.0}$M$_\odot$ to $\gtrsim 1$ dex for massive galaxies (M$_*>10^{10.5}$M$_\odot$).
The median SFHs across all models agree well at $z<1$ in the lowest mass bin. These trends are not easy to interpret since, as we discuss in Section \ref{sec:get_sfhs_from_models}, these SFHs are tracing the SFR of the main progenitor as well as of all the accreted systems. This means that this late time divergence is probably a combination of how the various models implement quenching as well as the SFH of the accreted systems. Although a full treatment studying the cause of these differences is outside the scope of this analysis, quantifying the differences between the PSDs for these SFHs begins to illustrate how differing strengths of SFR fluctuations across a range of timescales could shape the overall SFHs over the next few sections.

Additional plots showing the distributions of SFH parameters such as stellar mass, sSFR, SFH peak and width for the various models can be found in Appendix \ref{sec:sfh_prior_dists}.

\subsection{Comparing PSDs across different models}
\label{sec:sfh_diversity_psd_slopes}

\begin{figure}
    \centering
    \includegraphics[height=180px]{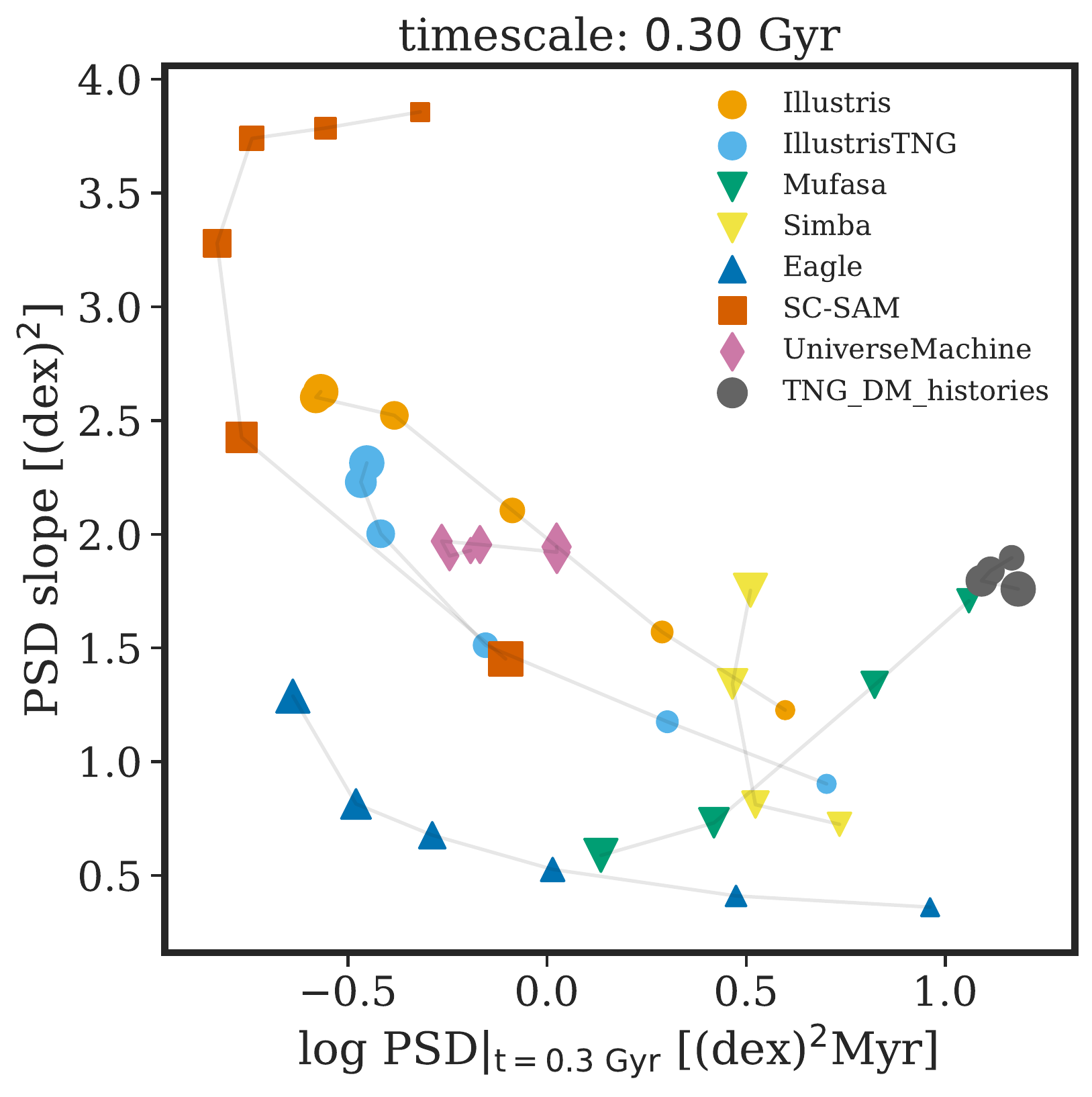}
    \includegraphics[height=180px]{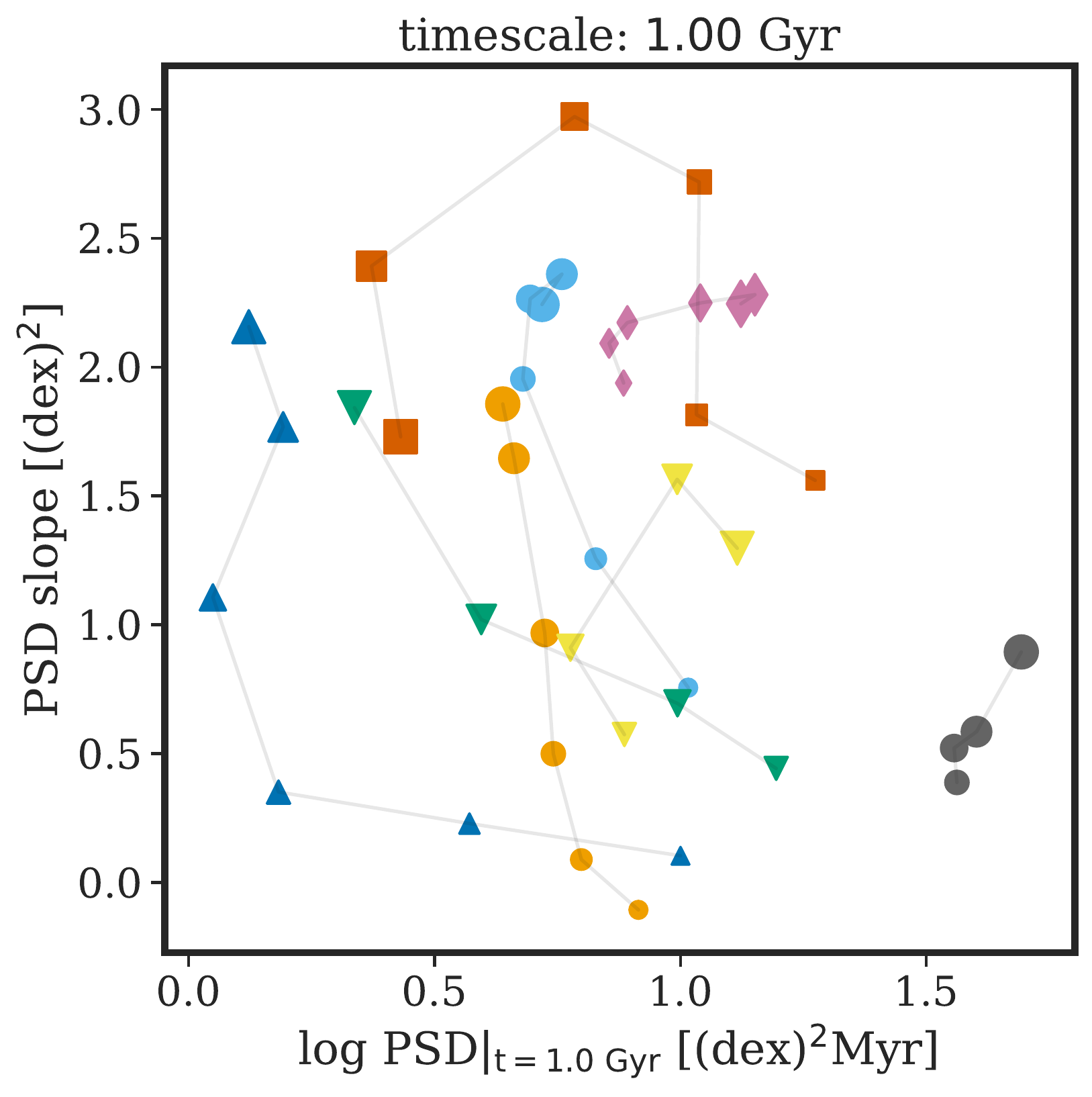}
    \includegraphics[height=180px]{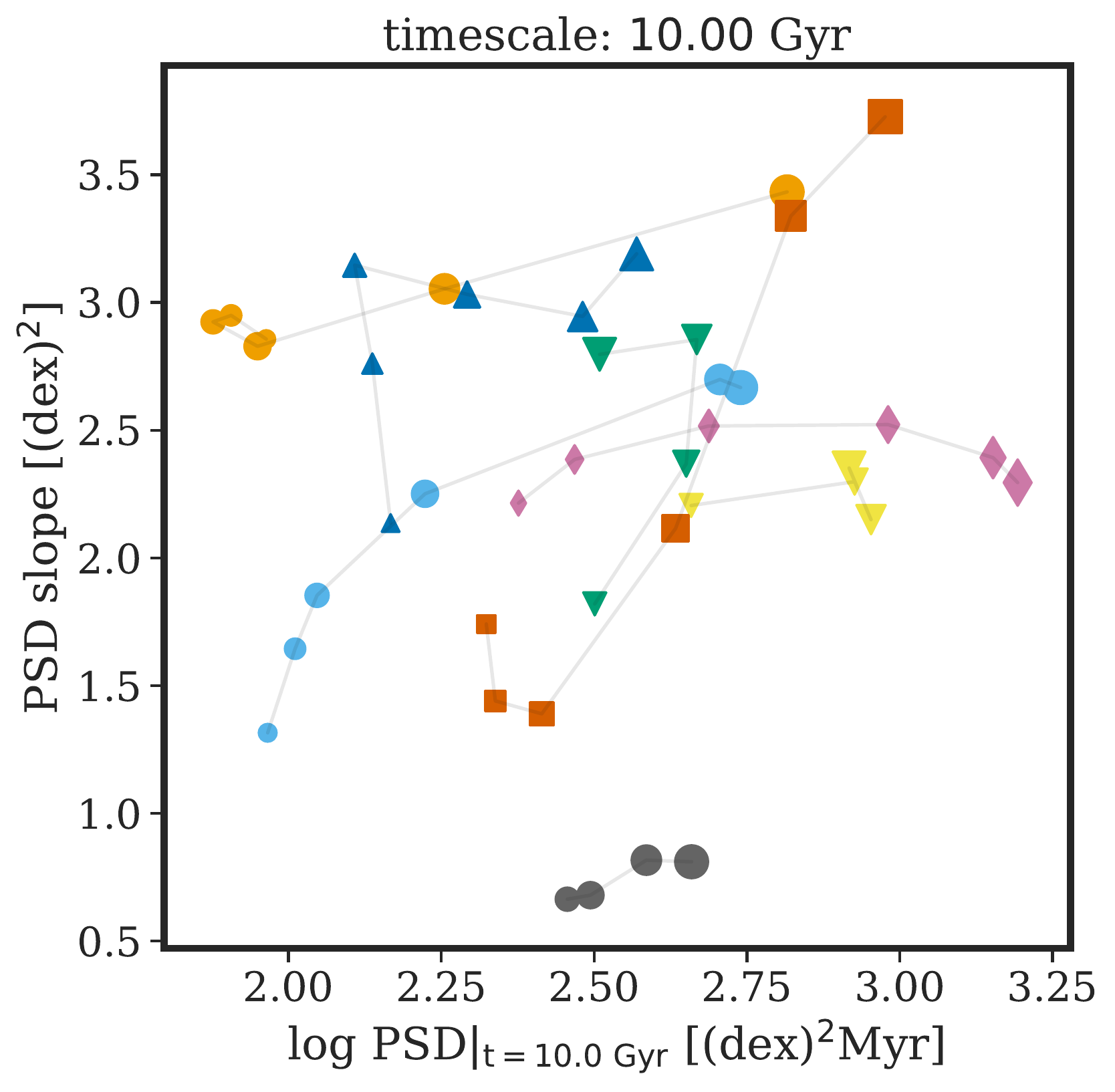}
    \caption{Quantifying the behaviour of the PSDs in slope-power space at different timescales. This amounts to taking cross-sectional slices of the PSDs in Figures 5,6,7 at 300 Myr (top) and 1 Gyr (middle) and 10 Gyr (bottom). The circle size increases with stellar mass, using the same 0.5 dex stellar mass bins as previous figures. The x-axis shows the overall power in the PSDs at different timescales and masses, while the slope indicates how tightly the timescales are coupled. Moving towards higher power and lower slope (bottom-right) increases how `bursty' the SFR is. While the PSDs inhabit a similar locus in slope-power space at shorter timescales, they show varied behaviour at timescales of $\sim 1$ Gyr. An interactive version of this plot can be found online at \url{https://kartheikiyer.github.io/psd_explorer.html}. }
    \label{fig:psd_slope_norm_comparison}
\end{figure}

\begin{figure}
    \centering
    \includegraphics[height=180px]{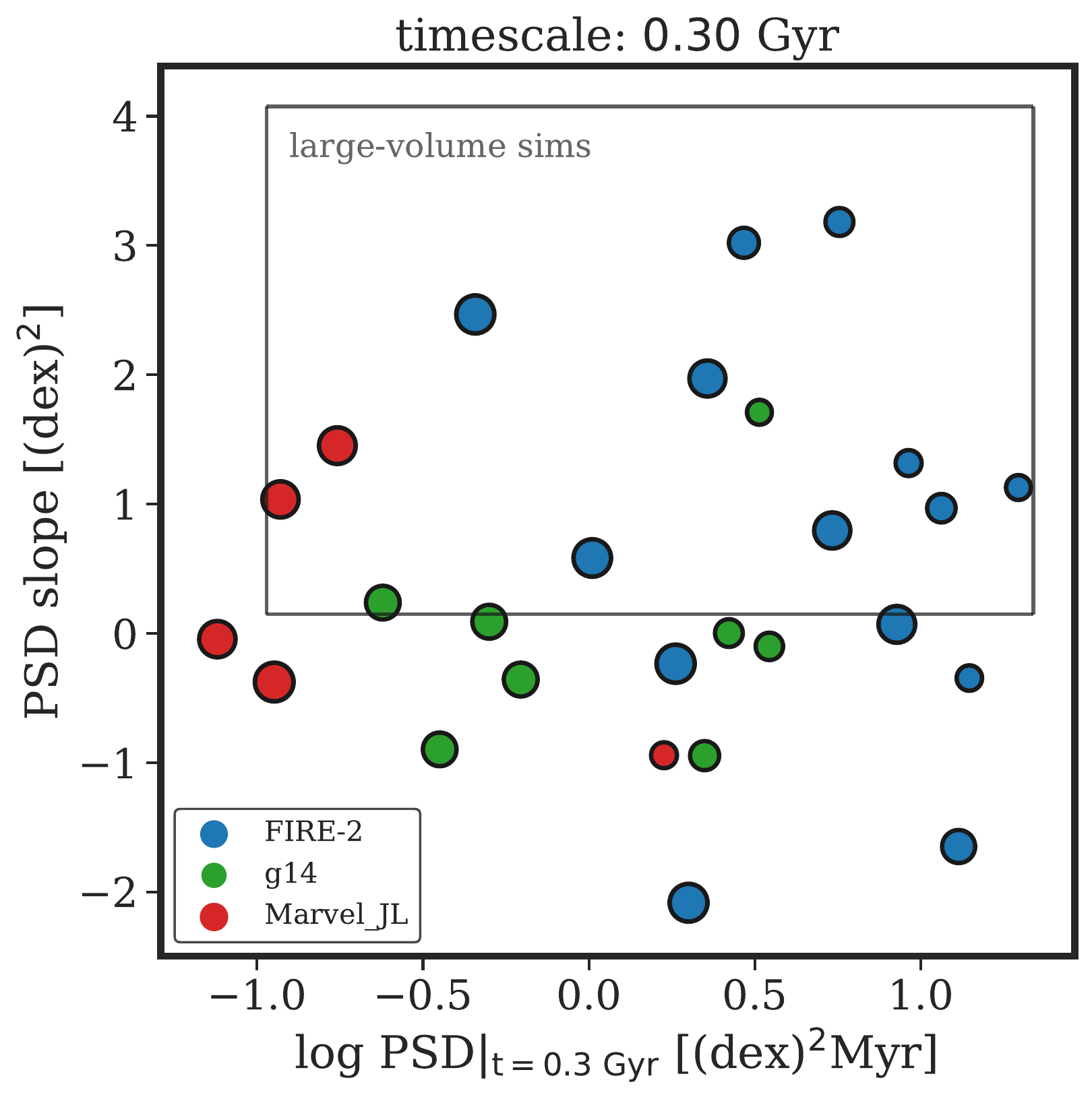}
    \includegraphics[height=180px]{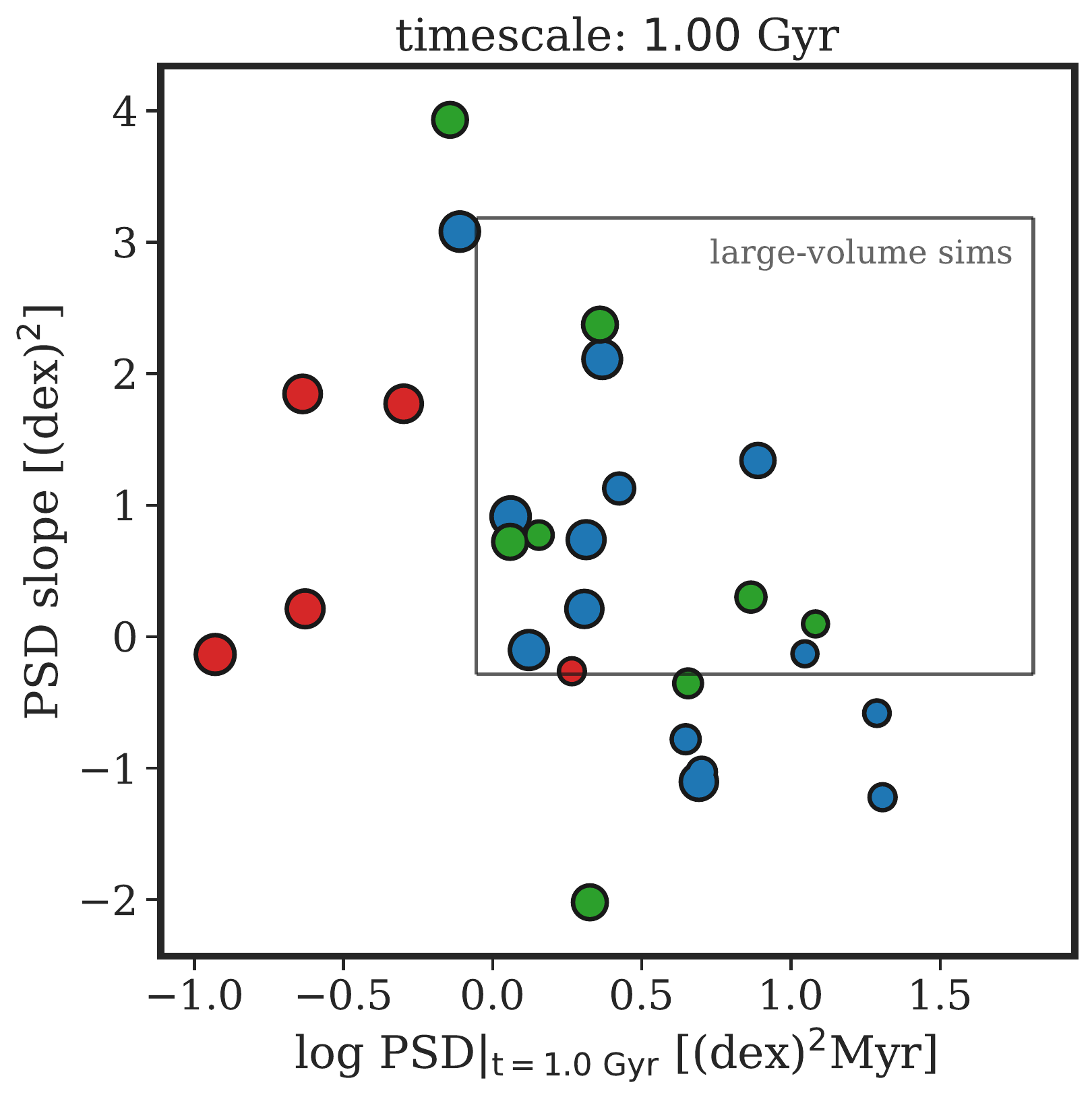}
    \includegraphics[height=180px]{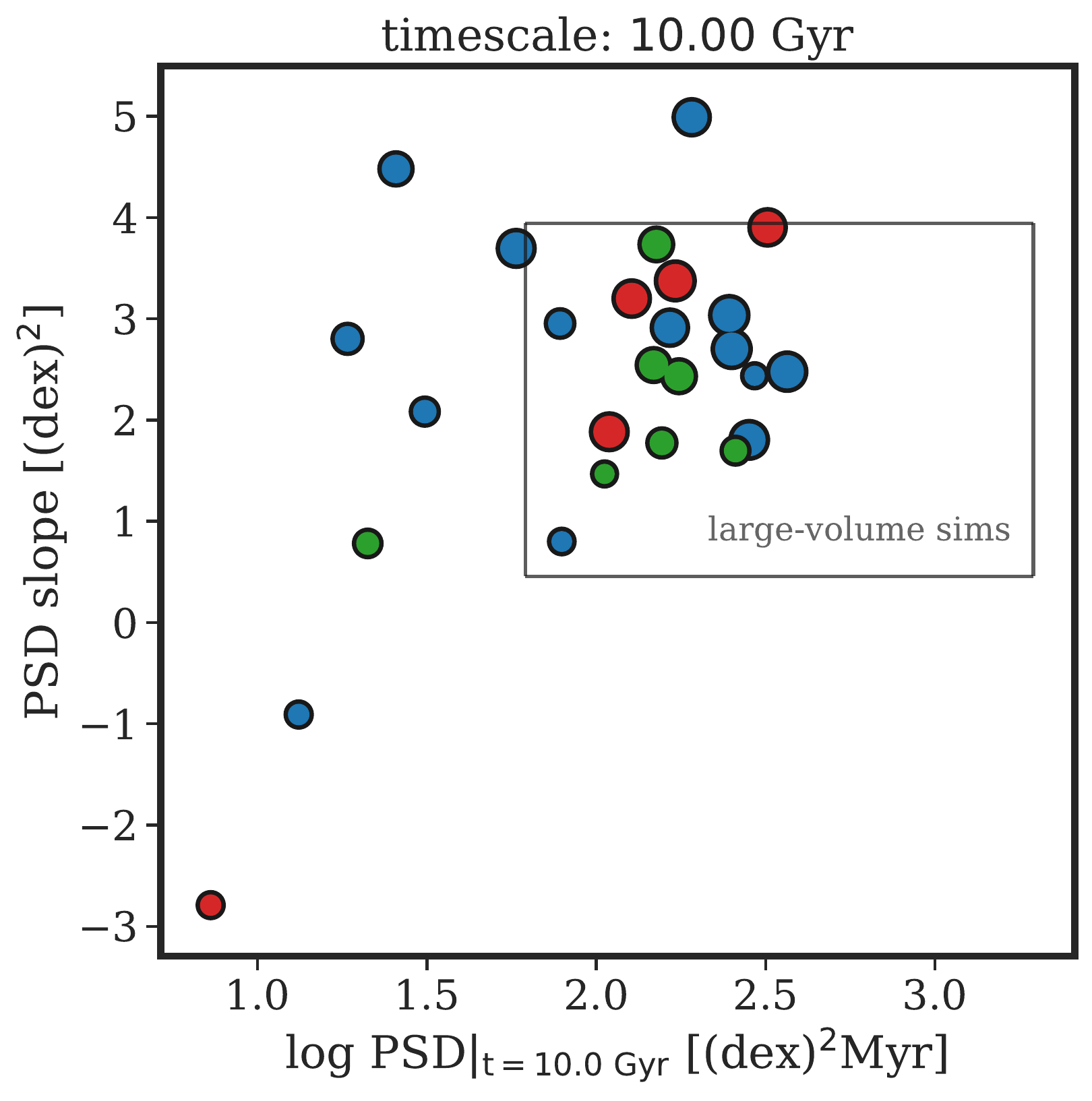}
    \caption{Similar to Figure \ref{fig:psd_slope_norm_comparison}, but for individual galaxies from the zoom simulations. The bounding boxes correspond to the edges of the corresponding panels in Figure \ref{fig:psd_slope_norm_comparison} for the median PSD slope and power from the large volume models. There is a notable trend towards increasing power with decreasing stellar mass on $\sim 300$ Myr timescales. While a similar trend is also seen in the large-volume hydrodynamical simulations, the lack of PSD contamination on the shortest timescales due to the significantly higher resolution of the zoom simulations makes this a more robust result, albeit with a much smaller sample.}
    \label{fig:zoom_psd_slope_norm_comparison}
\end{figure}

Section \ref{sec:psd_sim_results} describes some of the overall trends in the PSDs - the distribution of power across a broad range of timescales, with an increase in power towards longer timescales / shorter frequencies. However, each model shows unique trends in how the PSDs evolve with stellar mass, as well as the actual strength of the PSD at different timescales.

Since there are a range of modeling assumptions and numerical recipes used across the various models we consider, a comparison in PSD space serves to highlight the differences in the resulting variability of their SFHs on different timescales. Figure \ref{fig:psd_slope_norm_comparison} shows where the median PSDs of galaxies from the various models (as shown in Figures \ref{fig:psd_hydrosims}, \ref{fig:psd_hydrosims2}, and \ref{fig:psd_sam_um}) lie in PSD slope vs PSD power space at three representative timescales ($300$ Myr, $1$ Gyr and $10$ Gyr), and an interactive version of this plot spanning timescales ranging from $\sim 200$ Myr to $13$ Gyr can be found online\footnote{\url{https://kartheikiyer.github.io/psd_explorer.html}}. An equivalent plot showing individual galaxies from the zoom simulations is shown in Figure \ref{fig:zoom_psd_slope_norm_comparison}. The PSD power is the strength of SFR fluctuations or `burstiness' at a given timescale. The local slope of the PSD at a given timescale is computed using the PSD within a log timescale of $\pm 0.1$ dex, and is a measure of how tightly coupled the PSD is to adjacent timescales. Changing this interval while computing the slope does not affect the overall trends across the models.
A slope of 2 can be found in models of stochastic star formation described by random walks \citep{caplar2019sfhpsd, kelson2020gravity}. \citet{tacchella2020stochastic} find this to emerge naturally within the framework of the gas regulator model \citep{lilly2013gas} and in modeling stochasticity due to GMC formation and destruction.
Most high-mass and low-sSFR galaxies across the different models show a slope $\sim 2$, while UniverseMachine shows this at all stellar masses. Individual points for each model show the median slope and power of the PSDs in the same $0.5$ dex bins of stellar mass that are used in Figures \ref{fig:psd_hydrosims}, \ref{fig:psd_hydrosims2}, and \ref{fig:psd_sam_um}, highlighting evolution in PSD space as galaxies grow more massive.

\begin{figure*}
    \centering
    \includegraphics[width=\textwidth]{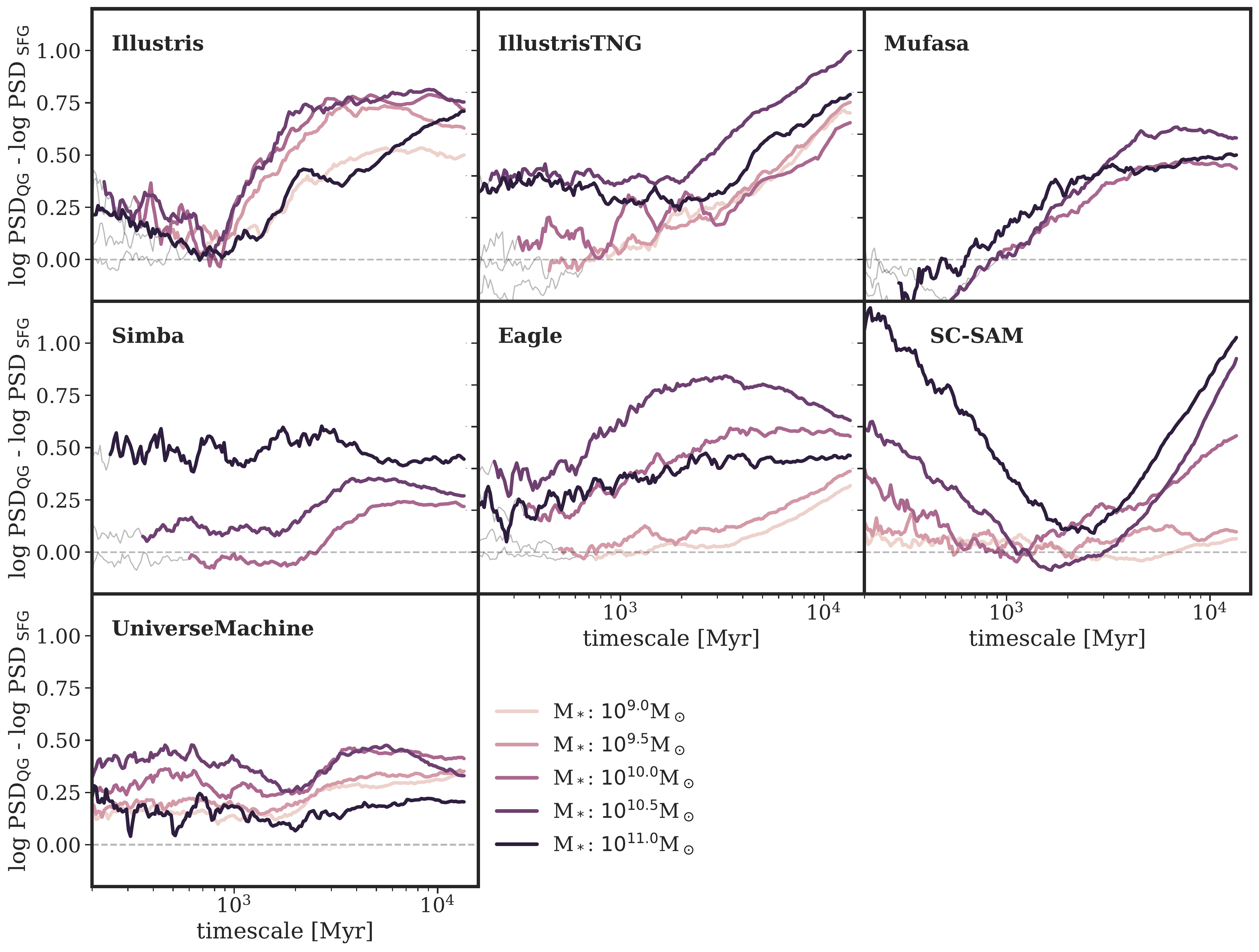}
    \caption{The difference between the median log PSDs of quiescent and star forming galaxies in $0.5$ dex bins of stellar mass. The bins are the identical to those in Figures \ref{fig:psd_hydrosims}, \ref{fig:psd_hydrosims2}, \ref{fig:psd_sam_um}, starting from $10^9$M$_\odot$ for all the large-volume models we consider except Mufasa and Simba, which start at $10^{10}$M$_\odot$ due to lower resolution.
    Coloured lines represent different mass bins, while grey curves denote regions where we expect resolution-dependent shot-noise to contaminate the PSDs.
    The PSDs of quiescent galaxies are notably greater than those of star forming galaxies on long timescales, with some models showing mass-dependent trends on shorter timescales.}
    \label{fig:sfg_quenched_psd_difference}
\end{figure*}

A key point to note is that the various models are extremely diverse in (i) the region of PSD space they occupy at a given timescale, and (ii) their evolution with stellar mass. At $\sim 300$ Myr timescales, there seems to be an overall attractor toward increasing slope and decreasing power as the stellar mass increases, although this does not hold for all the models. The short timescale power generally increases as a function of decreasing stellar mass, indicating that lower mass galaxies are generally more bursty across a variety of models. This trend is not as prominent for the SAM and empirical model. Meanwhile, at $\sim 1$ Gyr timescales the models seem to follow a range of different behaviours, although most models seem to converge on a PSD slope of $\beta \sim 2$ at high stellar masses. On the longest timescales, both slope and power tend to increase with increasing stellar mass, in part due to the increased contribution from quenched galaxies to the long-timescale PSD power. Depending on how the individual models implement quenching, however, the rate and extent of this effect can vary greatly (see Section \ref{sec:psd_quenching}).

It should be noted that although these trends are shown using the median values for the PSD slope and power in a given mass bin, there is a large amount of variance in the range of slope and power values possible for individual galaxy PSDs due to features that may be present in individual galaxy SFHs based on stochastic events like halo accretion fueled star formation and major mergers. The variance in slopes is from $\sigma ($PSD slope$) \sim 0.7-1.0$ (dex)$^2$ and in power is $\sigma ($log PSD$) \sim 0.2 - 0.7$(dex)$^2$Myr, corresponding to the shaded regions in the individual PSD plots and generally increasing with increasing stellar mass. While the large variance indicates that individual galaxies in a given mass range exhibit a large diversity in behaviour, trends across stellar mass are generally robust since they trace the behaviour of the entire population.

Given that the models span such a wide range in PSD slope and power at any given mass and timescale, observational constraints in this space \citep{caplar2019sfhpsd, wang2020var} would provide strong constraints on modelling galaxy physics.

\subsection{The PSDs of star-forming vs quiescent galaxies}
\label{sec:psd_quenching}

Quenching becomes an increasingly important phenomenon as we consider galaxies with higher stellar masses. This phenomenon can be driven by a range of different physical processes acting on different timescales. Since the quenching of galaxies is an observably measurable phenomenon, it is therefore possible to get observational constraints on quenching timescales and connect them to the relevant physical processes. Here we explore the differences in the PSDs of actively star-forming and quiescent galaxies at $z=0$ to determine what, if any, differences they show at different stellar masses.

To perform this analysis, we first need to select galaxies that are quiescent at the time of observation. There exist multiple ways of performing this selection, depending on the definition of quenching (e.g., through a cut in UVJ space, in specific SFR, or a threshold distance from the SFR-M$_*$ correlation, among others, see for example \citealt{donnari2018star, hahn2018iq}). In the current analysis, we separate galaxies into star forming vs quiescent using a commonly used threshold in sSFR ($\mathrm{sSFR} < 0.2/ \tau_{\mathrm{H}} \sim 10^{-10.83} yr^{-1}$ for quiescent galaxies at $z\sim 0$, see \citet{pacifici2016timing, carnall2018measure}). This approach is motivated by two reasons: (i) since we already have access to the SFHs, this allows us to avoid the systematic assumptions of forward modeling rest-frame UVJ colours and the degeneracies of separating quiescent galaxies in that space, and (ii) we avoid the systematics of accounting for different SFR-M$_*$ correlations across the different models \citep{hahn2018iq} and use a uniform threshold for comparison across the models.

Having identified quiescent galaxies across the various models, we then compare the PSDs of quiescent galaxies to those of star-forming galaxies at different stellar masses.
Since quenching distinctively alters the shape of a galaxy's SFH, we expect the PSDs of quiescent galaxies to show more power on long timescales.
Figure \ref{fig:sfg_quenched_psd_difference} shows the difference in the median log PSDs of quiescent and star forming galaxies in the same $0.5$ dex stellar mass bins used in the rest of this work. We exclude the highest mass bin (M$_*\sim 10^{11.5}$M$_\odot$), since there are not enough star forming galaxies in all the models to perform this analysis.

We see that at low stellar masses, the quiescent galaxy PSDs generally show greater power on long timescales, with the exact timescale varying across models, ranging from $\sim 900$ Myr to $\gtrsim 3$ Gyr. As we go to higher stellar masses, we find that there is an excess of power
\updated{across a range of shorter timescales in the IllustrisTNG, Simba, EAGLE, and SC-SAM models}. This could be explained by processes driving quenching also driving variability in SFRs on other timescales. For example, multiple short episodes of feedback due to (i) AGN-driven outflows leading to the eventual quenching of galaxies, as seen in the implementation of jet mode AGN feedback \citep{rodriguez2019mergers} or (ii) X-ray feedback rapidly evacuating the star forming gas in  the central regions \citep{appleby2019impact} could lead to increased short-timescale variability in Simba. In contrast, since Mufasa implements quenching primarily through a 'maintenance mode' feedback, it does not show a strong evolution with stellar mass.
Another explanation for this increase in power on short timescale for quiescent galaxies could be that the quiescent galaxies assemble their mass earlier, when SFHs in general were more bursty \citep{muratov2015gusty, hayward2017stellar}.
The phenomenon of quiescent galaxies assembling their mass earlier can be seen the correlation between t$_{50}$ and sSFR for quenched galaxies among the various models shown in Appendix \ref{sec:sfh_prior_dists}. However, the nature of this correlation is uniform across all the models and can not fully account for the variations in the difference between star-forming and quenched galaxy PSDs observed between the models.

In more detail, the Illustris and IllustrisTNG models both show sharp breaks above which the power in quiescent galaxies rises, with the break occurring on longer timescales in IllustrisTNG. The increase in power on short timescales with increasing mass is also more prominent in TNG compared to Illustris. The updated winds and AGN feedback in IllustrisTNG also show a noticeable increase in power on short timescales above masses $ 10^{10.5} $M$_\odot$, where AGN feedback becomes most effective. EAGLE shows a much broader range of timescales in comparison, similar to Simba albeit with high power at a given mass. The SAM shows a significant increase in power on timescales below $\sim 2$ Gyr, with this trend increasing with stellar mass. This seems to be primarily associated with stochastic starbursts on short timescales triggered by mergers, with more massive galaxies experiencing these events to a larger extent. UniverseMachine shows a moderate increase in power with quenching over all timescales.

In summary, PSDs across the different models show a range of behaviours when galaxies quench, with strong mass dependence in some models (IllustrisTNG, Simba, EAGLE, SC-SAM) and a range of timescale-specific breaks in the PSD ($\sim 900$ Myr in Illustris, $\sim 2-3$ Gyr in IllustrisTNG, $\sim 3-4$ Gyr in Simba, and $\sim 2$ Gyr in the SC-SAM). Observational constraints in PSD space for star forming and quiescent galaxy populations will provide sensitive probes of discriminating between the range of quenching mechanisms implemented across these models.

\subsection{How dark matter accretion shapes PSDs}
\label{sec:dmfh_psds}

\begin{figure*}
    \centering
    \includegraphics[width=420px]{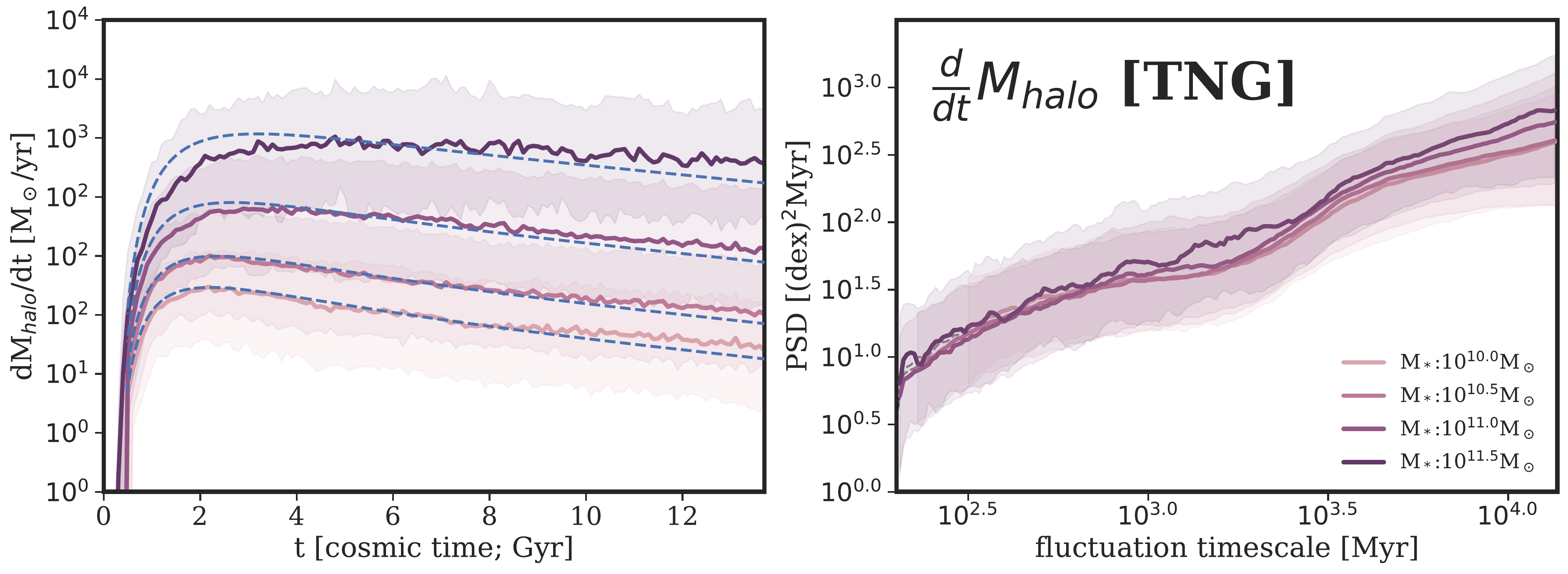}
    \caption{Equivalent to Figure \ref{fig:psd_hydrosims}, halo mass accretion histories and corresponding PSDs for the parent halos of galaxies in the IllustrisTNG simulation, in bins of stellar mass. In comparison to the SFH PSDs, the halo accretion history PSDs show a remarkable self-similarity for galaxies in different bins of stellar mass. The dashed blue lines in the left panel provide a comparison to the median DM accretion histories computed using the EPS formalism as outlined in \citet{correa2015accretion}, calculated using the median DM halo mass for each stellar mass bin.}
    \label{fig:DMhist}
\end{figure*}

Upon examining the PSDs of galaxy SFHs across different models, we find that most of the power resides in the long timescales on which SFHs rise and fall.
At early cosmic times, several models find the SFHs of galaxies to be correlated with the dark matter accretion histories (DMAHs) of their parent haloes \citep{2018ARA&A..56..435W}.
\citet{diemer2017log} model galaxy SFHs as log-normal curves, and find that the peak and width of SFHs in Illustris correlate strongly with the properties of their DMAHs with an offset between the formation times of haloes and galaxies that increases with stellar mass, along with a tight relation between the BH mass and peak time.
Similarly, \citet{qu2017chronicle} show that the SFHs of galaxies in EAGLE increasingly decorrelate from the halo accretion histories at increasing masses, by plotting the formation time vs accretion time for haloes and galaxies across stellar mass bins. They find this to be due to AGN feedback, which suppresses in-situ star formation and causes the stars in massive galaxies to form early and the galaxies to grow subsequently by mergers (i.e., the majority of star formation finished early), while haloes continue accreting mass until late times (i.e., massive haloes form late) \citep{neistein2006downsizing}.

From an analytical standpoint, \citet{kelson2014decoding} models star formation as a stochastic timeseries, with the `long-timescale memory' encapsulated by a Hurst parameter of $\sim 0.98 \pm 0.06$.
In \citet{kelson2016origin}, this model is extended to derive stellar mass functions at early times, explicitly relating the variance of the SFRs for an ensemble of galaxies to the DM haloes and their ambient matter densities at the epoch when star formation begins. \citet{kelson2020gravity} analytically estimate the slope of the DMAH PSD to be $\sim 1$.

With this in mind, it would therefore be instructive to (i) compute the PSDs of DMAHs and study their behaviour, (ii) study the extent to which variability in galaxy SFHs is tied to the variability in the DMAHs of their parent haloes, and (iii) examine if SFHs and DMAHs are coherent, to understand if dark matter accretion drives star formation.

\subsubsection{The variability of dark matter accretion histories}

We compute the PSDs for a sample of dark matter accretion histories from IllustrisTNG, defined as $\Delta M_{\mathrm{halo}}$ from one time step to another with the same $\Delta t = 100$ Myr bin width. The halo accretion histories are computed using the Friend-of-Friend (FOF) and SUBFIND algorithms \citep{davis1985evolution, springel2001populating, dolag2009substructures}, by selecting galaxies with $M_* > 10^9$M$_\odot$ at $z\sim 0$ and tracing them back in time to find all the dark matter particles associated with the halo of the main progenitor at each snapshot from $z\sim 20$ to $z=0$, described in detail in  \citet{pillepich2017simulating}. While this does not correspond directly to the full SFH that we have been considering so far, it is possible to relate it to the in-situ SFH of the central progenitor, and then connect the in-situ SFH to the full SFH. Since the halo accretion histories are only accessible at the discrete timesteps of the IllustrisTNG snapshots, they have been interpolated to match the uniform time-grid used throughout the rest of this work. Comparing the computed PSD after this interpolation to periodograms computed using the original uneven snapshot timesteps do not show any significant differences. We also repeated the analysis with different halo mass definitions based on the DM mass within certain fractions of $R_{\mathrm{crit},200}$ or within fixed distances of $10$, $50$ and $100$ kpc from the center of the halo potential, and found that the resulting trends do not change significantly.

Figure \ref{fig:DMhist} shows the dark matter accretion histories (DMAHs) of galaxies in IllustrisTNG across four bins in stellar mass. For each stellar mass, the median DM halo masses in the $0.5$ dex bin are: M$_\mathrm{halo} \sim 10^{11.61}$, $10^{11.89}$, $10^{12.37}$, and $10^{12.89}$M$_\odot$, corresponding to stellar masses of M$_*\sim 10^{10}$, $10^{10.5}$, $10^{11}$, and $10^{11.5} $M$_\odot $respectively. The dashed blue lines show the average DMAHs based on the extended Press-Schechter (EPS) formalism \citep{press1974formation, bond1991eps, lacey1993eps}, which provides an approximate description for the hierarchical growth of DM haloes from an initial Gaussian density field as a stochastic process.
Specifically, the accretion histories were computed using an analytic model derived from the EPS formalism described in \citet{correa2015accretion}. The analytic curves are a good match to the IllustrisTNG DMAHs, and show a rise and a slight subsequent decline described by the relation M$_\mathrm{halo}(z) = M_{0} (1+z)^{af(M_{0})}e^{-f(M_0)z}$, where $M_0$ is the mass of the halo at $z\sim 0$, $a$ depends on cosmology and $f(M_0)$ is related to the linear (spatial) matter power spectrum.
We find that the PSDs show a remarkable self-similarity, with a slight increase at the longest timescales corresponding to the overall normalisation of the halo mass. The PSDs also show a `plateau'-like behaviour at $\sim 1-3$ Gyr timescales, i.e., a sharp break toward increasing PSD slope from $\beta \sim 0.3$ to $\beta \sim 1.6$, followed by a break towards slopes of $\beta \sim 0.6-1$ on long timescales. However, this trend is weak in the highest-mass bin. Although outside the scope of the current work, the physical origin of this feature could be independently verified by comparing against PSDs from DM-only simulations. Such an analysis will necessitate a slightly different sample selection approach, since here we simply computed the PSDs for the DMAHs of the parent halos of galaxies in fixed stellar mass bins.

The slopes also increase to $\beta \sim 1.6$ as we approach the shortest timescales probed. Overall the median PSD slopes are $\sim 1$, consistent with the analytical derivation of \citet{kelson2020gravity}. On long timescales, haloes are thought to grow by smooth accretion, while on shorter timescales they grow by merging \citep{dekel2013toy}. Understanding the origin of this plateau, and whether it can be derived within the EPS formalism\footnote{i.e., relating the \textit{spatial} matter density power spectrum to the \textit{temporal} mass accretion history power spectrum, see \citet{kelson2020gravity}.} is therefore an interesting challenge for models of halo growth.

\begin{figure}
    \centering
    \includegraphics[width=230px]{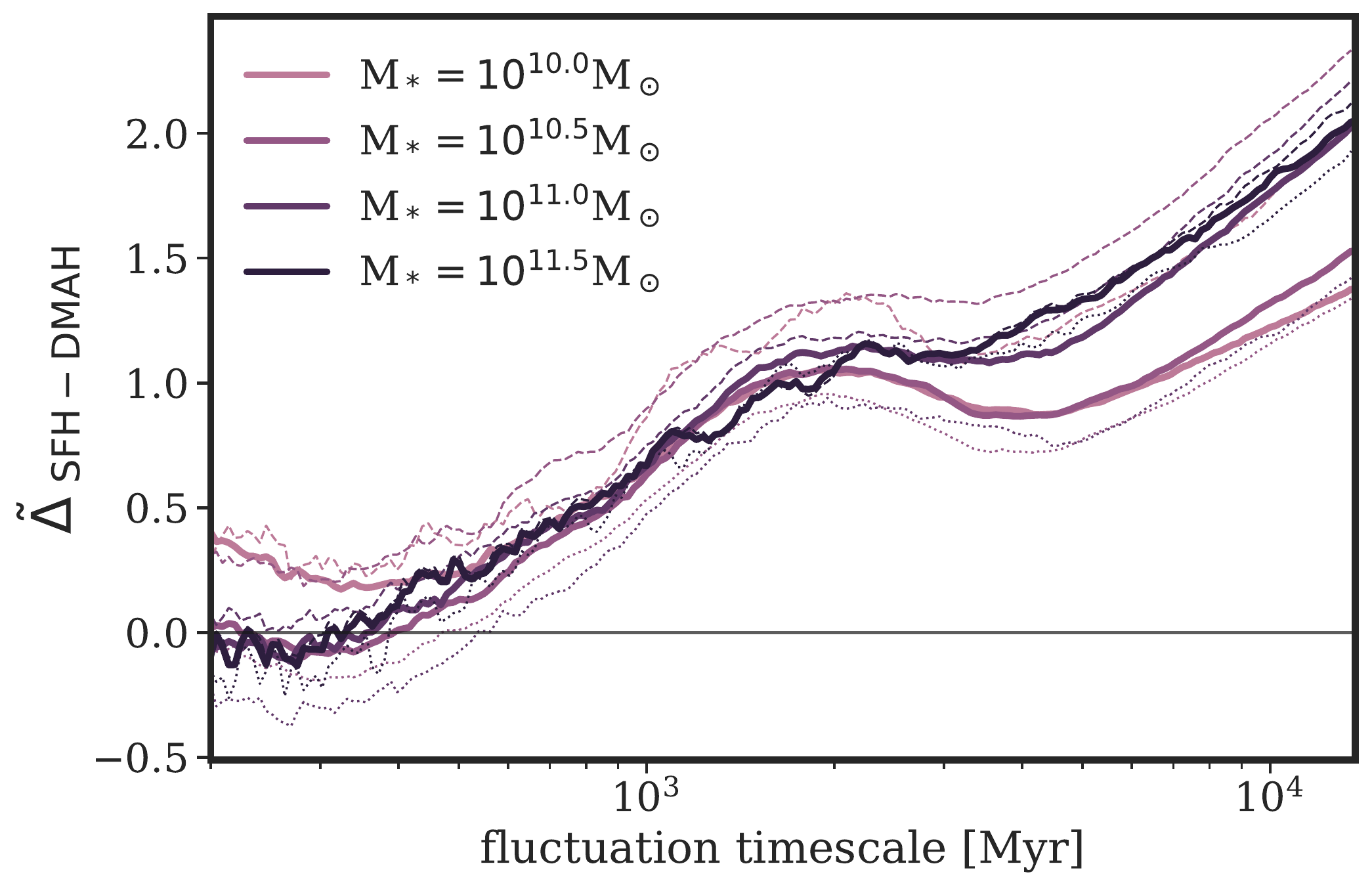}
    \caption{The difference between the median dark matter accretion history (DMAH) and SFH PSDs for IllustrisTNG ($\tilde{\Delta}_{\rm SFH-DMAH} = \log {\rm PSD}_{\rm SFH} - \log {\rm PSD}_{\rm DMAH. scaled}$). Since the DM accretion rates are generally higher and have more variance than their corresponding SFHs, the PSDs for each halo are scaled by a factor of $M_*/M_\mathrm{halo}$ prior to computing the median DMAH PSDs in a given mass bin. Thick solid lines show the median difference in PSDs corresponding to 0.5 dex mass bins centered at the values shown in the legend. Dashed lines show the difference in PSDs for quiescent galaxies, while dotted lines show the PSD difference for star forming galaxies.
}
    \label{fig:dm_excess_power}
\end{figure}

\begin{figure}
    \centering
    \includegraphics[width=230px]{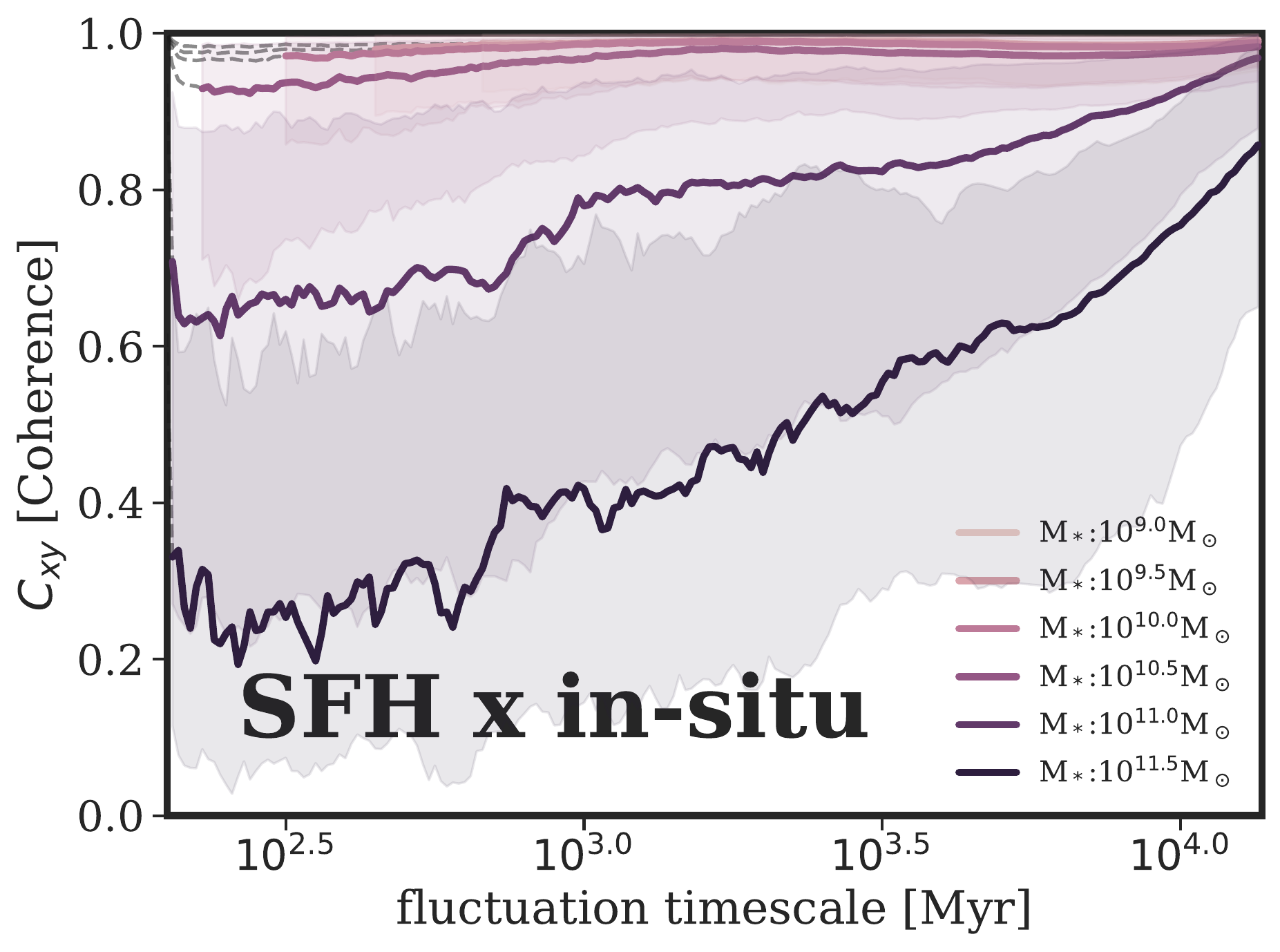}
    \includegraphics[width=230px]{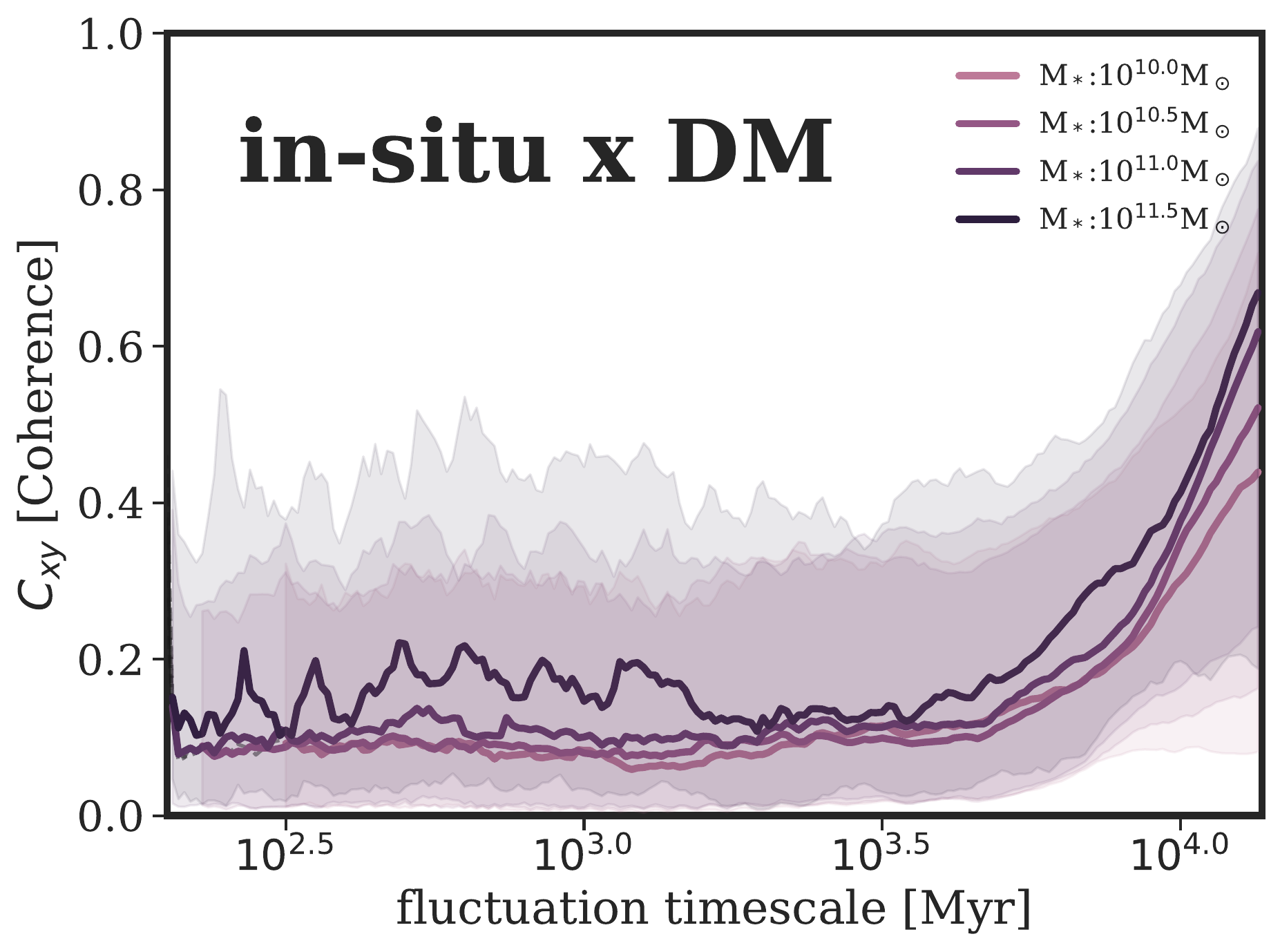}
    \caption{The coherence between the in-situ component of the SFHs to the full SFHs (top), and that of the in-situ SFHs with the dark matter accretion histories (bottom) of their parent haloes at different timescales. Coherence is defined as $C_{xy} = P_{xy}^2/|P_x P_y|$. As more massive galaxies grow a greater fraction of their mass ex-situ due to mergers, they increasingly decohere from their in-situ SFHs on shorter timescales. The PSDs of dark matter and in-situ SFR are largely mass invariant and only weakly related at short timescales, where baryonic processes dominate. The slightly higher coherence in the highest-mass bin on shorter timescales could be due to short-lived bursts of star formation induced by mergers.
}
    \label{fig:sfh_dm_coherence}
\end{figure}

\subsubsection{Comparing the PSDs of DMAHs to SFHs}

Having computed the PSDs of DMAHs, we would now like to compare them to the PSDs of SFHs computed earlier. To this end, Figure \ref{fig:dm_excess_power} shows the excess power in the median PSDs of galaxy SFHs from IllustrisTNG compared to those of DMAHs across bins of stellar mass $0.5$ dex wide.

Since the DMAHs generally have a higher overall normalisation and correspondingly larger fluctuations due to considering the accretion and mergers of the entire haloes instead of just their baryonic component, we normalize the PSD for each central galaxy - parent halo pair by the ratio of their stellar mass to halo mass, in effect bringing the DMAHS to the same scale as the SFHs. Doing so allows us to compare their PSDs on a similar footing. Note that this is not a perfect comparison, since it assumes that the $M_\mathrm{halo}-M_*$ ratio is roughly constant throughout cosmic time. However, this assumption only needs to hold for ensembles of haloes, and is motivated by studies that find a only a mild evolution of baryon fraction with redshift \citep{crain2007baryon} in conjunction with extensions to central galaxies \citep{kulier2019evolution}.

Figure \ref{fig:dm_excess_power} finds that the excess power in SFHs in comparison to DMAHs lies mostly on longer timescales, which can also be inferred from the steeper slopes of their PSDs - IllustrisTNG SFHs have median slopes of $\beta \sim 2 \pm 0.4$, compared to DMAHs, whose PSDs have median slopes of $\beta \sim 1\pm 0.4$.
On the shortest ($\sim 200$ Myr) timescales, the DMAHs have comparable power to the IllustrisTNG SFHs. While this does not imply that DM accretion is driving the variability on these timescales, it is a helpful coincidence that accounts for why \citet{mitra2016equilibrium, rodriguez2016main, kelson2020gravity} get the right scatter for the SFR-M$_*$ correlation using models that correlate SFRs with DM accretion rates, without having to invoke arguments of SFR-regulation by feedback.
There is a noticeable plateau in the DMAH PSDs that translates to a coherent feature at $\sim 1-3$ Gyr in Figure \ref{fig:dm_excess_power}, although the prominence of this feature decreases with stellar mass.
A portion of this excess power on long timescales appears to come from quenching, which decorrelates when galaxies form their stars from when haloes assemble their mass. This can be seen from the difference between the median PSD difference between SFHs and DMAHs for star forming (dotted) and quiescent (dashed) galaxies in a mass bin, and in Section \ref{sec:timescale_sim}.
Even with quenching accounting for up to $\sim 1$ dex of power on timescales $\gtrsim 3$ Gyr, there still remains an excess of about $\sim 0.8-1$ dex of power on timescales above a Gyr with a tail toward shorter timescales, that needs to be accounted for by mergers and dynamical processes within galaxies.

\subsubsection{The coherence of in-situ SFHs and DMAHs}

It is important to keep in mind that the mass assembly histories of galaxies are different from their SFHs, since mergers bringing in already-formed stars would be counted in the former at the time when the merger occurs, but in the latter when the ex-situ stars first formed. Since the contribution from ex-situ star formation is known to correlate strongly with stellar mass across different models \citep{rodriguez2015merger, qu2017chronicle, behroozi2018universemachine, moster2018emerge, tacchella2019morphology}, it would be instructive to understand the timescale dependence of correlations between the in-situ star formation and the full SFH, as well as the correlations between the in-situ star formation of the central progenitor and the DMAH of its parent halo. We quantify this by computing the cross power-spectrum, given by $P_{xy} = \int (\int x(t) y(t+t') dt') e^{-ikt} dt$, and using it to find the coherence, $C_{xy} = P_{xy}^2/|P_x P_y|$ for these two sets of timeseries, where $P_x, P_y$ are the PSDs of the two timeseries x and y (in this case SFHs and DM accretion histories, or full and in-situ SFHs) and $P_{xy}$ is the cross-power spectrum. The coherence is therefore the normalized excess in power compared to each series taken in isolation.

The top panel of Figure \ref{fig:sfh_dm_coherence} shows the coherence computed for the full SFHs compared to just the in-situ SFH of the central progenitor for IllustrisTNG galaxies. We see that while the coherence is high on long timescales, which means that the shape of the two SFHs cannot be too different, the coherence on shorter timescales falls off on shorter timescales with increasing mass. \citet{rodriguez2015merger, tacchella2019morphology} showed that more massive Illustris and IllustrisTNG galaxies assemble an increasing fraction of their mass ex-situ, due in part to an increased number of major and minor mergers. Mergers bring in lower-mass galaxies, which typically have more power on shorter timescales. This leads to the full SFH decorrelating from that of the central progenitor on shorter timescales.
The bottom panel of Figure \ref{fig:sfh_dm_coherence} shows the coherence computed between the DM accretion histories and the in-situ SFH of the central progenitor, which most closely tracks the parent halo. This plot quantifies the effect of baryonic physics on regulating SFR on short timescales, as the two quantities are linked on the longest timescales, but fall off rapidly at timescales below $\sim 3$ Gyr. Similar to the DM accretion history PSDs, there is only a weak trend with increasing stellar mass.

In summary,
(i) the variability of DMAHs, quantified using their PSDs, is self-similar across different masses and has a median slope of $\approx 1$;
(ii) the DMAHs do not contribute significantly to the overall variability of their SFHs, except at the shortest ($\lesssim 400$ Myr) timescales where their variability is similar to those of SFHs. Quenching can account for a significant fraction of the excess power in SFHs on the longest timescales;
and (iii) The DMAHs are coherent with the in-situ star formation of galaxies on long timescales ($\gtrsim 3$ Gyr). Therefore, they may set the overall shape of the in-situ mass assembly histories of their central galaxies.

\section{Discussion}
\label{sec:discussion}

The PSD formalism provides a useful way to quantify the variability in galaxy SFHs across different timescales. Applying this to a variety of different models of galaxy evolution, we find that the PSDs of galaxy SFHs generally show broken power-law shapes,
with a tendency to grow more featureless and tend to a single power-law with slope $\beta \sim 2$ toward higher stellar masses.
The PSDs also show a wide diversity between the models in terms of slope and power at any given stellar mass and timescale.
In Section \ref{sec:timescale_sim}, we relate these observed PSD features to existing estimates for the timescales on which different physical processes are expected to act, with a table reported in Appendix \ref{sec:lit_timescales_table}.
In Section \ref{sec:timescale_obs}, we discuss observational measurements and techniques that can be used to obtain constraints in PSD space.
Section \ref{sec:psd_resolution} demonstrates the effects of lower resolution on PSDs using additional runs of the IllustrisTNG simulation.
Finally, Section \ref{sec:future_work} considers possible directions for extending the analysis presented in this work.

\subsection{The characteristic timescales of physical processes in simulations}
\label{sec:timescale_sim}

There exist a range of estimates in the literature for timescales associated with different physical processes, some of which are shown in Figure \ref{fig:literature_timescales} and listed in Appendix \ref{sec:lit_timescales_table}. In this Section, we briefly summarize the current state of our understanding regarding which physical processes can contribute to SFR fluctuations at various timescales. By doing this, we can begin to connect the different features seen in the median PSDs of SFHs in Section \ref{sec:sfh_psds_diffsims} to the underlying physical processes responsible. It also serves as a useful starting point for future analyses looking at these features in greater depth within specific models. Starting with processes that act on the shortest timescales, we gradually work our way to the longer timescales that are the focus of the bulk of this paper.

\textbf{GMC formation and destruction:} Star formation on small spatiotemporal scales occurs in GMCs, whose lifetimes are sensitive to a variety of factors including cloud collisions and mergers, feedback from supernovae, cosmic rays and photoionisation, turbulence in the ISM, and the growth of magnetic fields \citep{dobbs2012giant, dobbs2015frequency, kim2017three, semenov2017physical, pakmor2017magnetic, benincasa2019live}. Current upper bounds on theoretical predictions for GMC lifetimes range between $\sim 7-20$ Myr  \citep{tasker2011star, benincasa2019live}, with estimates for the timescales of individual processes that influence GMC lifetimes reported in Appendix \ref{sec:lit_timescales_table}. Analytical models can also provide an understanding of when star formation in this regime can be bursty \citep{faucher2017model}.

Considering the rate of GMC formation and destruction to be a stochastic process, we would therefore expect a power-law PSD with slope $\beta \sim 2$ at these timescales \citep{kelson2014decoding, tacchella2020stochastic}. Although the large-volume models do not probe these timescales, the three suites of zoom simulations allow us to test this hypothesis. In fact, we do find the PSD in this timescale to be well-described by power-laws, and the \brooks galaxies show slopes of $\beta \sim 1.6^{+0.4}_{-0.1}$ uncorrelated with stellar mass, while the FIRE-2 galaxies show slopes of $\beta \sim 1.8^{+0.5}_{-0.4}$, with a mild trend of increasing slope with stellar mass over timescales of $\sim 10-20$ Myr.

\textbf{Dynamical processes within galaxies:} A range of physical processes act to influence the state of the ISM on galaxy dynamical timescales ($\sim 10^8$ yr). These processes include turbulence in the ISM, molecular gas encountering spiral arms and bars, galactic winds, and the rapid cycling of ISM gas between star forming and non-star forming regions, in addition to the exponential growth of magnetic fields, and stochastic inflows of CGM gas\footnote{The last two extend to longer timescales as well.}. Analytical models account for these processes through a range of timescales, including timescales for gas accretion and cooling, as well as star formation, turbulent crossing  and effective viscous timescales that describe how long it takes for accreted gas to reach the center of the galaxy \citep{dekel2009formation,  krumholz2010dynamics, forbes2014balance}.
For modeling these processes, resolution plays an extremely important role since resolving the ISM allows simulations to capture the effects of turbulence driven by feedback, as well as model the feedback self-consistently while relaxing the need for sub-grid recipes.
Most of our knowledge in this regime comes from small-volume simulations (e.g., a slice of a galactic disk, \citealt{kim2017three}, or an idealized disk \citealt{semenov2017physical}) or zoom simulations focusing on individual galaxies \citep{hopkins2014galaxies, ceverino2014radiative, christensen2016n}. Bursty star formation has been noted on timescales of $\sim 45$ Myr in the TIGRESS framework \citep{kim2017three}, and on $\leq 100$ Myr timescales in FIRE-2 \citep{sparre2017starbursts, hung2019drives}.

Since there are many competing factors at play, we expect the PSDs in this regime (and beyond) to be complicated, and this is what we generally see in all the zoom simulation suites. Overall, while the PSDs can still be approximated with a power-law, several PSDs show minor peaks\footnote{$\sim 35-60$ Myr for Sandra in Justice League, Rogue in Marvel and in m12m, m11e, m11d, m11i, m11v, and m12i in FIRE-2} or breaks\footnote{$\sim 60-90$ Myr for h986 from g14 and for m12b, m11f, m11i, m11g, m11c in FIRE-2} with average slopes in the $\sim 30-100$ Myr range of $\beta \sim 2.0^{+0.8}_{-0.7}$ for \brooks and $\beta \sim 1.3^{+0.4}_{-0.5}$ for FIRE-2. All three suites of simulations show increased scatter in the power-law slopes, along with a trend of increasing slope with stellar mass in this timescale range, perhaps correlated with decreasing dynamical timescales as galaxies grow more massive.

\textbf{Mergers:} Mergers between galaxies bring in a combination of stars that have already formed and gas that can fuel a burst of subsequent star formation, with timescales ranging from $\sim 100-500$ Myr \citep{hernquist1989tidal, barnes1991fueling, barnes1996transformations, mihos1996gasdynamics, robertson2006merger, hani2020interacting}.
The effect on SFHs comes from mergers as a  a primary mechanism for driving starbursts in galaxies (in addition to disk instabilities) and as a controversial trigger for quenching, depending on a variety of factors including the mass ratio, relative alignment, how gas-rich the merger is, and even if the merger triggers a central AGN \citep{hopkins2006unified, governato2009forming}.
Zoom simulations also predict that mergers or counter-rotating streams can lower the angular momentum of the gas disk within galaxies, leading to a compaction of the gas phase, which results in an enhancement of the SFR \citep{zolotov2015compaction}. These phases can last for one to a few hundred Myr and move galaxies to the upper envelope of the star-forming sequence \citep{tacchella2016confinement}.
\citet{rodriguez2019mergers} find that major mergers cause enhanced SFR at all masses below a threshold of $\sim 10^{11}$M$_\odot$ in Simba. \citet{tacchella2019morphology} find trends consistent with centrally enhanced star formation due to ex-situ star formation for intermediate mass ($10^{10-11}$M$_\odot$) galaxies, with mergers responsible for over two thirds of the ex-situ component toward the high-mass portion of that range. A notable consequence of this is the increasing loss of coherence between the in-situ and full SFHs of galaxies in Figure \ref{fig:sfh_dm_coherence} with increasing stellar mass.

In addition to the timescale of SFR enhancement following a merger, we also need to consider the fact that mergers themselves are stochastic events, and therefore carry an additional implicit timescale.  Estimates of merger timescales are generally $\mathcal{O}(1)$ Gyr \citep{boylan2008dynamical, lotz2011major, snyder2017massive}, and can vary significantly depending on assumed definitions and factors like pair separation and angular momentum of the system. Due to these factors, it can be difficult to isolate the effects of mergers on galaxy PSDs.

\textbf{Baryon Cycling:} The global efficiency of how galaxies are able to convert their gas into stars is almost an order of magnitude different from local efficiencies in star forming regions. \citet{semenov2017physical} tie this to the cycling of ISM gas between regions that are star forming and those that are not. In addition to this, gas that leaves the galaxy due to ejective feedback and returns also contributes to prolonging the period over which a galaxy continues to form stars (\citealt{christensen2016n, hopkins2018fire}, also see review by \citealt{tumlinson2017circumgalactic} and references therein). The lifetimes and dynamics of cold clouds in the halo are also subject to a variety of timescales \citep{forbes2019hydrodynamic}.

Estimated  timescales for the cycling of baryons span a wide range, from $\sim 100$ Myr to about 3 Gyr \citep{oppenheimer2010feedback, christensen2016n, mitra2016equilibrium, angles2017cosmic, grand2019gas}. Some studies find the timescales to scale with halo or stellar mass \citep{oppenheimer2010feedback, mitra2016equilibrium}, while other studies find it to be largely independent of mass \citep{christensen2016n}.
While we find evidence for peaks and breaks in the PSDs of individual galaxies on these timescales, especially in the zoom simulations (for example, in Figure \ref{fig:psd_fire2_single_galaxy}), the broad range of timescales and the dependence on galaxy properties other than stellar mass results in these peaks being washed out in the median behaviour for an ensemble of galaxies. However, it is possible that breaks in the PSD could be correlated with baryon cycling processes, and bears further investigation in future work. In particular, the evolution of the break timescales with stellar mass in different models could help us understand why some studies show a significant mass-dependent trend while others do not. However, since mergers and other factors also play a role at these timescales, their effects also need to be accounted for in such an analysis.

The $\sim 1$ dex excess in the power of SFHs compared to DMAHs after accounting for quenching could correspond to contributions from baryonic processes like mergers and baryon cycling occurring on halo dynamical timescales as the galaxy grows, leading to imprints in the PSD on timescales $ \propto 2\pi\tau_\mathrm{dyn} \sim 2\pi(0.1\tau_{\mathrm{H}}) \sim 2.1- 8.6$ Gyr over the past $\sim 10$ Gyr\footnote{Although it is outside the scope of the current work, it would be an interesting exercise to model the excess in the SFH PSDs as an aggregate effect of baryonic processes across a range of redshifts using a broken power-law model  with $\tau_\mathrm{break} \sim 2\pi \tau_\mathrm{dyn}(z)$, based on the formalism described in  \citet{caplar2019sfhpsd} and \citet{tacchella2020stochastic}.}. Since the dynamical time grows with decreasing redshift, the resulting contribution to the PSD would end up being smoothed out over a broad range of timescales.
The plateau in simulations like Illustris and IllustrisTNG and individual galaxies in the zoom simulations at $\sim 1-3$ Gyr are also indicative of a decorrelation timescale that naturally arises in damped random walk models of star formation \citep{caplar2019sfhpsd, tacchella2020stochastic}.

\textbf{Quenching:} Quenching in central galaxies can happen due to a lot of different factors - the shock heating of virial halo gas preventing cold-mode accretion \citep{dekel2006galaxy}, energy from AGN jets that heat gas and prevent it from forming stars \citep{somerville2008semi} and outflows that could remove cold gas from the galaxy \citep{di2005energy}. Observational scaling relations like the $M_{\mathrm{BH}}-\sigma$ correlation tie behavior on large (galaxy-wide) scales to sub-kpc scales on which SMBHs grow, leading to a unique scenario where sub-grid recipes for implementing BH growth and feedback affect when and how galaxies quench. In addition to this, recipes for how simulations implement cooling and star formation, and the strength of winds that blow gas out of galaxies all contribute to the overall trends seen in galaxy quiescence. Finally, the haloes of galaxies set the inflow rate of gas into the central galaxy, as seen through the correlation on long timescales between the DMAHs and in-situ SFHs. This dependence could tie the fueling of the central AGN to that parent halo, ultimately determining when the onset of quenching occurs \citep{chen2019quench}.

Figure 6 in \citet{wright2019quenching} shows a broad, unimodal distribution of quenching timescales in EAGLE galaxies extending out to $\tau_{\mathrm{H}}$ with a median of $\sim 2.5-3.3$ Gyr for low-mass centrals and at shorter timescales (median of $\sim 1.7-2.1$ Gyr) for high-mass centrals depending on the definition of quenching timescale. Longer quenching timescales at low masses are associated with stellar feedback prolonging star formation activity, while shorter timescales at high masses are associated with AGN activity.
Simba, on the other hand, shows a bimodal distribution of quenching timescales \citep{rodriguez2019mergers}, with a slow mode acting approximately over a dynamical time ($t_Q \sim 0.1 \tau_{\mathrm{H}}$) that is more numerous overall for central galaxies, and a fast mode ($t_Q \sim 0.01 \tau_{\mathrm{H}}$) that dominates at stellar masses of $M_* \sim 10^{10} - 10^{10.5} $M$_\odot$. The fast quenching mode is associated with AGN jet quenching causing a rapid cessation of accretion, since it becomes active at this mass range, and  merger rates are not preferentially elevated at these masses.
Additionally, X-ray feedback can rapidly evacuate the central regions of galaxies \citep{appleby2019impact} and contribute to short-timescale variability.
\citet{sales2015colours} find a quenching timescale of $\sim 2-5$ Gyr for galaxies in Illustris. \citet{nelson2018first} find the colour-transition timescale, a tracer of the quenching timescale, to be $\sim 0.7-3.8$ Gyr for IllustrisTNG galaxies.
Additionally, \citet{joshi2020fate} find that morphological transformations in IllustrisTNG clusters occur on timescales of $\sim 0.5-4$ Gyr after accretion, with a control group showing a broader distribution. They also find that morphological transformation lags $\sim 1.5$ Gyr behind quenching for gas-poor disks, while it precedes quenching by $\sim 0.5$ Gyr for gas rich cluster galaxies, and by $\sim 2.5$ Gyr for gas-rich control galaxies.

In studying the excess PSD power on different timescales and stellar masses due to quenching, we find that the excess variability in IllustrisTNG on short timescales rises strongly at M$_*\geq 10^{10.5}$M$_\odot$, correlated with the onset of strong kinetic-mode AGN feedback at M$_{\rm BH}\sim 10^{8.5}$M$_\odot$ \citep{weinberger2018supermassive}.
While we are not in a position to speculate about timescales of $0.01\tau_{\rm H}$, we do find a tail of excess variability extending to the lowest timescales in Simba that could be related to the jet-mode AGN feedback.
In EAGLE, \citet{wright2019quenching} find that galaxies at low masses quench primarily due to stellar feedback on long timescales, consistent with the excess power we see on timescales $\geq 2$ Gyr. As galaxies grow more massive (M$_* \geq 10^{10.3}$M$_\odot$) mergers and black hole activity increase sharply, leading to overall shorter quenching times, and additional variability on all timescales, as seen in Figure \ref{fig:sfg_quenched_psd_difference}. The excess power in the SAMs at high masses seems to be primarily due to increased stochastic starbursts triggered by mergers and AGN activity \citep{somerville2008semi}.
The increasingly featureless (scale-free) nature of the PSDs toward high stellar masses, where the fraction of quenched galaxies is the largest, could be due to the contribution to the PSD from quenching dominating all other contributions.

Since a combination of multiple processes is responsible for quenching at different stellar masses, it is difficult to constrain their relative strengths with observational measurements of quenching timescales. However, since these different processes also induce varying amounts of short-timescale variability, constraints in PSD space might be able to distinguish between processes and allow for better constraints on their relative strengths.

\subsection{Observational constraints in PSD space}
\label{sec:timescale_obs}

For the different galaxy evolution models we consider in Section \ref{sec:results}, we see a large diversity
in the amount of power in SFR fluctuations on a given timescale, the coupling between adjacent timescales, and the existence and location of breaks in the PSD. This makes the PSD a sensitive probe of both the strengths of physical processes and their numerical implementation in these models. Observational constraints in this space are therefore extremely important, and will allow us to better constrain the relative strengths of different processes for a population of galaxies at a given stellar mass and epoch. These observational constraints can come in three forms: (i)  constraints in PSD space obtained by measuring the SFR variability of ensembles of galaxies, which can be compared to the models we study, (ii) constraints on the timescales for observed phenomena like quenching or rejuvenation, which can be tied to breaks or peaks in the PSD, and (iii) constraints on timescales of physical processes, which can be used to isolate the effects of different processes contributing to the PSD on a given timescale. In this section, we will briefly discuss each of these.

\textbf{Ensemble constraints on SFR variability:}
The spectral energy distributions (SEDs) of galaxies are composed of spectrally distinct contributions from stellar populations formed at different ages relative to the time of observation. Interpreting these contributions gives us access to star formation rates averaged over different timescales. Nebular emission from the regions near short-lived O- and B-type stars provides constraints on SFR over the most recent $\sim 4-10$ Myr \citep{madau2014cosmic}. The rest-UV portion of the SED contains contributions from young stars that probe the SFR out to $\sim 30-100$ Myr, with a similar timescale probed by the rest-FIR portion of the SED, which contains the re-emitted light from the young stellar light absorbed by dust.  In addition to this, features like the strength of H$\delta$ absorption and the 4000\AA~ break are sensitive to SFR within the last $\sim 1$ Gyr and to the light-weighted age within $\sim 2$ Gyr, respectively \citep{kauffmann2006gas, wang2020var}.

To get constraints in PSD space,
it is useful to consider the SFR distributions for populations of galaxies and compare these distributions on different timescales to quantify a relative change in burstiness\footnote{This makes an inherent assumption of ergodicity, that the PSDs obtained from a population of galaxies can be connected to the PSDs of individual galaxy SFHs over time. This assumption is explored in detail in \citet{wang2020vartwo}.}. \updated{While this straightforward to forward-model for the large-volume models, the small number of zoom galaxies make this a more involved procedure while considering those models. In these cases, a workaround is possible by realising samples from the PSDs of zoom galaxies, similar to the procedure followed in \citep{tacchella2020stochastic}.}
In terms of the PSD formalism, this is equivalent to an observational constraint on the slope of the PSD between two timescales. This has been done for select timescales and populations of galaxies in \citet{guo2016bursty, broussard2019stochastic, emami2018closer}.
More recently, \citet{caplar2019sfhpsd} and \citet{wang2020var, wang2020vartwo} have performed analyses motivated by the PSD formalism to constrain the slope of the PSD and other features in its shape.
Most relevant to the current work, \citet{caplar2019sfhpsd} fit broken power-law models to galaxy fluctuations around the star forming sequence at $z\sim 0$ with M$_*= 10^{10}-10^{10.5}$M$_\odot$. With degeneracies due to current observational uncertainties, they find that they cannot constrain both a slope and break timescale, but find a break timescale of $\sim 200$ Myr assuming a slope of $\beta=2$. \citet{wang2020vartwo} extend this analysis in a spatially resolved direction and find PSD slopes of $\beta \sim 1-2$ in the timescale range $\sim 5$ Myr to $\sim 800$ Myr, assuming no break in the PSD (which implies that SFHs are correlated over the the age of the universe). They also find that the slopes generally decrease with stellar mass for M$_*>10^{9}$M$_\odot$, and are correlated with estimated gas depletion timescales in galaxies.
Going forward, these novel measurements can be used to constrain free parameters in the different models, and existing models can be used to make predictions for future observations with upcoming facilities like JWST and WFIRST.

\textbf{2. Constraints on the timescales for observed phenomena:}
Combining the spectral features from distant galaxies across a range of wavelengths in a full SED fitting code allows us to estimate the  star formation histories of individual galaxies with uncertainties \citep{moped, vespa, pacifici, pacifici2016timing, smith2015deriving, leja2017deriving, iyer2017reconstruction, carnall2018inferring, leja2018measure, iyer2019gpsfh}.
While these observationally derived SFHs are not sensitive to variability on short timescales, they can be useful for measuring the timescales for morphological transformations, mergers, quenching and rejuvenation, and even recent starbursts. These timescales can then be linked to features in the PSDs of galaxy SFHs, such as peaks or breaks.

\citet{pacifici2016timing} analyse a sample of quiescent galaxies from CANDELS at $0.2<z<2.1$ and find quenching timescales to be $\sim 2-4$ Gyr, with a strong mass dependence. \citet{carnall2018inferring} study a sample of quiescent galaxies from UltraVISTA at $0.25<z<3.75$ and find that the majority of galaxies quench on timescales of $\sim 0.4 \tau_{\mathrm{H}}$, with a rising set of galaxies towards the lower redshift portion of their observations with quenching timescales $\sim 0.6 \tau_{\mathrm{H}}$. \citet{iyer2019gpsfh} analysed a sample of CANDELS galaxies at $0.5<z<3.0$ and found that $\sim 15-20\%$ of galaxies showed evidence for multiple strong episodes of star formation, with the median timescale separating multiple peaks to be $\sim 0.4\tau_{\mathrm{H}}$, which matches the predictions using cosmological simulations by \citet{tacchella2016confinement}. The study also found that the SFHs of galaxies were correlated with their morphological classification, with an elevation in SFR on timescales over the last $\sim 0.5$ Gyr in galaxies classified as mergers and interactions, and with a longer period of SFR decline for spheroids compared to disks.

A number of studies \citep{lotz2011major, snyder2017massive, duncan2019observational} also use statistical estimates of the physical properties of galaxies to constrain merger rates and observability timescales. \citet{pandya2017massivetransition} uses a similar statistical approach to quantify the timescales on which galaxies experience quenching and rejuvenation by studying the relative number of galaxies that are star forming, quiescent, and transitioning between the two states at a given epoch.

Current observational techniques require a certain set of modeling assumptions, such as a choice of IMF, stellar population synthesis (SPS) model, and dust attenuation law. Combined with state-of-the-art observations, this leads to uncertainties of $\sim 0.2$ dex in estimating stellar masses, and $\sim 0.3$ dex in estimating star formation rates from SED fitting, with fractional uncertainties in SFR growing large as we go to lower values of SFR and older stellar populations. Caution should be exercised in analysing the variability across different timescales using these derived physical properties, with care taken in propagating measurement uncertainties and instrumental effects in observations to uncertainties on their estimated physical properties. One example of this procedure is in accounting for the difference between the observed and intrinsic scatter in the SFR-M$_*$ correlation due to measurement uncertainties \citep{kurczynski2016evolution, boogaard2018muse}, which would potentially affect the PSD slope described earlier in this section.

That being said, techniques to model and extract SFH information from galaxy SEDs are growing increasingly sophisticated \citep{leja2018quiescent, iyer2019gpsfh}, and are (i) better at estimating the older star formation in galaxies, (ii) using fully Bayesian techniques accounting for possible covariances between parameters, and (iii) implementing well motivated priors being used to break degeneracies between parameters like dust, metallicity and SFH. With this in mind, it is hoped that in addition to timescales, SFHs from upcoming surveys will also be able to provide direct constraints on PSD slope and power on the longest timescales. Functionally, this provides a way to infer the same information as the first class of constraints on these timescales, although in this case the timescales are estimated from the histories of individual objects as opposed to recent burstiness of ensembles of galaxies. This would, in principle, allow us to independently verify estimated timescales, and test the assumption of ergodicity inherent to constraints on the PSD obtained using ensembles of galaxies.

\textbf{3. Constraints on the timescales of physical processes:}
In addition to the two approaches described above,
observations can also directly measure timescales for gas depletion \citep{kennicutt1998global, wong2002relationship, bigiel2008star}, stellar winds \citep{sharp2010three,ho2016sami}, disk formation \citep{kobayashi2007simulations}, bulge growth \citep{lang2014bulge, tacchella2015evidence}, black hole growth \citep{hopkins2005physical} and more, albeit for limited samples of galaxies.
On short timescales, a large body of work also exists studying GMC lifetimes ($\sim 10-30$ Myr) \citep{zanella2019contribution, kruijssen2019fast, chevance2020lifecycle}, measuring the extent to which this depends on environment, and the extent to which it is decoupled from galactic dynamics.
\citet{krumholz2017dynamical} also find episodic starbursts lasting $\sim 5-10$ Myr with intervals of $\sim 20-40$ Myr in a ring around the Milky-Way's central molecular zone. Equivalent behaviour in the zoom galaxies would therefore manifest as a local peak in the PSD on those timescales.
In the local universe, resolved observations of stellar populations allow us to constrain the SFHs of nearby galaxies using colour-magnitude diagrams \citep{weisz2011acs}.
For galaxies where stellar populations can be resolved, additional timescale information can be obtained from chemical abundances, since the production of heavy elements by different types of supernovae trace a range of intermediate timescales \citep{kobayashi2007simulations, kobayashi2009role}. However, the masses of these galaxies are often too low to compare against the large-volume models considered in the current work.
Another interesting study along these lines uses the fact that supernovae are produced at a certain rate after an episode of star formation, to compute the delay time distributions of SN Ia using SN Ia yields in conjunction to observationally measured galaxy SFHs \citep{strolger2020delay}.
These constraints on the timescales of physical processes allow for better modeling of the individual components that contribute to the full PSD of a galaxy's SFH, and sometimes provide an independent check of behaviour predicted using the PSDs.

Using the PSD formalism as our basis, it is therefore possible to constrain the PSD power on certain timescales or the PSD slope on certain timescale ranges using observations, with currently available data already starting to provide initial estimates of PSD slopes on $\sim 4-800$ Myr timescales. Taken together, the three types of constraints outlined above, i.e., (i) estimates of SFR variability using ensembles of galaxies, (ii)  observationally measured timescales for phenomena like quenching, starbursts, and rejuvenation, and (iii) timescales for physical processes like gas depletion and GMC formation and destruction will allow us to compare features in the PSD such as slopes and peaks across different galaxy populations.

\subsection{Effects of resolution}
\label{sec:psd_resolution}

\begin{figure*}
    \centering
    \includegraphics[width=0.9\textwidth]{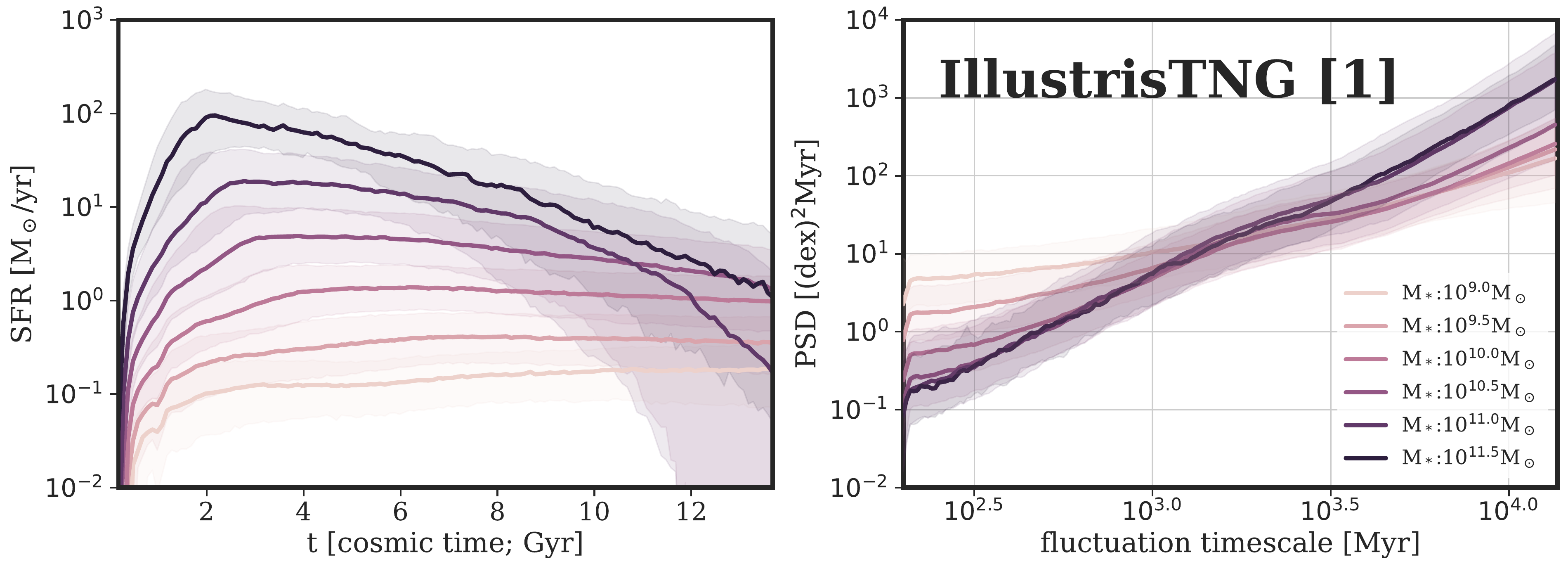}
    \includegraphics[width=0.9\textwidth]{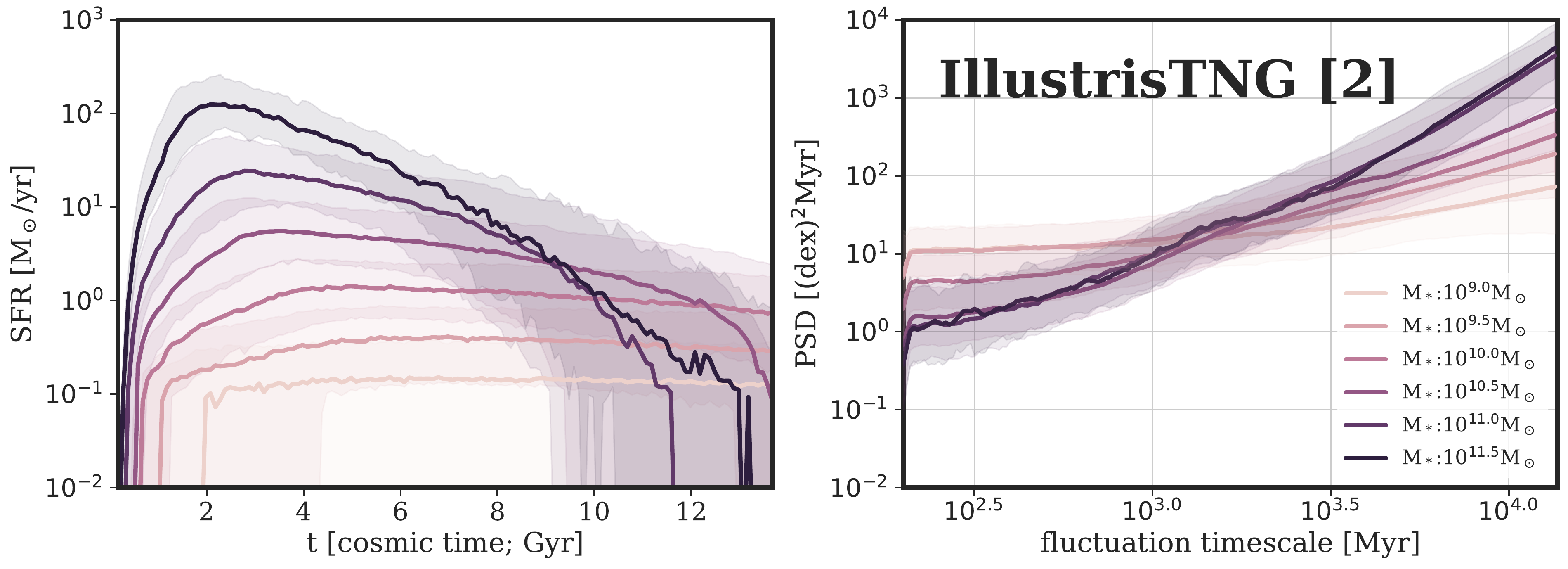}
    \includegraphics[width=0.9\textwidth]{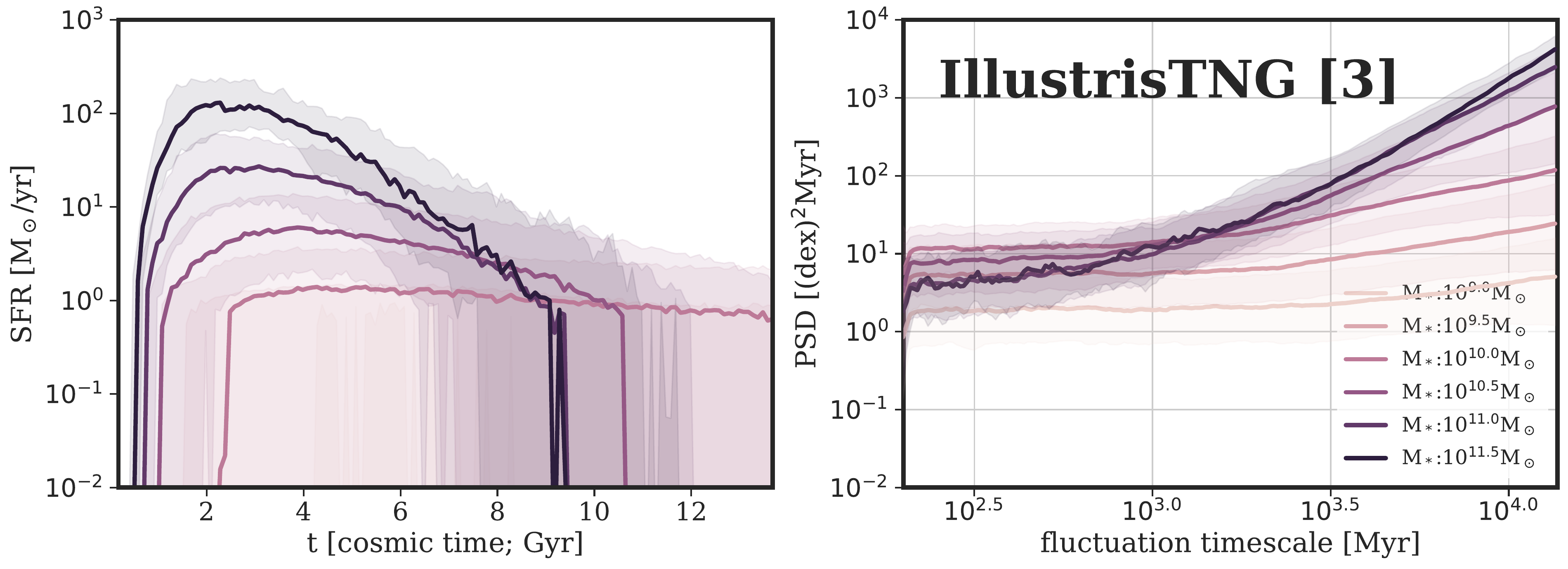}
    \caption{Exploring the effects of decreasing resolution (increasing star particle masses) on the SFHs and corresponding PSDs of galaxies using the IllustrisTNG simulation (M$_* = 9.44 \times 10^5 $M$_\odot\updated{/h}$, M$_* = 7.55 \times 10^6 $M$_\odot\updated{/h}$, and M$_* = 6.04 \times 10^7 $M$_\odot\updated{/h}$ respectively for the three).
    Decreasing the resolution leads to a boost of power on short timescales due to increased contribution of shot noise from discrete star particles. This manifests as a `white noise' floor in the PSDs that prevents us from probing the PSD to shorter timescales.}
    \label{fig:tng_resolution_effects}
\end{figure*}

In order to investigate the effects of resolution of the numerical simulation, we consider three realizations of the TNG100 simulation (TNG100-1,2 and 3), which are identical except for resolution. These are described in further detail in \citet{pillepich2017simulating}, and contain star particles with initial masses of $9.44 \times10^5 $M$_\odot/h$, $7.55 \times 10^6 $M$_\odot/h$, and $6.04 \times 10^7 $M$_\odot/h$, respectively. Mufasa and Simba therefore fall somewhere between TNG100-2 and TNG100-3 in terms of resolution, while EAGLE is comparable to TNG100-1. We show the SFHs and corresponding PSDs for these runs in Figure \ref{fig:tng_resolution_effects}.

In SFH space, we see that the different resolutions have a large impact on SFHs across all masses, especially in the portions with low SFRs. For the three simulations, with our adopted $100$ Myr time bins, the lowest SFRs we can probe are $\approx 10^{-2.02}$, $10^{-1.12}$ and $10^{0.68}~\mathrm{M}_\odot \mathrm{yr}^{-1}$ neglecting mass loss.
We see this in effect as the median SFHs for low-mass galaxies grow increasingly dominated by shot-noise and in the case of TNG100-3, completely drop off the plot. This resolution effect also affects high-mass quiescent galaxies, which leads to the apparent more rapid quenching of the median SFHs in the highest-mass bins. This is simply because the SFRs can only drop to their minimum value from the quantum of SFR given the resolution, leading to a steeper apparent drop in the SFHs.

In PSD space, we see that the effect of lower resolution is to increase the amount of white-noise in the PSDs, which manifests as a flattening to spectral slopes of 0 towards shorter timescales.
In addition to affecting the PSDs to higher masses, the white noise also increases in magnitude proportional to the mass of the star particles, leading to contamination at longer timescales in a given stellar mass bin. This effect is quantified in the analysis of Appendix \ref{sec:timescale_tests}.

\updated{While resolution can be a limiting factor in any analysis of small-scale features in hydrodynamical models, convergence tests on individual simulations  \citep{genel2018quantification, keller2019chaos}. and forward modeling the effects of discrete star particles as in Appendix \ref{sec:timescale_tests} allow us to understand and account for these limitations.
In this case, resolution effects mostly prevent us from studying the behaviour of the PSDs on small timescales, which can be circumvented using the zoom simulations, which have much higher resolution. It should also be noted that the location of the breaks in the PSD listed in Section \ref{sec:results} are robust to resolution, although their strengths can be affected by the amount of white noise. Therefore, breaks and peaks in the PSD of simulations should be carefully compared to observations (see also Section \ref{sec:timescale_obs}).}

\subsection{Going forward: physics vs numerics}
\label{sec:future_work}

The considerable differences between the PSD slopes and power across the different models seen in Figure \ref{fig:psd_slope_norm_comparison} are caused in part due to the different modeling assumptions for physical processes, for example, AGN seeding, growth and feedback, star formation and stellar feedback and processes governing galactic winds. These differences are also in part due to the resolution of the simulations, as seen in Section \ref{sec:psd_resolution} and numerical techniques used to implement gravity and magnetohydrodynamics (MHD), ranging from no explicit treatment of MHD in empirical and semi-analytic models, to differences between smoothed particle hydrodynamics, adaptive mesh refinement schemes and other hybrid techniques in hydrodynamical simulations (e.g., \citealt{vogelsberger2012moving, sijacki2012moving, kerevs2012moving}) using codes like AREPO \citep{springel2010moving} in Illustris and IllustrisTNG, GIZMO \citep{hopkins2014gizmo} in FIRE-2, Mufasa and Simba, a forked version of GADGET-3 \citep{springel2005modelling} in EAGLE, Gasoline \citep{wadsley2004gasoline} for the g14 suite, and ChaNGa \citep{menon2015adaptive} for the Marvel/Justice League suite of zoom simulations.

While the current work serves to illustrate the cumulative differences between models due to choices of numerical techniques and physical models, it is outside the scope of the current work to break down the individual contributions.
Building on the current work, there are three directions in which we can begin to better connect individual physical processes to their relevant timescales in a model-independent way:

(i) \citet{tacchella2020stochastic} propose an analytical model in PSD space using the gas regulator model of galaxy evolution \citep{lilly2013gas}, extending the model to account for the creation and destruction of GMCs on short timescales. Using this model, they derive the PSD as a broken power-law with multiple breaks that characterise the equilibrium timescale of gas inflow and the average lifetime of GMCs. Applied to PSDs from the different models we consider, this can explain the effective timescales for these processes across the various models.

(ii) In a slightly different direction, many of the models we consider have run additional simulations varying the input physics. For example, there is a set of $25$ Mpc$^3$ boxes run for Illustris-TNG varying a single parameter per run, including stellar and black hole feedback mechanisms, galactic wind scalings, and aspects of star formation \citep{pillepich2018simulating, nelson2018abundance}. The Simba model contains additional simulations varying the AGN feedback model \citep{dave2019Simba}.
\citet{crain2015eagle} describe model variations within the EAGLE suite varying stellar and AGN feedback.
\citet{choi2017physics} contains a suite of zoom simulations that are run with and without AGN feedback. The Santa Cruz SAM and other semi-analytic models are also capable of being run multiple times varying model parameters.

Using all of this data, it should be possible to characterise the effects of varying physical modeling assumptions with individual models, and use this across several models to understand the general trends and timescales for physical processes like stellar and AGN feedback and baryon cycling. However, as \citet{pillepich2018simulating} note, `the optimal choices for wind as well as black hole feedback strongly depend on the \textit{whole ensemble} of galaxy formation mechanisms incorporated into the model.' What holds for a given model need not generalise across all models, and extreme caution should be exercised while extrapolating trends from individual models, using the full available range of observational constraints described in Section \ref{sec:timescale_obs} as benchmarks.

(iii) Further studies will also be needed to investigate the link between the well-studied effects of spatial turbulence on star formation \citep{larson1981turbulence,nakamura2005quiescent,krumholz2005general, padoan2011star} and the natural emergence of power-law spatial correlation functions \citep{guszejnov2018universal} to its temporal manifestations studied in this work. Studies like \citet{di2015spatio}, which look at the joint spatio-temporal power spectra for numerical simulations of turbulent flows to identify the signatures of physical mechanisms, provide a useful starting point in this regard.

\section{Conclusions}
\label{sec:conclusions}

A range of physical processes acting on different timescales regulate star formation within galaxies.
Processes that act concomitantly over an overlapping range of timescales have complicated effects, and render it impractical to estimate the timescale of one process independently of the other.
The resulting process of galaxy growth is therefore diverse, and understanding the impact of the underlying processes across all timescales simultaneously can help explain this diversity.

Using the power spectral density (PSD) formalism, we quantify the variability of galaxy SFHs on different timescales for a wide range of galaxy evolution models and find:
\begin{enumerate}
    \item \textbf{Overall trends:}
    The PSDs of galaxy SFHs are well described by broken power-laws characteristic of stochastic processes, in line with theoretical descriptions by \citet{kelson2014decoding, caplar2019sfhpsd, kelson2020gravity} with most of the power lying on long ($\gtrsim 1$ Gyr) timescales.
    Across the full range of timescales investigated in this work ($\sim 200$ Myr $-10$ Gyr), the PSDs of galaxies with M$_*\sim10^{10}$M$_\odot$ show a median slope of $\beta \sim 1.6\pm 0.84$, increasing smoothly with mass to $\beta \sim 2.1 \pm 0.68$ at M$_*\sim10^{11.5}$M$_\odot$.
    \item  Although most models show comparable mass functions and similar overall behaviour in their SFHs, the specific PSD slope and power at any timescale can vary considerably across the different models. The PSD power can vary by up to an order of magnitude at a given timescale and stellar mass. Similarly, the local PSD slope can vary by $\sim 1.5$ at a given timescale and stellar mass\footnote{An interactive plot allowing the user to explore the PSD slope and power for the various models at different timescales can be found online at this link: \url{https://kartheikiyer.github.io/psd_explorer.html}}. Steeper slopes result in a larger fraction of the overall SFH power being concentrated on longer timescales. Interestingly, some models show a flattening of the slope at intermediate ($\sim 1-3$ Gyr) timescales, indicating that the SFHs decorrelate (i.e., lose memory) on these timescales.
    \item \textbf{PSD shape between models:} IllustrisTNG shows more variability on intermediate timescales compared to Illustris, as does Simba when compared to Mufasa. Updated feedback models (particularly for AGN feedback) in both of these simulations likely account for this. The UniverseMachine PSDs look quite self-similar, since the SFHs are closely tied to the DM-accretion histories of their parent haloes. The FIRE-2 simulations, with their significantly higher resolution and \updated{more explicit}
    feedback, show greater contributions at shorter timescales compared to the semi-analytic and empirical models. The \brooks simulations, which have comparable resolution to FIRE-2, show less variability across a range of timescales and a sharper trend for increasing burstiness with decreasing stellar mass.
    \item \textbf{Breaks in the PSD:}
    Illustris, IllustrisTNG, the SC-SAM and the zoom simulations show distinct breaks in their PSDs at several timescales across the different stellar mass bins. These breaks become less prominent with increasing stellar mass, with the PSDs approaching a scale-free power-law with slope $\beta \sim 2$. Mufasa, Simba, EAGLE and UniverseMachine show a smooth increase in slope toward longer timescales, with the slope being constant at timescales $\gtrsim 3$ Gyr.
    These breaks could stem from physical processes acting within galaxies, such as GMC lifecycle, dynamical processes, and gas regulation \citep{tacchella2020stochastic}.
    \item \textbf{Dark matter accretion histories:} The DMAHs of galaxies show self-similar behaviour across different stellar mass bins with a median power-law slope of $\beta\sim 1$, consistent with the analytic derivation by \citet{kelson2020gravity}. DMAHs do not contribute significantly to the overall variability of SFHs, except on the shortest timescales ($\lesssim 400$ Myr). The excess power in SFH PSDs compared to those of the DMAHs increases to long timescales, and is likely due to a combination of mergers, baryon cycling and AGN feedback.
    \item Studying the coherence between the PSDs of the full vs in-situ SFHs shows that mergers are responsible for decorrelating the two at short timescales. Since mergers are effectively stochastic events, there is no preferred timescale for this decorrelation, with the coherence falling off smoothly toward shorter timescales. Since higher mass galaxies experience more mergers, the decoherence is also a function of the stellar mass.
    \item The in-situ SFHs are coherent with the DM accretion histories of their parent haloes on long timescales ($\gtrsim 3$ Gyr), independent of stellar mass. This coherence is likely due to the growth of a galaxy's parent halo determining the fueling and therefore
    the subsequent star formation in its central galaxy. The decline in coherence quantifies the increasing importance of baryonic physics in regulating SFR on shorter timescales.
    \item \textbf{Variability on short timescales:} A number of models display a trend of increasing power on short timescales ($\sim 200-300$ Myr) with decreasing stellar mass, i.e. lower mass galaxies are burstier. This is notable for some of the large-volume hydrodynamical simulations (Illustris, IllustrisTNG, Mufasa, and EAGLE) and the zoom simulations (\brooks and FIRE-2), while short timescale power in the Santa-Cruz SAM, UniverseMachine and DM accretion histories is largely invariant as a function of stellar mass. Since the latter three models are most closely linked to halo merger trees, their lack of burstiness suggests that this shorter timescale behaviour is due to hydrodynamical feedback mechanisms that are not adequately captured by these models.
    \item \textbf{Zoom simulations:} The zoom simulations are a good test-bed to study the time-dependent variability of SFHs on shorter timescales, with their higher resolution allowing us to probe their PSDs to the much shorter timescales on which GMCs are created and destroyed. Studying galaxies from the FIRE-2 and \brooks zoom suites, the broken power-law behaviour of the PSDs is found to continue to nearly an order of magnitude below the timescales studied in the rest this work. The power in the zoom PSDs on long timescales is generally lower than their large-volume counterparts. Galaxies in FIRE-2 generally have more power on short timescales compared to galaxies in \brooks, although the trend of increasing short-timescale `burstiness' to lower masses is stronger in the latter.
    \item \textbf{The effects of quenching on PSDs:} Separating galaxies into star forming and quiescent populations in a given mass bin allows us to quantify the excess strength in the PSD due to the physical processes responsible for quenching. This excess power at a given timescale can be nearly an order of magnitude, with the dependence on stellar mass, the existence of a quenching timescale, and the behaviour of the PSDs below this quenching timescale varying widely across the different models.
\end{enumerate}

The PSD formalism allows us to quantify the strength of SFR fluctuations on different timescales. Studying the SFHs of galaxies from different models of galaxy evolution shows large differences in PSD space, due to differences in resolution and the implementation of sub-grid recipes. In conjunction with these models, observational measurements of SFR variability on different timescales will provide a useful new constraints on the relative strengths of the different physical processes that regulate star formation in galaxies.

\section*{Acknowledgements}

\updated{We would like to thank the anonymous referee for their insightful comments.} We would like to thank Phil Hopkins, Ena Choi, Viraj Pandya, Yuan Li, Harry Ferguson, Gwen Eadie, and Bryan Gaensler, and the entire the KSPA 2018 cohort for productive discussions and comments, and Peter Behroozi for making the UniverseMachine SFHs publicly available.
K.I. gratefully acknowledges support from Rutgers University and from the Dunlap Institute for Astronomy and Astrophysics through the Dunlap Postdoctoral Fellowship.
The Dunlap Institute is funded through an endowment established by the David Dunlap family and the University of Toronto.
S.T. is supported by the Smithsonian Astrophysical Observatory through the CfA Fellowship.
This work was initiated as a project for the Kavli Summer Program in Astrophysics held at the Center for Computational Astrophysics of the Flatiron Institute in 2018. The Flatiron Institute is supported by the Simons Foundation. K.I. and S.T. thank them for their generous support.
We acknowledge the Virgo Consortium for making their simulation data available. The EAGLE simulations were performed using the DiRAC-2 facility at Durham, managed by the ICC, and the PRACE facility Curie based in France at TGCC, CEA, Bruy\`eres-le-Ch\^atel.
Resources supporting the g14/Marvel/JL simulations were provided by the NASA High-End Computing (HEC) Program through the NASA Advanced Supercomputing (NAS) Division at Ames Research Center.
Support for Program number HST-AR-14564.001-A was provided by NASA through a grant from the Space Telescope Science Institute, which is operated by the Association of Universities for Research in Astronomy, Incorporated, under NASA contract NAS5-26555.

\textit{Software}: Astropy \citep{astropy:2013, astropy:2018}, AstroML \citep{vanderplas2012introduction}, commah \citep{correa2015accretion}, matplotlib \citep{caswell2019matplotlib}, scipy \citep{virtanen2020scipy}, numpy \citep{walt2011numpy}, corner \citep{corner}.

\section*{Data Availability Statement}

The data underlying this article will be shared on request to the corresponding author.


\bibliographystyle{mnras}
\bibliography{db_refs}


\appendix

\section{Finding the shortest timescales that can be probed in hydrodynamical simulations} \label{sec:timescale_tests}

Hydrodynamical simulations, both cosmological (Illustris, TNG, Mufasa, Simba) and zoom (FIRE-2, \brooks) have limits on the lowest SFR possible in any given time bin that is set by the mass of the star particles they use in our archaeological approach. All the simulations listed above turn gas into a star particle probabilistically depending on whether certain temperature and density conditions are met. In practice, this introduces portions in the SFH where the $\mathrm{SFR} = 0$, punctuated by small spikes which contain $\mathcal{O}(1)$ star particles. The effect of this on the power spectrum is to introduce white noise on the timescales where the SFR is probabilistically populated by discrete star particles. Looking at the PSD of individual galaxies, we see the effects of this effectively Poisson-distributed `shot noise' as a flattening as we approach short timescales. This depends on the amount of time the SFH spends in the vicinity of the minimum SFR threshold, set by

\begin{equation}
\langle \mathrm{SFR}_{\mathrm{min}}\rangle = \langle M_{*,\mathrm{sp}} \rangle / t_{\mathrm{PSD}}
\end{equation}
where $M_{*,\mathrm{sp}}$ is the average stellar mass of the star particles in the simulation given in Table \ref{tab:sim_details}, and $t_{\mathrm{PSD}}$ is the timescale being probed. From this relation, we see that the effects of shot noise on the PSD are greater on short timescales, as well as for simulations that have more massive star particles. However, finding the amount of time SFHs at a given stellar mass spend below  $\mathrm{SFR}_{\mathrm{min}}$ is a nontrivial task, depending on the shape of the SFH itself, and the number of the fluctuations around the median shape that could take it below $\mathrm{SFR}_{\mathrm{min}}$.

In the simplest case, given an SFH that is simply a constant $\mathrm{SFR}_{\mathrm{const}} = \psi_{\mathrm{mean}}$ + stochastic fluctuations $\mathrm{SFR}_{\mathrm{fluct}} = (N(0,\psi_\sigma))$, the distribution of $\mathrm{SFR}(t)$ at any given time is simply given by a Gaussian $N(\psi_{\mathrm{mean}}, \psi_\sigma)$. $\psi_{\mathrm{mean}} = M_*/\tau_{\mathrm{H}}$ is set by the stellar mass of the galaxy, where $\tau_{\mathrm{H}}$ is the age of the universe at the epoch of interest. The amount of time any SFH  at a given stellar mass spends below a threshold SFR is then given by

\begin{align*}
&t(\mathrm{SFR} < \mathrm{SFR}_{\mathrm{min}}| M_*, z) \\
&= \tau_{\mathrm{H}} \int_{-\infty}^{\mathrm{SFR}{\mathrm{min}}} \mathrm{exp}\left(- \frac{(\mathrm{SFR}-\psi_{\mathrm{mean}})^2}{(\psi_\sigma)^2} \right) d\mathrm{SFR}  \\
&= \frac{\tau_{\mathrm{H}} \psi_\sigma \sqrt{\pi}}{2} \left( 1+ \mathrm{erf} \left[\frac{M_{\rm *,sp}/t_{\rm PSD} - M_*/\tau_{\mathrm{H}}}{\psi_\sigma}\right] \right)
\end{align*}
Using this, we can set a threshold on the amount of shot noise, and limit our analysis to timescales above that.

\begin{figure}
    \centering
    \includegraphics[width=\columnwidth]{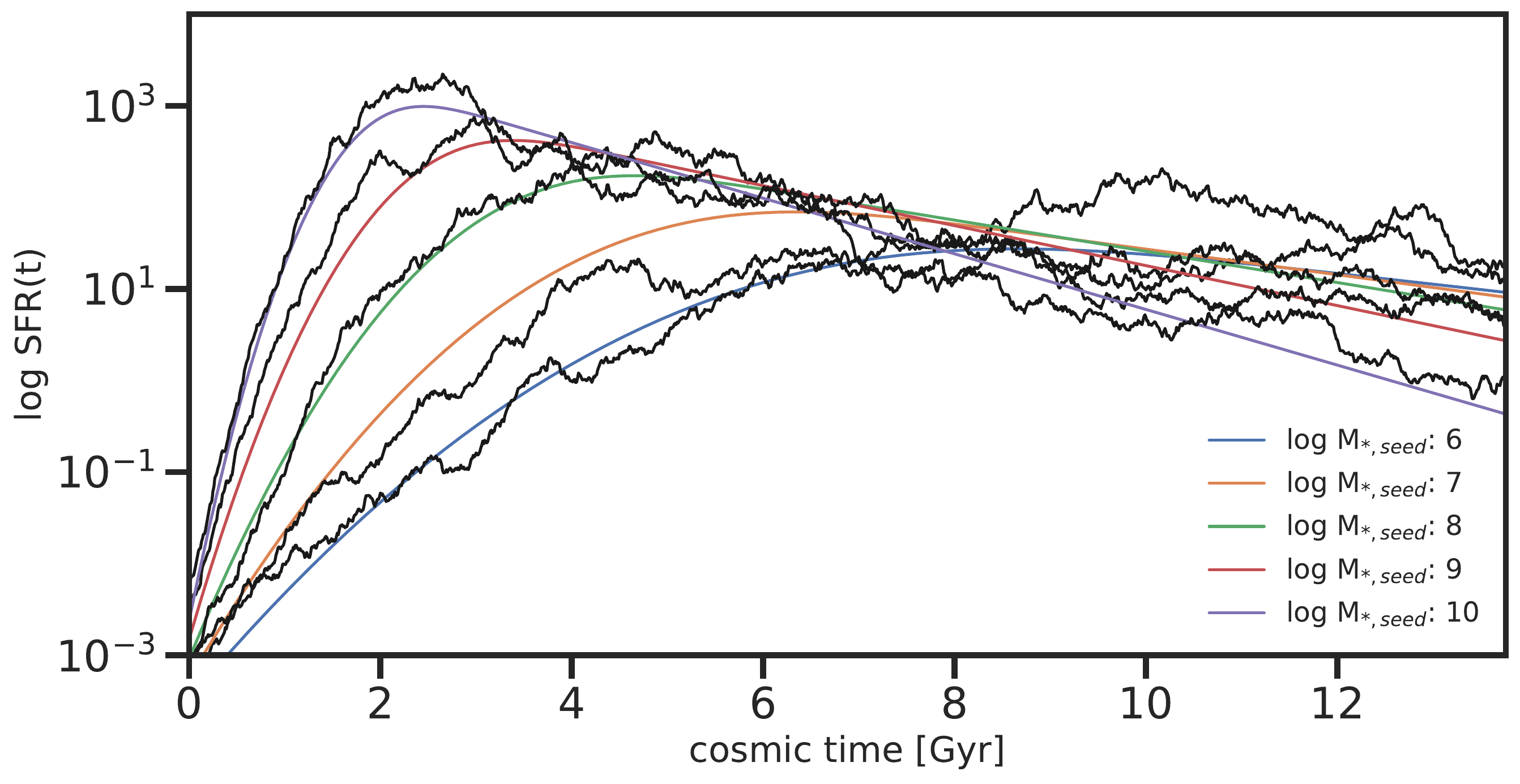}
    \caption{Generating SFHs for validation: For each test, we generate mock SFHs that follow the SFR-M$_*$ correlation from \citet{schreiber2015herschel} following the procedure in \citet{ciesla2017sfr} corresponding to different seed masses. We then realize physically motivated SFHs as perturbations around these smooth curves with a spectral slope of 2 and a scatter of $\approx 0.3$ dex. }
    \label{fig:validation_sfhs_example}
\end{figure}

Realistic SFHs are more complicated, however. From the evolution of the SFR-M$_*$ correlation and the cosmic SFRD, we know that SFHs tend to rise at high redshifts and plateau or fall at low redshifts. From our simulations, we also see that on long timescales the SFH perturbations can be described as a nontrivial power law $(\mathrm{PSD}(f) \propto f^{-2})$. We therefore consider the case where the median SFH is not stationary, generating median SFH curves for galaxies of different stellar masses using the procedure described in \citet{ciesla2017sfr}.  To this, we add perturbations of $\sim 0.3$ dex with a spectral power-law slope of 2, to create an ensemble of 10,000 mock SFHs. Examples of such SFHs are shown in Figure \ref{fig:validation_sfhs_example}. We also repeat our analysis for the cases where the power-law slope is 1-3, finding no significant difference in our results.

Using these mock SFHs, we model the effects of discrete star particles in the same way as the simulations. To do this, we discretize the mock SFH by rounding the SFR in each time bin to its nearest number of star particles, and consider the excess as the gas probability that a star particle will be formed in that time bin. Star particles are then added to each bin using a random draw with that probability.

\begin{figure}
    \centering
    \includegraphics[width=\columnwidth]{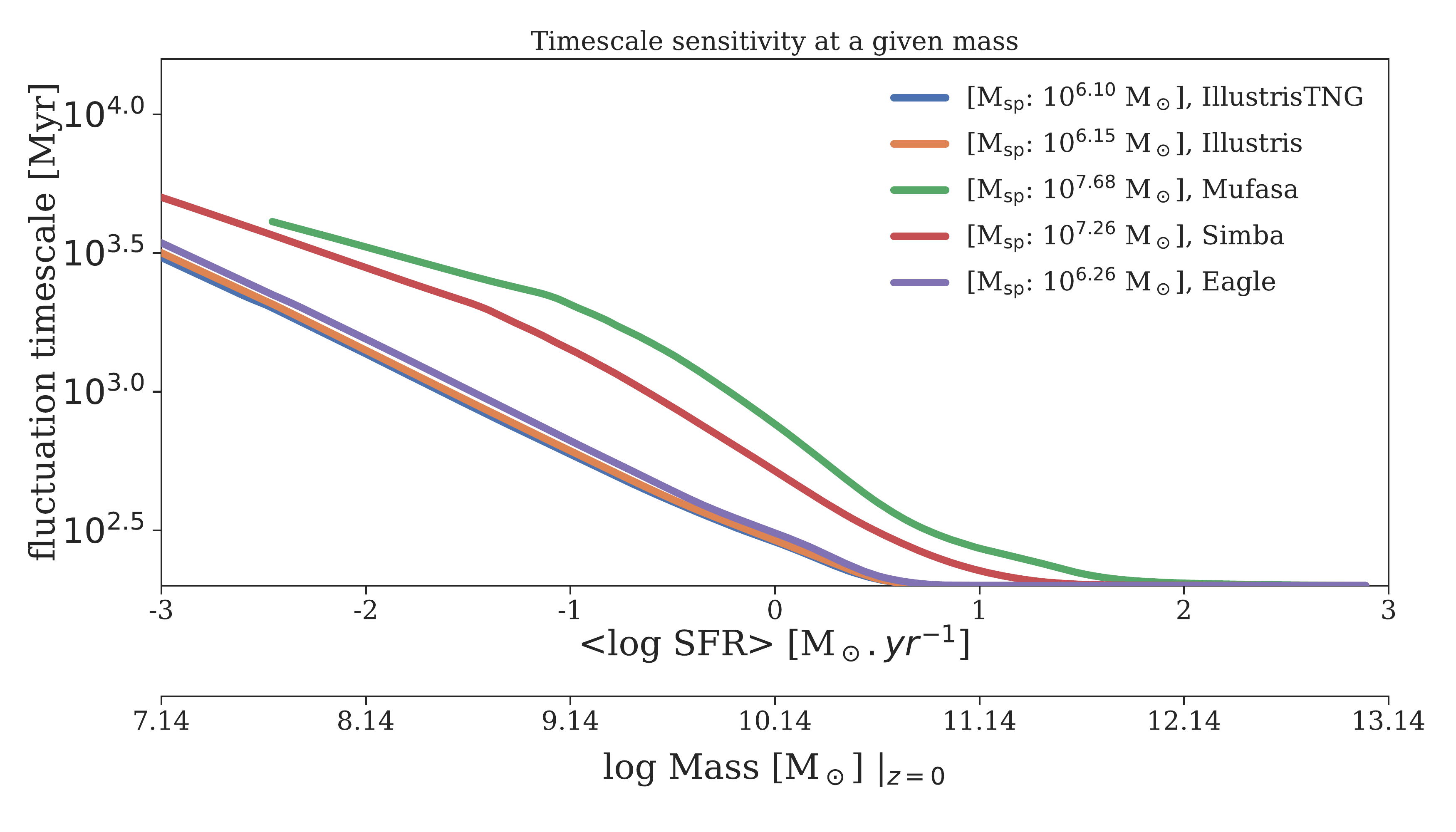}
    \caption{The lowest timescales we can probe as a function of stellar mass for galaxies from the different hydrodynamical simulations we consider. For each SFH realized using the procedure \updated{described} in \ref{fig:validation_sfhs_example}, we introduce shot noise proportional to the mass of the star particles for the different models. By comparing the pristine PSD to the PSD with shot noise, we determine the loss of sensitivity in the PSD as a function of lifetime averaged SFR and stellar mass at $z=0$.}
    \label{fig:validation_thresholds}
\end{figure}
We then compute the power spectra for these SFHs before and after the discretization procedure, and quantify the timescale at which the divergence from the original PSD exceeds a certain threshold (here 0.3 dex in PSD space). We also tried fitting the PSD corresponding to the discretized SFH with a broken power-law to quantify the timescale at which the transition from $\alpha = 2$ to white noise ($\alpha=0$) happens, and find that our results do not significantly change. Based on these numerical experiments, Figure \ref{fig:validation_thresholds} shows the thresholds for each simulation.
For all cases, the figures can be read in two ways:
\begin{itemize}
    \item Read horizontally, the figures give the minimum timescale to which we can study the PSDs for galaxies in a given stellar mass bin at a particular epoch. These have been used to set the thresholds in Figure \ref{fig:psd_hydrosims}.
    \item Read vertically, the figures give the minimum SFR (and therefore the minimum stellar mass) needed to probe a certain timescale or regime of the PSDs of galaxies.
\end{itemize}
Below these stellar masses (and timescales) the effects of discretization of the star particles begins to dominate the SFRs, and thus the PSDs.

\section{SFH diversity across models}
\label{sec:sfh_prior_dists}

\begin{figure*}
    \centering
    \includegraphics[width=\textwidth]{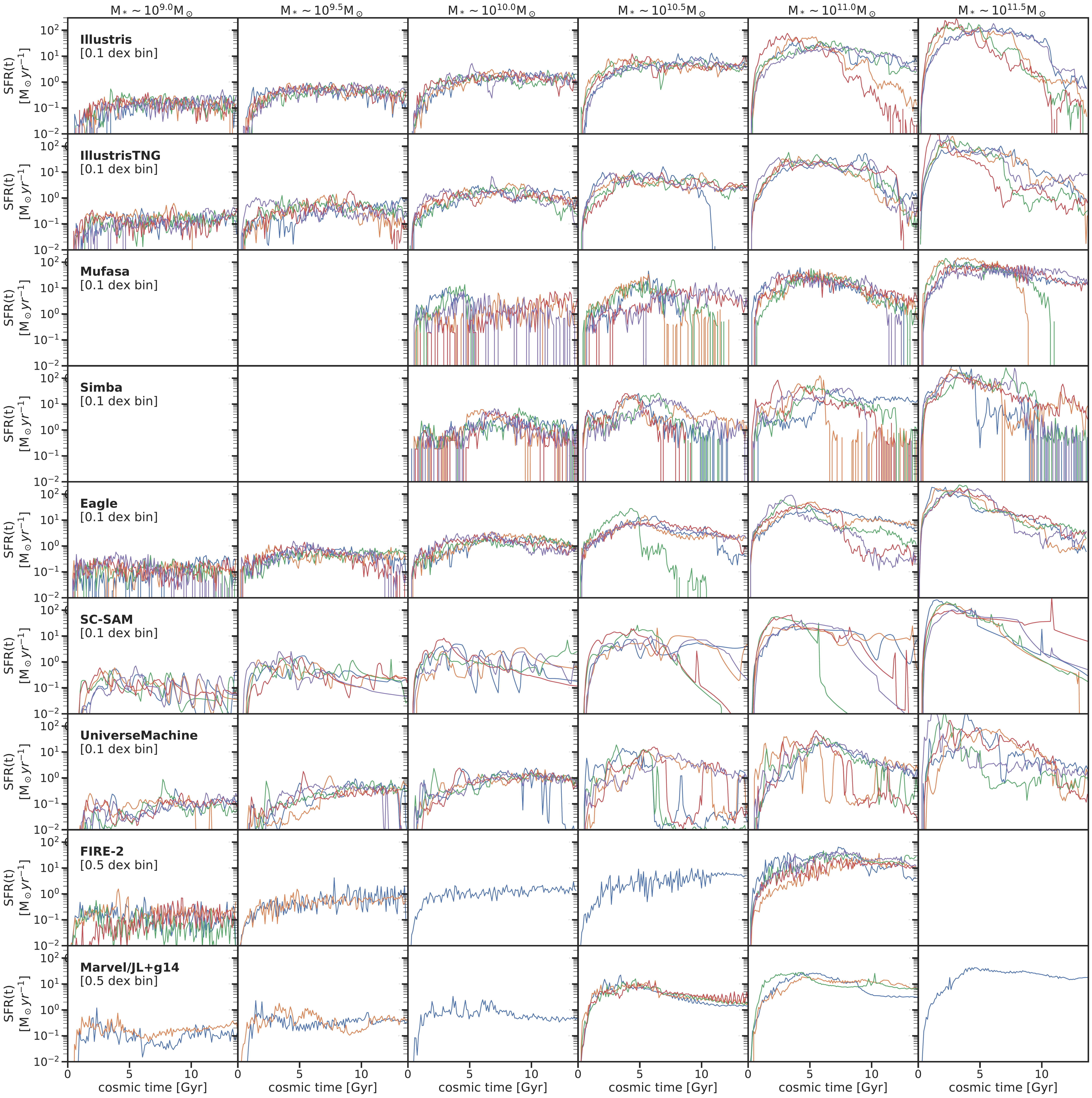}
    \caption{Representative SFHs from each model we consider across a range of stellar masses. For each model, we pick five SFHs randomly from galaxies that have stellar masses within 0.05 dex (0.25 dex for the zoom simulations) of the stellar masses reported at the top of each column. The panels display a large range of diversity in SFHs across mass, and among the different models.}
    \label{fig:sfhs_from_each_model}
\end{figure*}

\begin{figure*}
    \centering
    \includegraphics[width=0.3\textwidth]{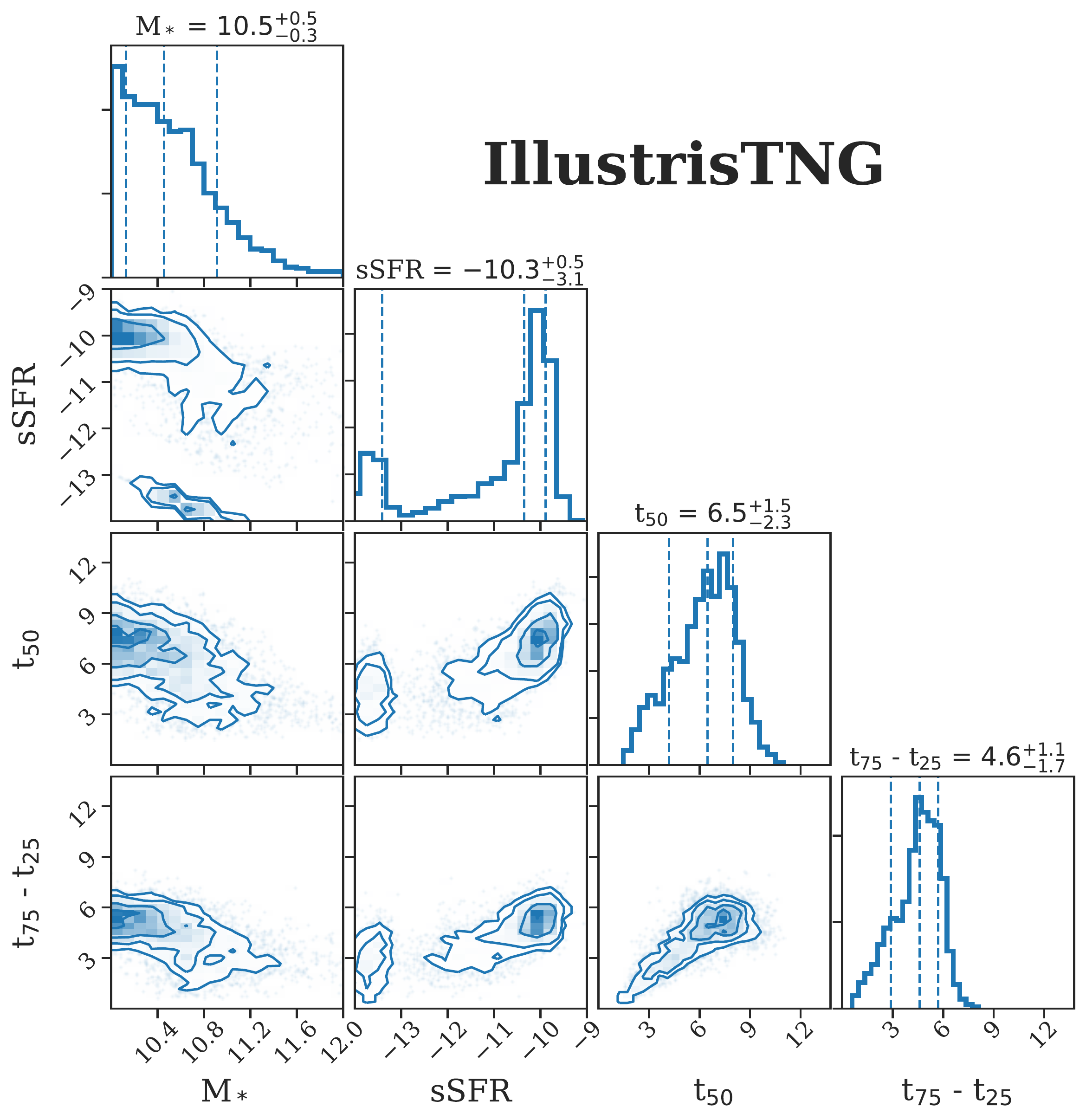}
    \includegraphics[width=0.3\textwidth]{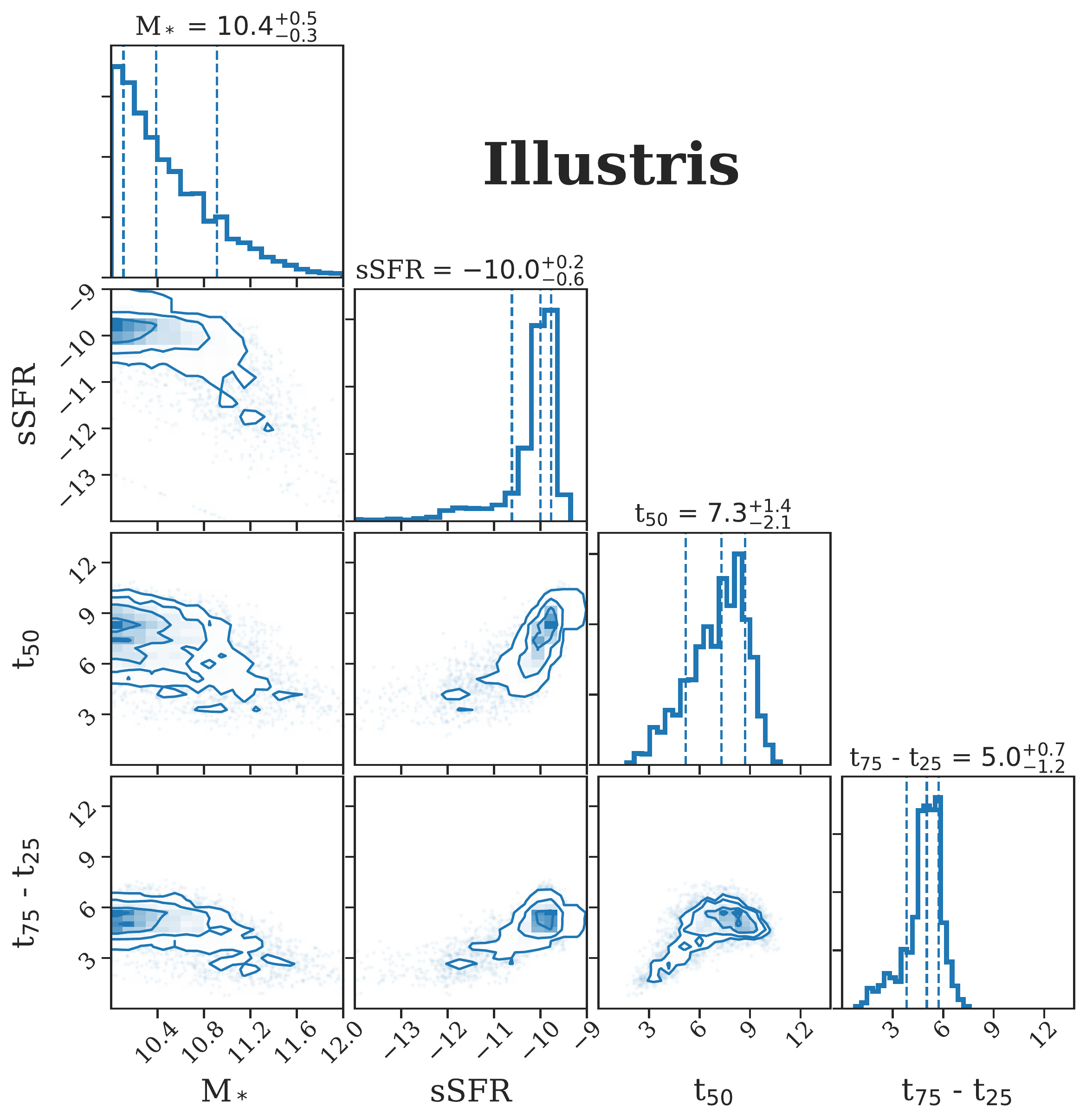}
    \includegraphics[width=0.3\textwidth]{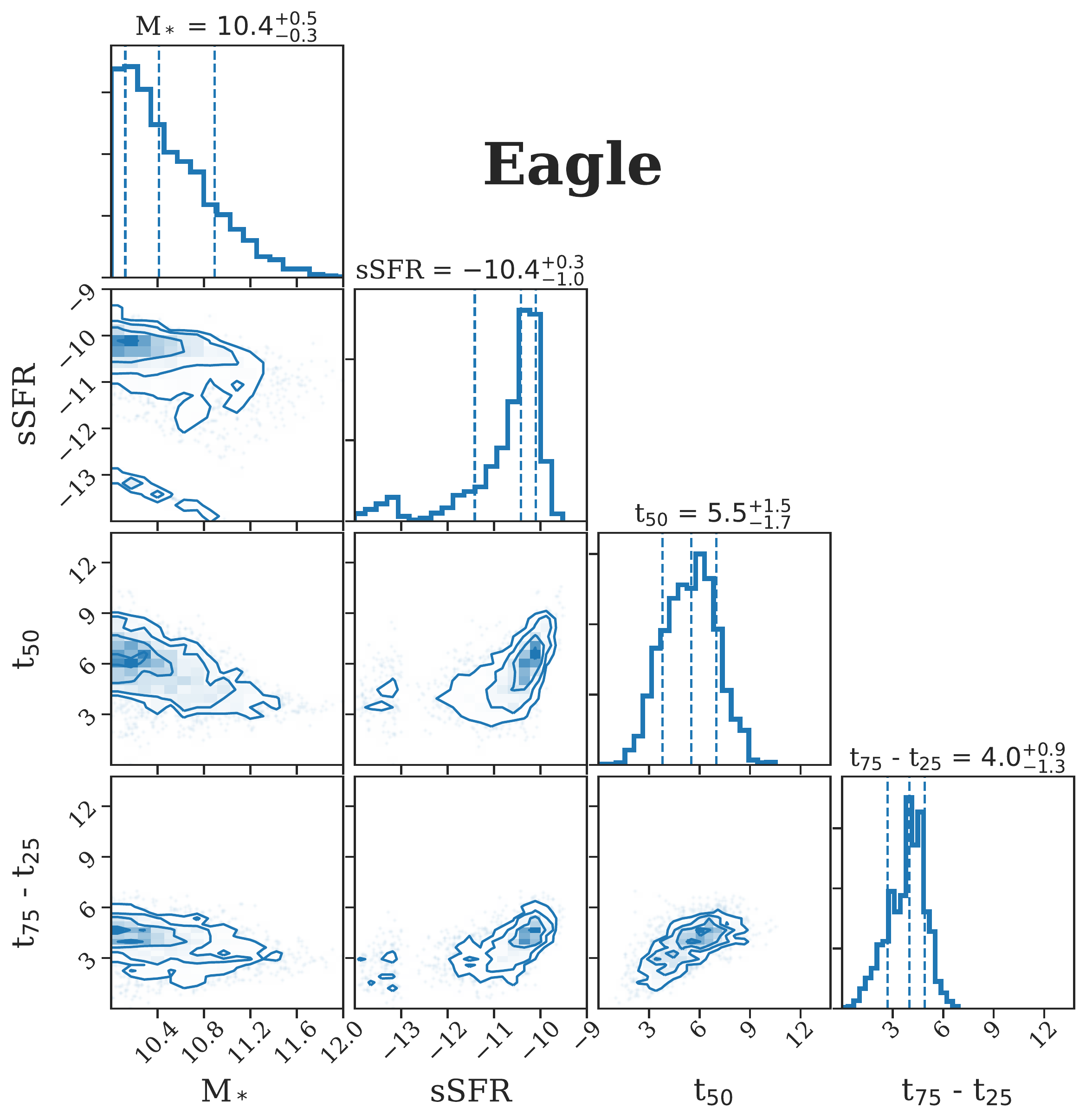}
    \includegraphics[width=0.3\textwidth]{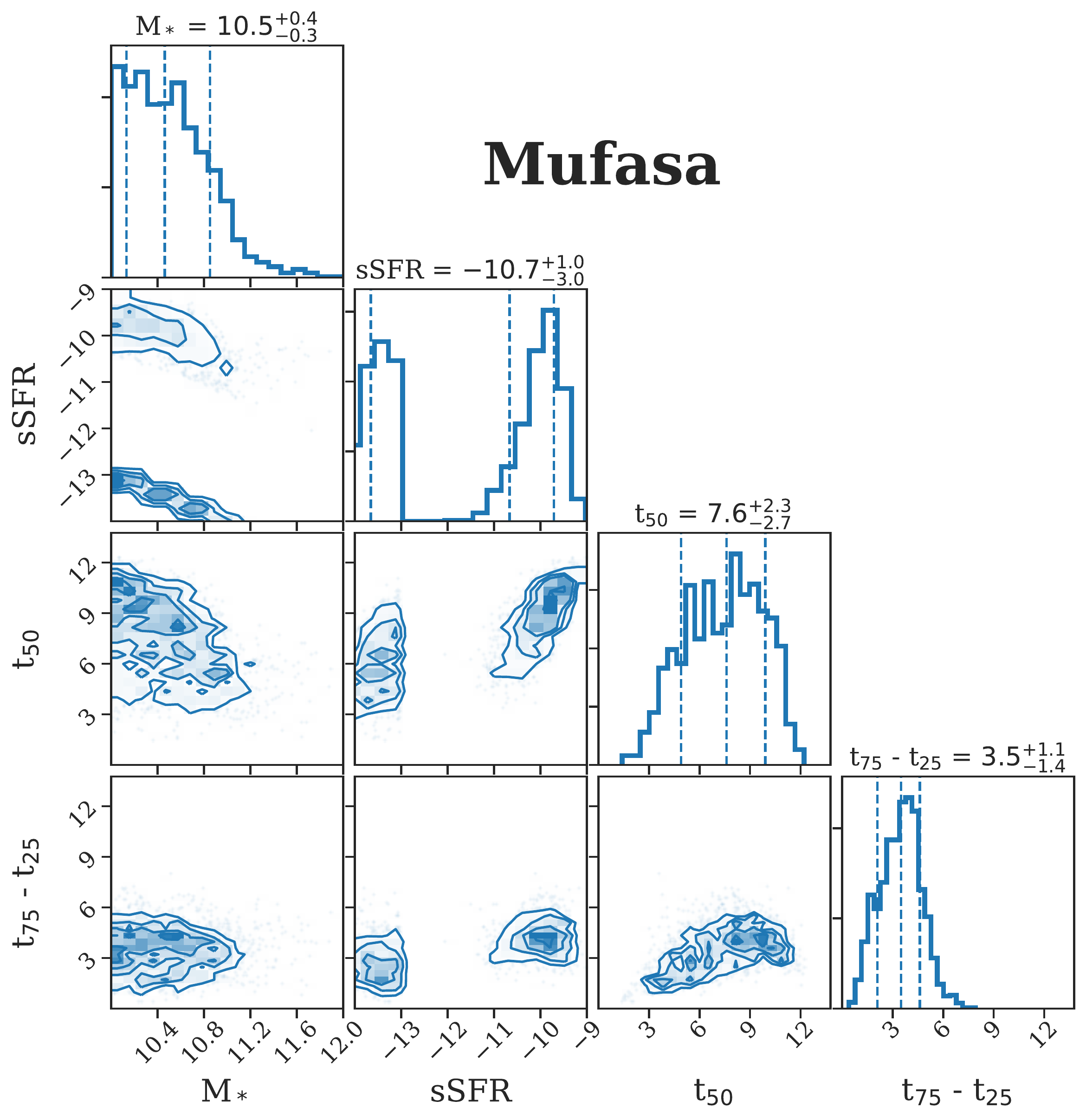}
    \includegraphics[width=0.3\textwidth]{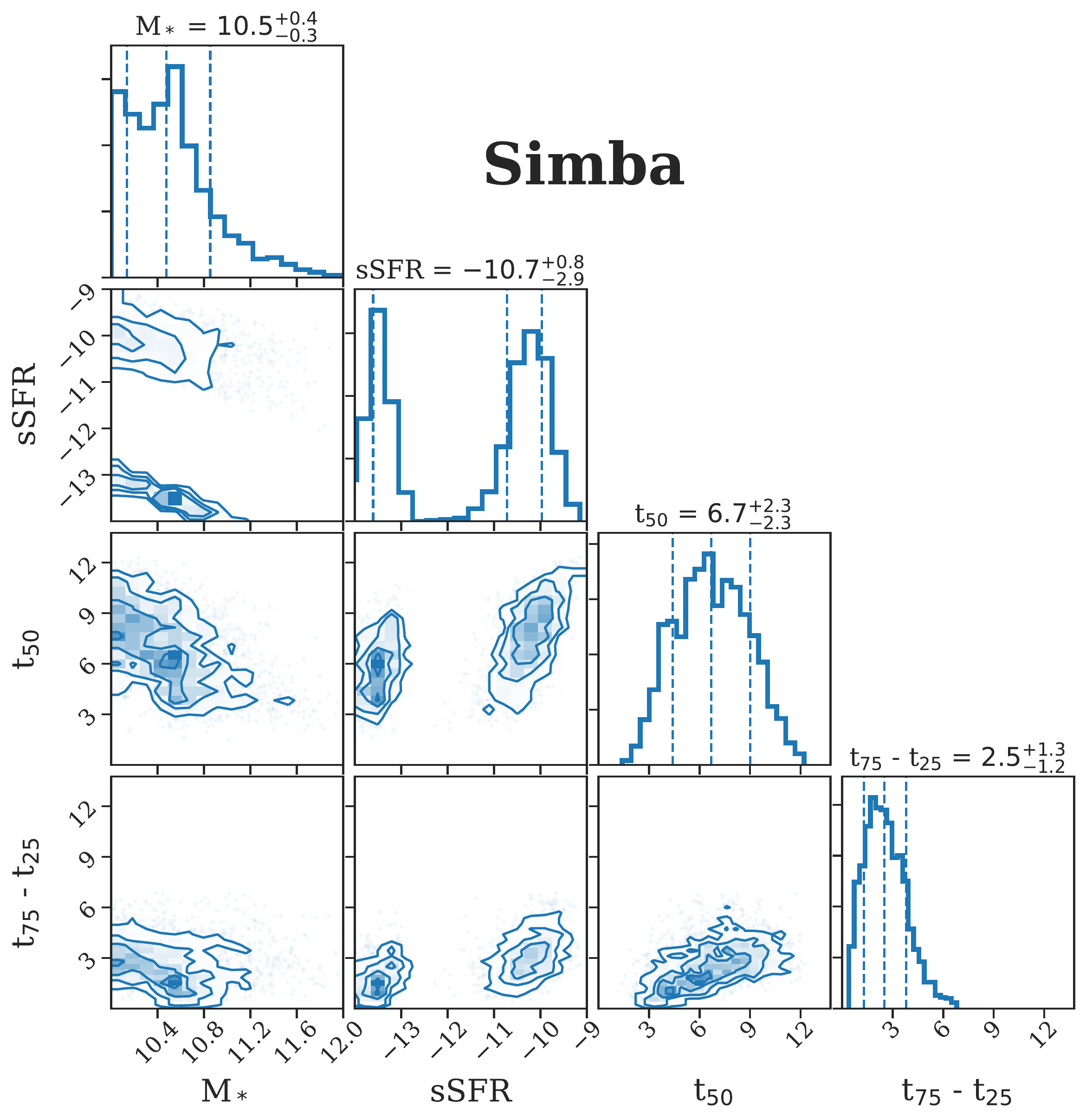}
    \includegraphics[width=0.3\textwidth]{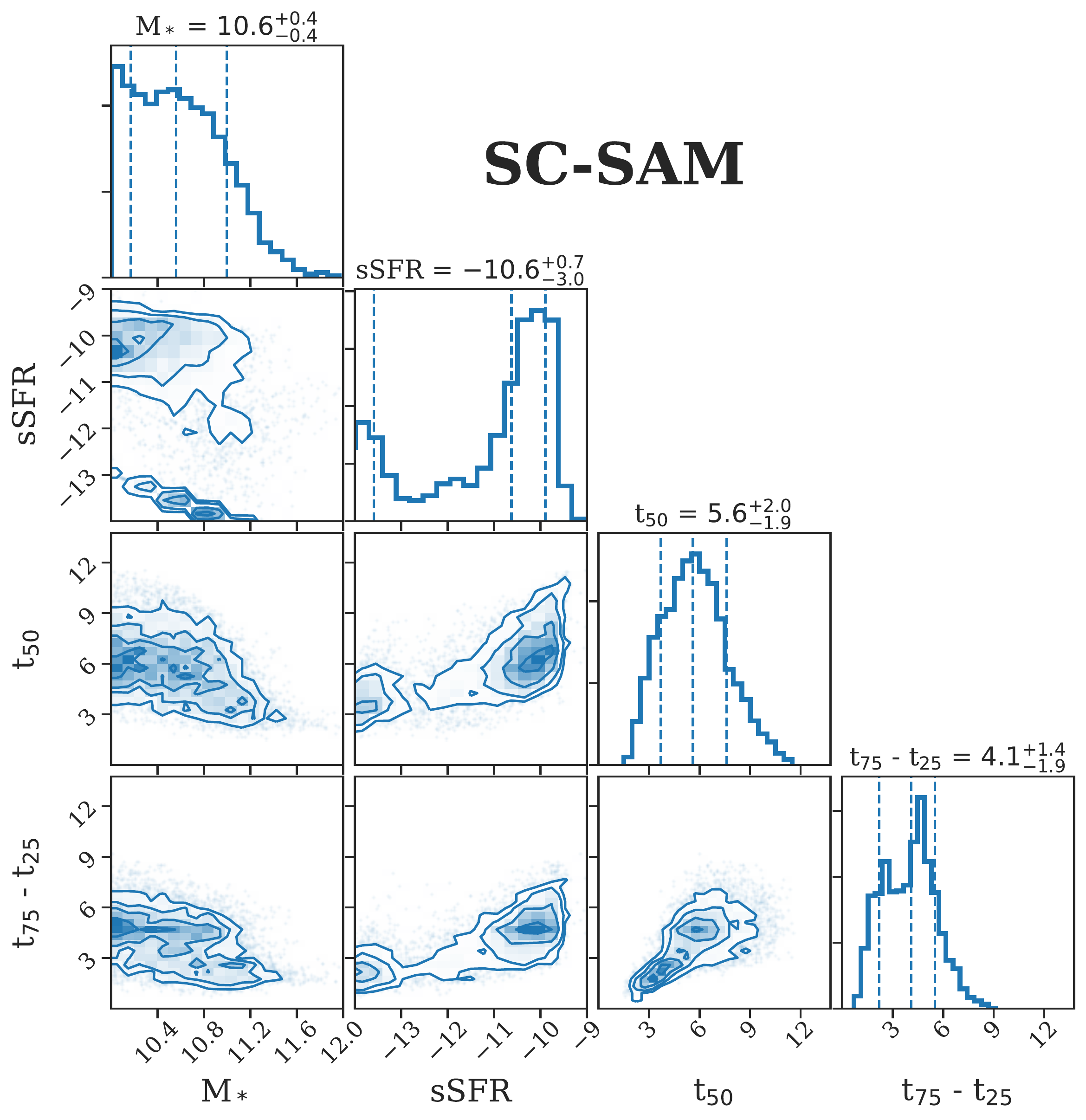}
    \includegraphics[width=0.3\textwidth]{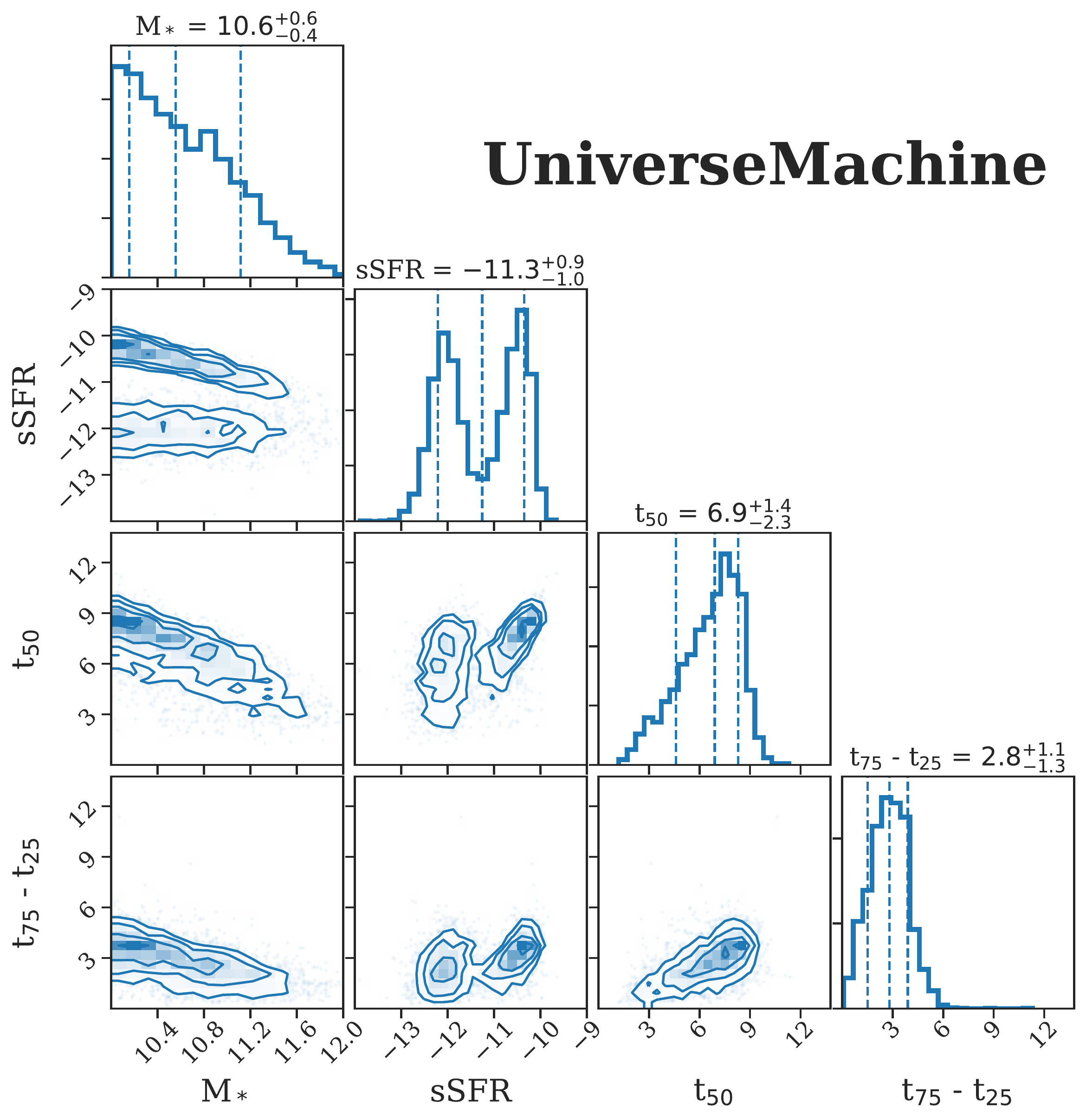}

    \caption{Distributions of SFH parameters at $z \sim 0$ for the different galaxy evolution models under consideration. The four histograms of the corner plot \citep{corner} show the distribution for log Stellar Mass ($M_*$, [M$_\odot$]), log specific star formation rate (sSFR, $yr^{-1}$), the half-mass time ($t_{50}$, [Gyr]), and the width of the galaxy's star forming period ($t_{25}- t_{75}$, [Gyr]). The remaining panels show the covariances between the different quantities. \updated{The numbers above each column show the median and 16-84$^{th}$ percentile values for each quantity across the various models}.
}
    \label{fig:sfh_priors}
\end{figure*}

Figure \ref{fig:sfhs_from_each_model} shows five randomly chosen SFHs from each model across a range of stellar masses. Mufasa and Simba are shown above $10^{10}$M$_\odot$ due to resolution limits, corresponding to the rest of the analysis in this work. As seen from their PSDs, the SFHs of these galaxies show a wide range of diversity in the strength of fluctuations on different timescales. The SFHs of lowest stellar mass galaxies from the hydrodynamical simulations often show shot noise due to discrete star particles. However, this noise mostly affects the PSDs on short timescales, which is computed in Appendix \ref{sec:timescale_tests} and accounted for while analysing their PSDs.

Figure \ref{fig:sfh_priors} shows distributions of the stellar masses, specific star formation rates, and SFH shape parameters $t_{50}$ and ($t_{75}- t_{25}$) as well as their covariances for the various large-volume galaxy evolution models we consider. $t_{50}$ is defined as the cosmic time in Gyr at which a galaxy forms half of its total mass in stars, and $t_{75}-t_{25}$ is the amount of time taken by the galaxy to go from having formed $25\%$ of its total mass to $75\%$ of its total mass. The zoom simulations do not contain enough points to robustly sample a distribution and therefore are not shown.

The mass-vs-sSFR plots show a variety of slopes for the star-forming sequence (SFS), ranging from roughly linear for IllustrisTNG, Illustris, EAGLE, and the SC-SAM to sub-linear for Mufasa, Simba and UniverseMachine. UniverseMachine in particular shows a remarkably strong quiescent population, in contrast with some models. It should be noted that the sSFR is computed using the number of star particles formed within the last 100 Myr, and might differ from the gas-based SFR. This effect is especially important for Simba and Mufasa, whose lower resolution decreases the probability that a star particle is formed in the last 100 Myr, leading to a much higher fraction of galaxies with sSFR hitting the lower boundary.

The $t_{50}$ quantifies the time in Gyr at which a galaxy formed half its total mass. A \updated{small} value for $t_{50}$ therefore indicates that the galaxy formed most of its mass at high redshifts. In conjunction with this, the ($t_{75}- t_{25}$) is the amount of time during which the galaxy formed the middle 50\% of its total mass, and serves as a proxy for the width of the period during which the galaxy was star forming. Although stellar masses and sSFR distributions among most models are similar, the distributions of $t_{50}$ and ($t_{75}- t_{25}$), which can now be observationally constrained through SED fitting \citep{iyer2019gpsfh}, vary significantly. It is interesting to note that most simulations have a tail of populations with \updated{low} $t_{50}$, usually corresponding to massive quiescent galaxies that formed most of their stellar mass in a short burst of star formation, as evidenced by the correlations between $t_{50}$ and mass, sSFR, and ($t_{75}- t_{25}$). Massive late bloomer galaxies as found by  \citet{dressler2018late} at $0.4<z<0.7$, which formed most of their mass in the last $\sim 1.5$ Gyr, thus do not feature prominently in any of these models.

\section{Estimates of timescales in the literature}
\label{sec:lit_timescales_table}

Table \ref{tab:sim_timescales} reports the estimated timescales for physical processes from current literature used in the rest of this paper and for generating the timescale ranges shown in Figure \ref{fig:literature_timescales}.

\begin{table*}
    \begin{tabular}{c|c|c}
    \hline
         Physical process & Timescale range & Reference \\
         \hline
        SNe, Cosmic rays, Photoionization from Starburst99 & $4-20$ Myr & \citet{starburst99} \\
        GMC lifetimes & $\sim 5-7$ Myr & \citet{benincasa2019live} \\
        molecular cloud formation timescale & $\mathcal{O}(10)$ Myr & \citet{dobbs2012giant, dobbs2015frequency} \\
        Turbulent crossing time & $\sim 10-30$ Myr & \citet{semenov2017physical} \\
        Free-fall time at mean density & $\sim 10-50$ Myr & \citet{semenov2017physical} \\
        Molecular cloud collision timescales & $\leq 20$ Myr & \citet{tan2000star} \\
        Cycling of ISM gas between SF regions and ISM & $\sim 20-100$ Myr & \citet{semenov2017physical} \\
        GMC lifetimes (MW-like disks) & $\leq 20$ Myr & \citet{tasker2011star} \\
        Bursty SF in TIGRESS & $\sim 45$ Myr & \citet{kim2017three} \\
        Molecular gas encounters spiral arms (MW-like)  & $\sim 50-100$ Myr & \citet{semenov2017physical} \\
        Galactic winds affecting ISM  & $\sim 50-200$ Myr & \citet{marcolini2004three} \\
        Galaxy wide gas depletion timescales & $\sim 2-10$ Gyr & \citet{semenov2017physical} \\
        Local gas depletion timescales (SF regions) & $\sim 40-500$ Myr & \citet{semenov2017physical} \\
        Exponential growth of B field  & $\sim 50-350$ Myr & \citet{pakmor2017magnetic} \\
        Merger induced starburst & $\sim 90-450$ Myr & \citet{robertson2006merger} \\
        Starburst timescale after major merger & $\sim 90 - 570$ Myr &  \citet{cox2008effect} \\
        Rapid fluctuations of inflow rates in FIRE  & $\lesssim 100$ Myr & \citet{hung2019drives} \\
        Exponential growth of B field & $\sim 100$ Myr & \citet{hanasz2004amplification} \\
        Fast quenching in Simba  & $\sim 0.01 \tau_H \approx 100$ Myr & \citet{rodriguez2019mergers}\\
        AGN feedback timescale & $\lesssim 0.2$ Gyr & \citet{kaviraj2010simple} \\
        Recycling timescale  & $\propto M_{\mathrm{halo}}^{-1/2} \sim 300$ Myr $-3$ Gyr & \citet{oppenheimer2010feedback} \\
        Crossing time & $\sim 300$ Myr $-1$ Gyr & \citet{bothun1998modern} \\
        Median recycling timescale (with large dispersion) & $\sim 350$ Myr & \citet{angles2017cosmic} \\
        Mean Depletion time  & $\sim 470-490$ Myr & \citet{tacchella2016confinement} \\
        Recycling timescale & $\propto M_{*}^{-0.19} \sim 400$ Myr $-1$ Gyr & \citet{mitra2016equilibrium} \\
        Exponential growth of B field (gas disk) & $\sim 500-800$ Myr & \citet{khoperskov2018global} \\
        Enhanced SF after merger (IllustrisTNG) & $\sim 500$ Myr & \citet{hani2020interacting} \\
        Median recycling timescale (galactic fountains)  & $\sim 500$ Myr & \citet{grand2019gas} \\
        Halo dynamical timescale & $\sim 0.1 \tau_H \approx 0.5-2$ Gyr & \citet{torrey2018similar} \\
        Morphological transformations in IllustrisTNG & $\sim 500$ Myr $-4$ Gyr & \citet{joshi2020fate} \\
        Galaxy mergers (fitting formula)  & $\sim 500$ Myr $-10$ Gyr & \citet{jiang2008fitting} \\
        Effective viscous timescale & $\approx 600$ Myr $f_g^{2/3} R_{10} V_{200}^{-1} \dot{M}_{*,100}^{2/3}$ & \citet{krumholz2010dynamics} \\
        Morphological transformation - quenching delay & $\sim 0.5$ Gyr (gas rich), $\sim 1.5$ Gyr (gas poor)& \citet{joshi2020fate} \\
        Quenching in IllustrisTNG (color-transition timescale)  & $\sim 700$ Myr $-3.8$ Gyr & \citet{nelson2018abundance}\\
        Recycling timescale (half the outflow mass)  & $\sim 1$ Gyr & \citet{christensen2016n} \\
        Slow quenching in Simba  & $\sim 0.1 \tau_H \approx 1$ Gyr & \citet{rodriguez2019mergers} \\
        Merger timescales (VELA) & $\mathcal{O}(1)$ Gyr & \citet{lotz2011major}\\
        Metallicity evolution timescale ($z\sim 0) $ & $\sim 1.8-2.2$ Gyr & \citet{torrey2018similar} \\
        Recycling times (weak feedback) & up to $\sim 3$ Gyr & \citet{ubler2014stellar} \\
        Merger timescales (dynamical friction)  & $\sim(1-10)$ Gyr & \citet{boylan2008dynamical}\\
        Oscillations around the SFMS & $\sim 0.2-0.5\tau_H \approx 2-5$ Gyr & \citet{tacchella2016confinement} \\
        Quenching in Illustris (satellites) & $\sim 2-5$ Gyr & \citet{sales2015colours} \\
        Quenching in EAGLE & $\sim 2.5-3.3$ Gyr, extending out to $\tau_H$ & \citet{wright2019quenching} \\
        Recycling times (strong feedback) & up to $\sim 11$ Gyr & \citet{ubler2014stellar} \\
    \hline
    \end{tabular}
    \caption{A summary of timescales estimated in different simulations and analytical models, assuming $\tau_{\rm H} \approx 10$ Gyr at $z\sim 0$ where timescales are reported in terms of the Hubble time.}
    \label{tab:sim_timescales}
\end{table*}

\label{lastpage}
\end{document}